# The Extreme Activity in Comet Hale-Bopp (C/1995 O1): Investigations of Extensive, Narrowband Photoelectric Photometry

David G. Schleicher
Lowell Observatory; 1400 W. Mars Hill Rd, Flagstaff, Arizona 86001; dgs@lowell.edu

Peter V. Birch
Perth Observatory

Tony L. Farnham
University of Maryland

Allison N. Bair
Lowell Observatory

**Abstract**

Conventional narrowband photoelectric photometry of Comet Hale-Bopp (1995 O1) was obtained on 99 nights from mid-1995 to early-2000, yielding gas and dust production rates over an unprecedented range of time and distance. The appearance of Hale-Bopp presented a prime opportunity for active comet studies and its inherent brightness and orbital geometry allowed the characterization of its long-term activity. Throughout the apparition Hale-Bopp released, by far, more gas and dust than any other comet ever measured. As a very high dust-to-gas ratio object, dust production was successfully measured throughout the apparition, with dust consistently slightly red in color. All five gas species including OH and NH were detected just inside of 5 AU inbound, while $C_2$ and $C_3$ were detected to just past 5 AU outbound and CN was followed until nearly 7.7 AU. Heliocentric distance dependencies ranged between -1.2 to -2.7 in log-log space, with the extremes magnified by the large extrapolations in Haser model parameters at large distances. Hale-Bopp's enormous size and associated extremely high outgassing resulted in a much larger collisional zone, which in turn yielded outflow velocities more than 2× higher than ever previously measured at comparable distances. Even so, volatile composition remained within the "typical" classification, consistent with most Oort Cloud comets, and water production follows the expected curve based on a standard water vaporization model. However, seasonal effects provided evidence for inhomogeneities among the major source regions on the surface of the nucleus. Preliminary modeling of the nucleus and coma successfully matches this seasonal behavior.

Unified Astronomy Thesaurus concepts: Comets (280); Long period comets (933); Comae (271); Comet volatiles (2162)

## Section 1 -- Introduction

Comet Hale-Bopp (C/1995 O1) was a unique, intrinsically bright object that consistently exhibited the highest gas and dust production of any measured comet except for a few rare cases, and was one of the most thoroughly studied comets in history. Despite never approaching within 1.3 AU of Earth, Hale-Bopp had an incredibly high apparent brightness near perihelion. Its brightness, combined with the long lead time between discovery and closest approach – as well the use of a new generation of instrumentation in the IR and millimeter regimes – resulted in a record number of new detections of molecular species for a comet (cf. Despois 1997; Crovisier 1997; Crovisier & Bockelee-Morvan 1999; Bockelee-Morvan et al. 2000; Bockelee-Morvan et al. 2004). Moreover, since it had already reached 11th mag when independently discovered by A. Hale and T. Bopp on 1995 July 23 at a record distance of 7.2 AU, the outgassing of the more common molecules could be followed over an unprecedented range of distances; see for instance the now famous "Christmas tree plot" by Biver et al. (2002) showing the evolution of production rates with heliocentric distance for several molecular species. Additionally, extensive imaging, both broad and narrow band, were also obtained during Hale-Bopp's lengthy apparition by multiple researchers around the world (cf. Kidger 1997; Vasundhara et al. 1997). In all, hundreds of papers and presentations, covering a wide range of topics and techniques, arose from "the Great Comet of 1997". A starting point regarding the many findings is provided in the summary from the "First International Conference on Comet Hale-Bopp" held in 1998 February (West 1997).

Given Hale-Bopp's overall importance, there is an unusually small number of published papers that utilize calibrated data obtained with narrowband filters. A major reason for this was the timing of the transition from the International Halley Watch (IHW) filters introduced prior to the arrival of 1P/Halley in the mid-1980s (and for which the UV filters in some sets had delaminated with age), to new filters designed in 1996 as replacements for use with Hale-Bopp. Calibrations for these new filters were not completed and published until 2000 (see Section 2 for more details regarding these NASA funded filter sets); as a consequence we only published our results that utilized the IHW filters for the earlier portion of its apparition, which are part of a special issue of Science magazine featuring studies conducted prior to perihelion (Schleicher et al. 1997). These measurements (and other reports in the special issue) highlighted the extreme nature of H-B as compared even to the most active comet previously studied – Halley itself. Note that in this same time frame, ESA was also developing new sets of comet filters in support of their ROSETTA mission planned for Comet 46P/Wirtanen (Schulz & Schwehm 1996), but apparently these were not yet available in time for use with H-B.

In the following quarter century, our analyses of Hale-Bopp (H-B) mainly concentrated on the imaging we obtained, with studies focusing on the bulk spatial distribution of gas and dust in the coma near perihelion (1997 April 1), in addition to its detailed spiral jet morphology at that time, which yielded H-B's rotational period. Images also exhibited multiple radial jets in 1996 and again in late 1997 and early 1998. In addition to a paper in the special issue of "Earth, Moon, and Planets" associated with the 1998 International Conference on Comet Hale-Bopp (Lederer et al.

1997/99), many of these findings were presented at various meetings over several years (cf. Farnham et al. 1997; Schleicher et al. 1999a; Farnham & Schleicher 2002; Schleicher et al. 2004). Earlier efforts to formally publish our many H-B results, however, were repeatedly delayed, first waiting for our photometry observations to be completed (as we continued to follow H-B to as large a distance as feasible in 1999/2000), needing to finish the new filter calibration and applying it to this large data set, and ultimately due to funding limitations with H-B having become "old news" and various new projects that became higher priorities. Most recently, updated results from our efforts associated with the narrowband photometry for H-B were given by Bair et al. (2018).

This paper presents all 332 sets of observations of Comet Hale-Bopp obtained by us with traditional photoelectric photometers at Lowell and Perth Observatories from 1995 to 2000, beginning only two nights following its discovery. Section 2 summarizes the observations and reductions. Results are presented in Section 3, including an overview, then sub-sections on gas, dust, and evidence for outbursts, and then concluding with water production and the results of preliminary nucleus modeling and the causes of some seasonal behaviors that were observed. In Section 4 we discuss the implications of these many results, and place H-B in a broader context. Note that upcoming papers will explore in detail the gas and dust jet morphology, the associated rotational period determination, along with our detailed model and our final model solution.

**Section 2 – Observations, Reductions, and Methodologies**

*2.1 Instrumentation*

All photometric data reported in this paper were obtained with one of three telescopes: the John S. Hall 42-in (1.1-m) and the 31-in (0.8-m) telescopes located at Lowell Observatory in Flagstaff, AZ, and the Lowell 24-in (0.6-m) telescope at Perth Observatory in Perth, Australia. Conventional photoelectric photometers with EMI 6256 photomultiplier tubes were employed, one using pulse counting electronics for all of the northern hemisphere observations, and a similar one but with DC electronics at Perth Observatory. Each photometer had an aperture wheel with a wide range of aperture sizes, many of which were used due to Hale-Bopp's high brightness. Note that the 42-in had swappable secondary mirrors during this era, and that the first night of observations were obtained at f/16 while all other nights with the 42-in used the f/8 secondary; the projected aperture sizes were computed appropriately.

As noted in the Introduction, new narrowband comet filters were designed and produced during the apparition of Hale-Bopp, for a variety of reasons. These included the development of larger sized CCDs since the creation of the International Halley Watch (IHW) filter sets in the early 1980s that in turn required larger filters, the degradation of some of the IHW UV filters due to manufacturing limitations in the early 1980s, and many researchers wanting to observe Hale-Bopp who had not been in the field at the time of Comet 1P/Halley. In spite of these issues, funding for new filters was uncertain until Comet Hyakutake (C/1996 B2) was discovered in late January of 1996 and it was determined that it would make a very close passage past Earth less than two months later. While it was impossible to obtain filters on such short notice, this event prompted NASA to move ahead with funding filters for Hale-Bopp, with author Schleicher leading the design and ordering process that resulted in the first new filters arriving in 1996

December, and all filters arriving during the following two months. The key point for this narrative is that only IHW filters were available for the first 17 months of the apparition, but even after the new HB filters (no hyphen to distinguish the filter set from the comet) arrived, the IHW filters were used in tandem throughout the perihelion time frame. This latter choice was due to a very practical aspect – the new filters were not yet calibrated, and would take more than a year to do so because of the need to measure all of the standard stars, and yet we wanted to continue to rapidly measure, reduce, and disseminate results. The specific filters used each night depended on which new filters had been delivered, the number of filter slots available, and the length of the observing window each night because H-B remained close to the Sun in early 1997. In all, we used both the standard IHW complement of five gas filters – OH, NH, CN, $C_3$, and $C_2$ – and two continuum filters – at 3650 Å and 4845 Å – along with the HB set of the same five gas species and three continuum filters – at 3448 Å, 4450 Å, and 5260 Å. The ion filters and the red continuum filter were not utilized from either set due to the lack of red sensitivity of the phototubes. Finally, note that the filter situation was made simpler for observations from Perth because Hale-Bopp was unavailable from 1996 November to 1997 July as it was too far north; IHW filters were used prior to this break while only HB filters were used afterwards. Finally, due to Perth's low elevation and correspondingly high atmospheric extinction in the UV, neither the OH nor the NH filters were used (from either filter set), nor was the 3448 Å continuum filter (in favor of the 4450 Å continuum filter).

*2.2 Observations*

A single observational set consists of a series of measurements of each available filter (from either the IHW or HB set) for a particular photometer entrance aperture. The unusually large range of aperture sizes (from 20 to 155 arcsec diameter) was due to both the large range of distances of H-B along with its high brightness near perihelion; up to five apertures were used on a single night. In total, 332 data sets were obtained over 99 nights, 211 from Lowell and 121 from Perth, with many additional scheduled nights weathered out. The first observation was taken on 1995 July 25, only 2 days following H-B's discovery and at a distance of 7.14 AU while the final data sets were obtained on 2000 March 3 at 10.56 AU. The observational circumstances on each night are summarized in Table 1, which includes the time from perihelion ($\Delta T$), the heliocentric and geocentric distances, $r_H$ and $\Delta$, respectively, and phase angle, $\theta$. The distribution of the nights of photometry throughout the apparition are shown in Figure 1.

**Table 1 – Observing Circumstances**

**Figure 1 – the Distribution of Observations**

Both early and late in Hale-Bopp's apparition the observing methodology followed our normal practices, but its high brightness near perihelion required several adjustments. First, and perhaps most obviously, sky measurements that are normally acquired at an offset of about ¼ to ½ degree from the comet, needed to be made up to a few degrees away. Second, even using small apertures H-B was much brighter than any of our standard stars during March through May 1997, resulting in the need to apply significant corrections for "dead-time", where pulse-counting systems miss an increasing number of photons because they arrive too close together. Because the brighter of our comet standard stars (4[th] and 5[th] magnitude) already require a 1-2% correction for dead-time, the coefficient for the Lowell photometer had already been determined to have a value of $7.0\times10^{-8}$ s, resulting in adjustments by 3.5% at 500k s$^{-1}$, and 14% at 2m s$^{-1}$, our highest measurement. Third, a highly unusual problem was encountered at H-B's brightest, with count levels plummeting from expected values and initially requiring us to avoid using the larger aperture sizes. Ultimately, we realized that the counter had rolled over, and once we confirmed that the value at which this occurred was indeed constant, we were able to simply add this amount to the recorded value, and then apply the deadtime correction.

*2.3 Reductions*

Flux standard stars associated with the appropriate filter sets were measured nightly to determine extinction coefficients and absolute calibrations. These determinations were, in turn, used to convert count rates of each gas species for H-B into emission band fluxes following continuum subtraction, and ultimately into column densities, $M(\rho)$, and production rates, $Q$. For each continuum filter the flux per Ångstrom was then converted into the now-standard quantity $A(\theta)f\rho$, a proxy of dust production first introduced by A'Hearn et al. (1984). This product of the albedo of the dust grains at a given phase angle, the filling factor, and the projected aperture radius, is aperture and color independent if the grains are grey in color and follow a canonical $1/\rho$ fall-off in their spatial distribution.

As usual, we used the standard procedures detailed in A'Hearn et al. (1995). Having utilized two different filter sets, the initial reductions to fluxes needed differing coefficients, and these were given for the HB filters by Farnham et al. (2000) and revised calibration for the IHW filters by Farnham & Schleicher (2005); in each case new modeling of the filter profiles combined with cometary spectra provided improved decontamination of the continuum filters. Fluorescence efficiencies (L/N) for each gas species were applied to compute the number of molecules within the aperture. For the three heterogeneous molecules the Swings effect significantly changes L/N as a function of the Doppler shift with respect to the Sun; these nightly values are given in Table 1 (see Schleicher & Bair 2011 for further details). In contrast, the homogeneous species have too many rotational lines to produce a significant Swings effect, and we continue to use the same values for $C_2$ and $C_3$ as given by A'Hearn et al. (1995). The reduced fluxes and final column densities for all observations are listed in Table 2. The further computation of production rates (Q) first applies the Haser model to extrapolate the column abundance for each species to a total number of molecules in the coma. We then divide by the lifetime of each daughter species to determine the equilibrium production rate; the specific scalelengths and lifetimes again are from A'Hearn et al., and the resulting production rates are presented in Table 3.

**Table 2 – Photometric Fluxes and Column Abundances**

**Table 3 – Production Rates and $Af\rho$, and Q(H$_2$0)**

Most comets in our database exhibit little or no variation in the derived production rates as a function of the photometer aperture size, except for our photometry of Comet Halley where large rotational variations resulted in strong aperture effects during the rotational cycle (Schleicher & Millis 1989). This general lack of a trend with aperture provides confidence that using the Haser model and our adopted scalelengths to extrapolate to the total coma abundance from the column abundance is valid. In the case of H-B, however, while much of the data behaves in a similar manner there are two regimes where this breaks down. The most prominent instance was during the first half of 1997, in particular in the months closest to perihelion. Large trends with aperture size were evident for all species and generally in the same direction – systematically larger Qs were calculated for larger aperture sizes – directly implying that the standard values for the scalelengths were no longer appropriate and larger values were needed. Consistent with this, the outward motion of the spiral jets we observed in our narrowband imaging of the gas species revealed velocities significantly greater than canonical values, again indicating that the scalelengths needed to be increased. Finally, our computed Qs during this interval were systematically lower than those reported in the mm and radio, whereas the results from various techniques were in good agreement earlier in the apparition.

The cause for all of this is clear: at very high production rates, the size of the collision zone in a comet's coma is much larger than normal and this, in turn, results in a corresponding increase in the size of the region of gas acceleration and a higher net outflow velocity (cf. Combi 2002). We initially hoped to determine adjustments to the scalelengths used with the Haser model but several issues prevented this. First, it became evident that each observed species required different degrees of adjustment as a function of production rate, because of large differences in parent scalelengths or lifetimes, and thus each daughter species was formed at varying characteristic distances from the nucleus, and differing collisional densities. For instance, the parents of OH and $C_2$ dissociate at much larger distances than the parents of $C_3$ and CN, altering how they compare to the basic assumptions regarding the Haser model. Second, the short viewing window each night in late 1996 and most of early 1997 generally prevented us from obtaining data over a wide range of aperture sizes, for which we might have tested possible adjustments to the scalelengths. Third, while we also had narrowband imaging, most of the imaging was obtained under non-photometric conditions, making accurate continuum subtraction – needed because of the high dust-to-gas ratio in H-B – very difficult to achieve. Ultimately, despite several attempts to compute appropriate scalelength adjustments throughout this time range (cf. Schleicher et al. 1999b; Baugh & Schleicher 2003), we abandoned this approach of obtaining improved production rates. As an alternative, we returned to spatial profiles extracted from wide-field imaging that we obtained on a few photometric nights near perihelion. Based on these radial profiles, we were able to estimate the maximum adjustment to the production rates themselves at the comet's peak activity. These values and estimates of production rate adjustments needed are discussed in Section 3.2.

Separately, because $H_2O$ is the undisputable parent of OH, and water is the dominant volatile species in comets, in addition to water being the standard comparison species in other comet studies, we have computed a vectorial equivalent of the water production rate throughout the apparition starting with the Haser $Q$(OH) values. This empirical relationship, $Q(H_2O) = 1.361\ r_H^{-0.5}\ Q(OH)$, was derived by Cochran & Schleicher (1993) and includes the branching ratio, the $r_H$-dependence of the water velocity, and the distance at which the two models yield equivalent production rates, and was discussed in Schleicher et al. (1998) regarding Comet 1P/Halley. As usual, we use this relationship here for H-B to compute water production rates, but the much larger collisional zone near perihelion again causes complications. However, unlike for the daughter species discussed above, we have recently been successful in utilizing data from several high $Q$ comets in our database to determine an additional empirical relation for making a first-order adjustment in high production rate situations to our standard $Q(H_2O)$ calculation:

$$\Delta \log Q(H_2O) = [\log(Q(OH) / r_H^2) - 28.6] / 6.$$

We defer a detailed discussion of this relation to an appendix in our forthcoming paper on our entire database (Schleicher & Bair, in progress), but simply note here that the value of the denominator, 6, is known to no better than a single digit, and corrections are only made when $\log(Q(OH)/r_H^2)$ is greater than 28.6. Also note that by applying a correction to the water production rates computed in the standard manner (rather than to the scalelengths themselves), we greatly simplify the process. The final, adjusted values of $Q(H_2O)$ are listed in the last column of Table 3.

Turning to the dust, as previously noted we continue to compute the now standard proxy for dust production, $A(\theta)f\rho$, as first defined by A'Hearn et al. (1984), which is the product of the dust albedo at a given phase angle, the filling factor for the aperture, and the projected aperture radius. In the canonical situation, with constant dust production and the dust moving radially outward from the nucleus at a constant velocity, yielding a $1/\rho$ radial distribution, the product is independent of aperture size and wavelength if the dust is grey in color. Thus, deviations from this ideal scenario reveal properties of the dust grains. With a range of phase angles ($\theta$) from 3° to 49° (see Figure 1), we also utilize our composite phase curve, first introduced by Schleicher & Bair (2011) based on a derived function from our Halley observations (Schleicher et al. 1998) combined with higher phase angle calculations by Marcus (2007), to normalize dust measurements to 0° phase angle. Note that this composite phase curve was consistent with measurements of Comet Lulin (2007 N3) between 0° and 51° (Bair et al. 2018), nearly identical to the range of H-B's phase angles. This phase angle adjustment factor is given for each night of observations in Table 1.

In the case of Lowell observations, the uncertainties listed in Table 3 associated with each data point are based on photon statistics and reflect the 1σ values derived from the propagation of the observational uncertainties. Due to H-B's brightness, these formal uncertainties are generally quite small except at large heliocentric distances. However, because DC rather than pulse counting electronics were used for the Perth observations, there are no photon statistics and uncertainties were only estimated based the brightness of the comet in each filter, but these values ultimately proved much too small. In the case of the dust, actual uncertainties are probably 2-3× the listed values, while the gas uncertainties should be increased by even greater factors, due to large amounts of continuum subtractions, and even complications such as the higher sky brightness over the City of Perth when H-B was low in the west. Ultimately, we consider that the scatter in the final results on any given night, as evident in the figures, provides a much better estimate of the true uncertainties.

**Section 3 – Photometric Results for Comet Hale-Bopp**

*3.1 Overview*

In Figure 2 we provide the grand summary of all of our measurements for Hale-Bopp over the entire 4 and 2/3 years of its apparition. Logarithmic values of the production rates for the five gas species, along with logarithmic values of $A(0°)f\rho$ as derived from the green continuum filter (either at 4845Å or 5260Å, depending on which filter set was used), are plotted as a function of time from perihelion. Observations from Lowell are shown as circles while those from Perth Observatory are indicated with triangles (from which no measurements of OH and NH were obtained), and measurements obtained prior to perihelion are indicated by open symbols while those from after are filled. Finally, we color-code the data based on the projected aperture size, $\rho$, specifically by factors of two for each color step using the logarithm of $\rho$ in kilometers as given in the key, to show trends or lack of trends with aperture size at different times during the apparition.

**Figure 2 – Qs & $A(0°)f\rho$ vs ΔT**

Perhaps unsurprising given the extreme duration of observations and range of distances, production rates changed by nearly 2 orders of magnitude or more, varying with the interval over which each species was successfully measured. More interestingly, the absolute production rates in H-B surpassed any other comet at comparable distances with just one exception – Comet 17P/Holmes during the first weeks of its extraordinary outburst by more than 14mag (or ~500,000×) in 2007 while at a distance of 2.4 AU (Schleicher 2009). Regarding non-outbursting objects, the next most productive comet that we have measured in over 45 years was Comet 1P/Halley (Schleicher et al. 1998), and H-B gas production was about 20-fold greater at any given distance while dust was more than 100-fold greater; a direct comparison of the pre-perihelion measurements of the two comets was shown in Schleicher et al. (1997). It is thus clear that in addition to being the brightest comet in several decades, Hale-Bopp was truly exceptional in its levels of activity.

While the general quality of our measurements is quite good, resulting in formal uncertainties much smaller than the size of the data points for much of the apparition, it's quite apparent that there is considerable scatter at certain intervals among the observations. The primary interval takes place in the few months around perihelion and is completely due to the previously discussed aperture effects at very high production rates. While we defer detailed discussion until later, note that all five gas species suffer from this though by differing degrees, but that aperture effects on the derived values of $Af\rho$ are unexpectedly affected by a significantly *smaller* amount (see aperture size color-coding in Figure 2). The next pair of intervals showing significant scatter were obtained more than a year before, and more than a year after H-B reached perihelion, corresponding to distances greater than approximately 5 AU. Lower signal-to-noise (S/N) with larger $r_H$ is a major cause for this, especially in the case of the dust. However, the situation is compounded for the gas production rates. These more uncertain underlying dust continuum values must first be removed from the gas filter measurements, resulting in even greater uncertainties associated with the remaining gas emissions. We note that all dust measurements with the green continuum filters were positive while, following continuum subtraction, many of the net gas values at large $r_H$ became negative and thus are listed as undefined in Table 3 and shown by the downward arrows near the bottom of each panel in the Figure 2. Overall, we consider the apparent scatter at large $r_H$ provides a better estimate of the uncertainties than do the formal photometric uncertainties (listed in Table 3), for the reasons presented at the end of Section 2. Finally, the last and smallest interval for which there is significant scatter is during a few weeks in late May and early June 1996, between 4.3 and 4.1 AU. Not only do production rates vary from night to night, but we detect variable aperture effects, leading us to attribute these behaviors to at least two outbursts taking place that we will return to in Section 3.4.

We conclude this overview of our data by briefly examining the pre- and post-perihelion asymmetry (or lack thereof) in production rates. Unexpectedly, there was considerable variation among the three carbon-bearing species, with CN exhibiting the greatest asymmetry with clearly higher values inbound, while there is a lesser amount for $C_3$ and almost no asymmetry for $C_2$. Unfortunately, we have no information of our own regarding OH and NH due to the lack of data from Perth, but water observations by other investigators indicate asymmetries most similar to CN. Dust, normalized for phase angle, also shows a relatively small amount of asymmetry; further details regarding both gas and dust will be presented in the next two subsections, respectively.

### *3.2 Gas Production as Functions of Time and Distance*

In Figure 3, we present the logarithm of the production rates of each gas species as a function of the logarithm of the heliocentric distance ($r_H$). As in Figure 2, observations obtained from Perth Observatory are shown as triangles, and measurements taken prior to perihelion are given as open symbols while post-perihelion data use filled symbols. Error bars are always presented but are usually smaller than the point size except at large distances. Note, however, that the uncertainties presented on the Perth data are thought to be underestimates, especially for the gas species. While all data were shown in Figure 2, including cases given as downward arrows when the result was negative following sky and continuum subtraction, here in Figure 3 we have removed gas measurements at large distances that we do not think are meaningful. It is evident that by 1998 October, when H-B reached 6.7 AU outbound, that both $C_2$ and $C_3$ yielded negative values nearly as often as positive results and were dominated by noise, whereas the cut-off for CN was about 1 AU further out between February and March of 1999. We have correspondingly removed from this and later figures data points beyond these cut-offs. Inbound, an additional factor makes many of the earliest observations also very suspect – high airmass and near-twilight conditions throughout 1995 and into 1996 March from both sites. For Perth, there is clearly excessive scatter and/or negative gas fluxes from 1995 September through 1996 February, while the more limited Lowell data, including our very first night in 1995 July and the two sets in 1996 February/March lack consistency. Ultimately, we consider only observations beginning in 1996 April, i.e. inside of 5 AU, as being useful for further analyses. (Finally note that no observations of H-B were attempted in later March due to the very close encounter of Comet Hyakutake (1996 B2) on which all observing time was focused.)

**Figure 3 – Qs & $A(0°)f\rho$ vs log $r_H$**

As already noted regarding the $\Delta T$ plots from Figure 2, the production rates for each carbon-bearing species are generally very symmetric on either side of perihelion; again, neither OH nor NH were measured from Perth, so we have no direct information regarding possible asymmetries from our data. Perhaps more remarkable, the $r_H$-dependencies of each gas species are quite linear over the region with good S/N, excepting the near-perihelion regime with large aperture effects. In fact, the only exception to this is seen in the CN measurements between about 5 and 3 AU inbound, that are systematically higher than the overall trend line. The fitted log-log slopes for each species are listed in Table 4. To avoid the large aperture effects near perihelion, measurements from <1.2 AU were not included in these fits, and all data obtained at >4.0 AU were excluded due to generally much greater scatter, primarily because of the large amount of continuum subtraction in this very high dust-to-gas ratio comet (see Section 3.3). Note that because of our 1.2 AU cut-off, coupled with H-B's conjunction with the Sun shortly after perihelion, no post-perihelion data were included in the linear fits until H-B was beyond 2.16 AU.

**Table 4 – $r_H$ -Dependence of Production Rates**

As can be seen from the tabulated values, CN and $C_3$ are near the middle of the range of slopes for H-B before perihelion with values near -1.8. The slope for OH is steeper at -2.4 while $C_2$ is the steepest at -2.7; NH is the shallowest near -1.3. These two extremes are likely directly associated with the fact that using the basic 2-generation Haser model does not work well for species having more complicated parentage, thus generally yielding aperture trends and/or $r_H$-dependent trends. In the case of NH, it is the 3$^{rd}$ generation starting from $NH_3$, while $C_2$ has multiple parents – $C_2H_2$, $C_2H_6$, and $C_3$ – some of which also have their own parents. As is the case here for H-B, $C_2$ and NH have often exhibited opposite aperture trends, due to the specific choices made regarding the nominal parent and daughter scalelengths for these two species (see A'Hearn et al. 1995), thereby also usually resulting in opposite extremes for $r_H$-dependent slopes. Thus, we consider these apparent outliers in the $r_H$-dependencies to mostly be simple artifacts caused by the limitations of the adopted Haser scalelengths and of the model itself, amplified due to the extreme conditions associated with H-B. Of the carbon-bearing species having sufficient data to compute pre- and post-perihelion slopes, only CN shows significant asymmetry, consistent with the departure from a linear trend seen when H-B was inbound. Finally, note that the slopes of the gas species are similar to that observed for the majority of other long period, non-dynamically new comets, i.e. between about -1.5 to -3.0 (A'Hearn et al. 1995) and consistent with our expanded data base (Schleicher & Bair 2016), while Jupiter-family comets generally have steeper slopes.

As would be expected given the variations in slopes among the gas species, there are trends in several abundance ratios with heliocentric distance. Perhaps the ratio of most interest is $C_2$-to-CN, because this quantity is associated with the carbon-chain depleted compositional class, and there is evidence for a trend in this ratio with distance, that we attribute to the just-discussed Haser $C_2$ issues (A'Hearn et al.1995; Schleicher & Bair 2016). What makes this notable for H-B is the much greater range of distances over which it was observed, yielding a greater change in the apparent ratio. In Figure 4 we show this effect by plotting this and other interesting abundance ratios as a function of time when H-B was within 4 AU. (We again exclude values at larger distances due to larger uncertainties in the data, the effect of outbursts inbound, and the increased Haser model scalelength extrapolations.) Because of the asymmetric behavior of CN about perihelion, the $C_2$-to-CN ratio also has very different slopes before and after. Because any modeling issues can *not* yield asymmetries, we conclude that such differences are most likely caused by seasonal effects. We will return to this in Section 3.6.

**Figure 4 – Ratio Qs vs ΔT; NH/OH; CN/OH; C2/OH; C2/CN; C3/CN**

A more symmetric and much smaller trend is visible in Figure 4 for the $C_3$-to-CN ratio, while $C_2$-to-OH is essentially flat over time simply because the component rH-dependences are so similar. The abundance ratio of CN-to-OH significantly decreases as H-B approached the Sun, partly due to the smaller slope for CN inbound and partly caused by the general trend for this ratio with distance seen in several other comets and again thought to be an artifact of the specific scalelengths chosen for the Haser model; note that Haser scalelengths were based on spatial distributions for each species between about 0.7 and 1.5 AU, and significant extrapolations are required at much larger distances. Knowing that the canonical Haser scalelengths do not properly reproduce the observed spatial profiles at high production rates due to higher outflow velocities, and that these effects differ among the five gas species due to quite different parent and/or daughter lifetimes, we conclude that at least some of the measured drop in the CN-to-OH ratio at small $r_H$ values is an artifact of the model. This is confirmed, based on post-perihelion water measurements presented later in Section 3.5, where the equivalent CN-to-OH ratio would have an even steeper slope than seen pre-perihelion.

The mean values for its production rate or abundance ratios for all data inside of 4 AU place H-B within the "Typical" compositional class in all respects (A'Hearn et al. 1995; Schleicher & Bair 2016). Specifically, its log mean CN-to-OH ratio is -2.42, near the mean value of -2.48 for all typical comets over this same heliocentric distance range, while its mean $C_2$-to-CN ratio is +0.11, very close to the mean of +0.12 for the typical comets. Its mean log NH-to-OH abundance ratio of -1.86 is among highest in our database for comets observed < 4 AU; the mean typical value over this distance is -2.23.

Only two datasets are available for comparison with our measurements for the carbon-bearing species. The more extensive set is from Rauer et al. (1997, 2003) who obtained long-slit spectroscopy for CN, $C_2$, and $C_3$ (as well as $NH_2$, discussed further below). They have several nights of data for all species both pre- and post-perihelion, measured between 2.8 and 4.6 AU, and with $C_3$ detected out to 7.0 AU and CN successfully detected out to 9.8 AU following perihelion. Their production rates for all three C-bearing species show steeper slopes than our own, with individual production rates for CN and $C_2$ being quite similar to ours inside of about 3.5 AU but dropping lower than ours beyond this point, while their $C_3$ production rates are somewhat lower than ours at all measured distances. Note that, due to the long-slit spectra that they obtained, they were able to constrain Haser scalelengths through much of this range, and differences between their results and our own are largely due to the different values of these parameters. Regardless, their data similarly shows that only CN exhibits significant asymmetry between the pre- and post-perihelion production rates, being somewhat higher before perihelion than it was after. The second dataset is from Cudnik (2005) who used HB narrowband filters in the ~one month surrounding H-B's perihelion. Nearly all of their measurements fall within the range of production rates of our measurements in this same time frame, the exception being their three $C_3$ measurements that are 2-3× higher than ours during the few weeks after perihelion. While no other datasets with NH could be readily located, Rauer et al. (1997, 2003) additionally measured $NH_2$, which follows the same trends seen with their $C_2$ and $C_3$ data, having somewhat steep slopes and no offsets between pre- and post-perihelion production rates, and their values one-fourth of our own near 3AU. Note that an extrapolation of their steeper slope would result in a convergence with our NH values inside of 2 AU, suggesting that the increasing difference at large $r_H$ is again due to the choice of parameters, especially in our use of a two-generation Haser model for a three-generation molecule. Finally, the available datasets for comparisons of OH and water production are quite extensive and will be discussed in detail in Section 3.5.

Returning to the aperture effects seen in our photometry and briefly discussed in Section 3.1, these are easily seen in the near-perihelion time frame of the ratio plots in Figure 4, where we have again color-coded the aperture sizes. However, we emphasize that these various aperture trends are *not* sufficient to explain the different $r_H$-dependent slopes for the ratios but rather have a different explanation that we will return to in Section 3.6. The smallest aperture trends are evident for the CN-to-OH ratio, followed by the $C_3$-to-CN ratio. In contrast, consistently large but opposite trends are evident for ratios containing NH vs. those containing $C_2$, even at distances well beyond perihelion. For instance, most species already exhibit aperture trends in the derived production rates by about 150 days before perihelion and generally increase as the comet continued to approach the Sun, with larger apertures yielding higher values for OH, $C_3$, and $C_2$. The same but much smaller trend is observed for CN. NH shows a large trend, but opposite of all other species. As already discussed, this behavior of NH is thought to be primarily due to the inadequacies of the 2-generation Haser model for this 3$^{rd}$ generation species coupled with the very high production rates and the specific scalelengths originally chosen. To distinguish the trends for each individual component, in Figure 5 we return to separate panels for each species (similar to those shown in Figure 2) but now zoomed in to the near-perihelion time frame along with a few months preceding to demonstrate the transition from high to extremely high production rates. Unfortunately, H-B's solar conjunction in 1996 December to 1997 January greatly limited the observing window and the time to use multiple apertures during this transition and the magnitude of the variations with aperture size is only visible within 50 days of perihelion. By mid-February production rates had increased by 5× or more from November, and the derived answers vary by 2× or more with aperture size on a given night.

**Figure 5 –Qs and $A(0°)f\rho$ vs ΔT; near-perihelion only**

These aperture effects are a direct consequence of gas outflow velocities that were much higher than usual resulting in a much broader spatial distribution than computed for canonical Haser scalelengths. Thus, only a smaller fraction of the coma is contained within any particular aperture. The fundamental cause of the higher velocities is the outward acceleration of the gas due to the very large collision zone. This was modeled by Combi et al. (1997), with significant collisions continuing beyond $10^5$ km and outflow velocities up to 2-3 km s$^{-1}$. However, because the observed daughter (or granddaughter) species characteristically form at quite different distances due to the varying parent lifetimes, and survive for different distances due to their own distinct lifetimes, the amount of change for each scalelength will differ. In particular, a shorter-lived species, such as $C_3$ would not reach as a high of a velocity because it dissociates much closer to the nucleus. Thus the adjustment to the production rates computed using the canonical scalelengths will differ for each case, and we are unable to determine accurate correction factors as a function of increasing production. Based on the spatial profiles extracted from narrowband images obtained at perihelion, Schleicher et al. (1999b) determined that most species' parent and/or daughter Haser scalelengths needed to be increased by 2-3× the canonical values to reproduce the bulk radial fall-off for each gas species, in turn yielding perihelion production rates ~1.1-3.6× larger than otherwise calculated, with OH (i.e. water) exhibiting the largest correction, then CN and NH, while $C_3$ and $C_2$ would be adjusted the least. However, we emphasize that we have *not* attempted to apply these adjustments to the data presented here. Spatial distributions will be examined in detail in our subsequent paper focusing on our imaging datasets.

Also mentioned briefly in Section 3.1 is the issue of rotational variability in H-B's production rates. The evidence for this is simple: changes are clearly evident over the course of many nights near perihelion, and there are no credible claims of outbursts having taken place during this interval. Examining the situation more closely confirms this conclusion. It was determined that Hale-Bopp had a rotation period of about 11.3 hr, based on the outward motion of its spiral dust and gas jets that were detected by many investigators in the spring of 1997 (Lecacheux et al. 1997; Licandro et al. 1998; Farnham et al. 1998; Jorda & Gutiérrez 2002). Our own analyses revealed a projected outward dust velocity of ~1 km s$^{-1}$ at a distance of about 150,000 km, while the CN velocity was >2 km s$^{-1}$ at a comparable distance (Schleicher et al. 1999b). Median projected radii for our photometric measurements were typically 2-4×10$^4$ km, and thus the gas molecules remained within the photometer aperture only about one-half of a rotation cycle. In other words, with a single source region dominating the jet morphology, a diurnal variation in production rates is expected. Unfortunately, such variability is somewhat masked by two issues, the first of course being the strong aperture effects, combined with a differing range of apertures during each night of observations due to the specific telescope in use and the length of the observing window. Additionally, because H-B always had a small solar elongation in early 1997, we typically had coverage of less than 2 hours per night. Thus it took many nights to sample the entire rotational cycle, and the same rotational phase could only be observed after 8 nights. Note that even if more appropriate scalelengths had been applied, aperture trends largely remained, another confirmation that rotational variability was the primary cause. This is best seen by focusing only on the smallest aperture data (purple) of each species to minimize the dilution by older material within the column defined by the aperture. Somewhat unexpectedly, OH displays the greatest variations from one observing run to the next while CN exhibits the least. Finally, note that dust, discussed next, exhibits essentially no day-to-day variability due to its lower outflow velocity, resulting in material remaining in the aperture for an entire rotational cycle.

*3.3 Dust Production as a Function of Time and Distance*

In many respects, dust production in Hale-Bopp is better behaved than that of the gas species. In the dust panels of Figures 2 and 3, we give the log of *A*(0°)*fρ*, following the adjustment for phase angle as discussed in Section 2.3, thereby removing features associated with the minima in the phase curve (Figure 1). At perihelion, the peak value of log *A*(0°)*fρ* reaches 6.50, corresponding to nearly 3.2×10$^6$ cm, while our final measurements beyond 10 AU drop to about 5.0×10$^4$ cm. The derived $r_H$-dependent slopes for the green continuum are quite different before and after perihelion (see Table 4), again restricting observations to between 1.2 and 4.0 AU, although this change is almost entirely due to the general curvature evident in Figure 3 combined with the inbound data only existing between 1.2 and 2.1 AU. Also note that slope determinations for the other continuum positions are incomplete because the ultraviolet was not measured at Perth, and the blue continuum filter (from the HB set) only became available a few months prior to perihelion.

One of our most surprising discoveries was the absence of a large aperture effect for dust throughout the apparition, including both the long durations as H-B approached and receded from the Sun as well as the near-perihelion interval when the gas species exhibited large aperture effects. Even so, we did observe a consistent trend with aperture size beyond about 2 AU, with smaller apertures yielding larger values of *Afρ*, but the size of this effect – a less than 20% decrease for aperture sizes increasing by 2-3× – was somewhat smaller but otherwise consistent with what has usually been measured in our database (Schleicher & Bair 2016). This behavior, measured by Baum et al. (1992) in 10 of 14 comets, was given the designation "fading grains", and was presumed to be caused either by grains becoming darker with time or smaller with time due to the ice within the grains vaporizing away. Based on this, one obvious conclusion is that H-B's grains are not dissimilar to most comets and therefore likely contain a significant ice component.

However, the trend reverses near perihelion, with larger apertures now having larger *Afρ* values, by nearly the identical amount. Thus, the total change in the aperture trend is approximately 40% for a range of ~3× in projected aperture radius. While sizeable, and most likely caused by the higher outflow velocities that cause gas species' aperture trends, the appropriate adjustment that should be applied to obtain a corrected *Afρ* is not well constrained. A simple estimate, based on the gas values, would be ~2× near perihelion, yielding a peak, phase-corrected value for *Afρ* of ~6×10$^6$ cm. This can be compared to Halley's phase adjusted value of ~8×10$^4$ cm, or about 75 times more for Hale-Bopp, despite its significantly larger perihelion distance (0.91 AU vs 0.59 AU for Halley). Finally, while we see no evidence for rotational variations near perihelion, such variations would not be expected because the lower outflow velocities of the dust (compared to the gas) meant that more than one rotational cycle of material is contained within the viewing aperture, completely diluting any variability.

While the use of *Afρ* provides an excellent first-order comparison regarding dust production for an individual comet as a function of time or distance and remains useful in bulk comparisons among comets, the computation of a mass loss rate based on *Afρ* requires information regarding the basic properties of the dust particles that is often unknown. While some of these issues, including particle size distribution and porosity of the grains, may be constrained if an object has been visited by a spacecraft, in most cases including H-B, unknowable assumptions must be made. Additionally, the majority of the mass can be effectively hidden to measurements made in the visible regime by being in much larger chunks of material that are centimeters or even meters in size while the visible particle cross-section is dominated by micron-sized grains. With these caveats in mind, we apply an approximate conversion factor created by C. Arpigny decades ago (cf. A'Hearn et al. 1995) where an *Afρ* value of 10$^3$ cm is equivalent to a dust production rate of about 1 metric ton per second. Using this, we estimate that peak dust release from H-B's surface at perihelion was ~6×10$^6$ kg s$^{-1}$, while an integration throughout the apparition meant about 3×10$^{10}$ metric tons were released. For comparison, we note that Jewitt & Matthews (1999), based on sub-mm spectroscopy, found a value just 1/3 of this result at perihelion, and the identical value as ourselves for the entire apparition. Given the two quite different methodologies employed, and the number of assumptions made in each case, the agreement must be considered fortuitous, but it gives credence to Arpigny's empirical conversion factor. Independent of the various assumptions, based on our *Afρ* measurements more than one-half of the dust mass released from the surface took place in the two-month interval surrounding perihelion. In spite of this, there is no evidence of a surge in large, heavy grains being released near perihelion, which would be expected to linger and yield higher loss rates in the months following; rather, Jewitt & Matthews saw no asymmetry about perihelion.

At large heliocentric distances, however, $Af\rho$ begins to level out near a value of $10^5$ cm, and remains near $6$-$7\times10^4$ cm all the way out to 10 AU. Even allowing for the presumed decreasing outflow velocities of the dust as gas production diminished with distance, it is difficult to believe that dust production remained at such elevated levels. And unlike the situation of the gas species where measurements beyond 7-8 AU were noise dominated and discarded, no continuum measurements went negative following sky subtraction, and we therefore accept the value as is (except for noting that the real uncertainties are likely 2-3× greater than the tabulated values; see Section 2.3). We instead conclude that these coma $Af\rho$ results are increasingly contaminated by the dust tail of H-B. Note that the phase angle systematically decreases with distance as the comet recedes from the Sun (see Fig 1) just as $Af\rho$ begins to flatten out. This behavior is qualitatively consistent with an ever-increasing tail contribution in our photometric aperture as the tail becomes more aligned with our line-of-sight. Concurrently, the projected aperture size increased with distance, also increasing the proportion of tail material. Separating out the tail component, however, was not attempted due to the numerous assumptions that would need to be made to yield a quantitative result.

The color of the dust grains was unexpectedly unremarkable in nearly all respects. As usual for comets, the dust was slightly red, with the degree of reddening of less than 10% per 1000 Å between the green and blue filters and less than 20% per 1000 Å between the green and ultraviolet filters. The only significant variations with distance were the somewhat more neutral colors in the couple of months surrounding perihelion, where a small aperture effect was also evident with larger apertures exhibiting the least amount of reddening. While neutral colors could indicate icy grains, the ice component would be expected to decrease with distance from the nucleus, just the opposite of what we observe; we have no explanation for this at this time. Note, however, that the blue continuum filter was only introduced with the HB filter set, while the ultraviolet filters were not used at Perth, thereby limiting the coverage for some color pairs. Even so, the existing color pairings showed no unusual colors that might be associated with H-B's enormous dust release.

As expected from the differing $r_H$-dependencies, the dust-to-gas ratio, based on the phase corrected $Af\rho$ vs $Q$(OH), varies for H-B throughout the apparition. The largest values for the ratio (-23.4 to -23.2 cm s molecule$^{-1}$) occurred just beyond 4 AU inbound and again near perihelion (-23.7 cm s molecule-1). Using CN production as a proxy for OH outbound, the highest values are seen beyond about 5 AU, consistent with the dust behavior previously discussed. Inside of 4 AU, the average logarithmic value for $A(0°)f\rho/Q$(OH) is -24.2, with individual values generally remaining within 2× of this average; the summary value for the entire apparition is -24.1. The dust-to-OH and dust-to CN ratios behave similarly in the lead-up to perihelion with only a very small trend with heliocentric distance. In the several days prior to perihelion, however, the dust-to-OH ratio is somewhat higher than it was after, while dust-to-CN shows no offsets. At further distances after perihelion, where we only have dust-to-CN, the ratio is somewhat higher following perihelion than it was before, again presumably due to the presence of the dust tail. When compared to our entire database of comets, H-B is within the top 5%, with only 8 objects having greater dust-to-gas ratios, while the entire range of values covers more than 2 orders of magnitude (Bair & Schleicher 2019). What is unclear, however, is to what extent the dust-to-gas ratio for H-B is comparable to those measured for other comets. For instance, does it truly reflect the dirt-to-ice ratio of the nucleus or is it influenced by the ability of the high gas flow to lift heavier particles than is generally possible for smaller cometary bodies?

*3.4 Outbursts During the Apparition?*

The only interval for which we saw strong evidence of outbursts in our data was in late May/early June of 1996, corresponding to $\Delta T \sim$ -300 days and $r_H \sim$ 4.3-4.1 AU. First, production rates varied considerably during this interval. CN, for instance exhibited higher values on June 3, 9, and 10, while significantly lower values on May 28 and 29. Meanwhile, $Af\rho$ was much higher on June 3 and on June 17 as compared to nearby nights. Second, there was a greater dispersion in values with aperture size than was generally seen at other times in 1996. Moreover, trends with aperture size reversed from June 1 and 3 to June 9 and 10. Each of these characteristics is consistent with a comet undergoing outbursts, because a short release of excess gas and dust will propagate outwards through the coma directly yielding changing aperture trends along with the obvious increase in the total amount of material. Specific details will vary by species, depending on lifetimes and, in the case of the dust, slower outflow velocities as compared to the gas; however, as already was seen for measurements at large distances, there is a large amount of scatter for all gas species except for CN due to the very low photon count levels and/or the poor contrast of the gas emission to the large underlying continuum that needed to be removed. Note that while the characteristics of the variability described here are also consistent with rotational effects, we discard this possibility for the following reason. At a distance of about 4 AU, the lifetime of each species is 16× that at 1 AU. Thus, even short lived $C_3$ would have a lifetime of about 5 days while CN's lifetime is greater than a month. For projected aperture radii ranging between 30,000 and 100,000 km, much of each species' molecules within the observing column will have been diluted by material from multiple rotational cycles ($P \sim$ 11.3 hr), and no rotational signature is expected to be detectable.

In contrast, and as previously discussed, the variability we observed near perihelion was ascribed to nucleus rotation partly because material (gas or dust) would remain within the observing column for less than a single rotational cycle, as well as the simple fact that no one else reported any evidence of outbursts despite "everyone" looking at the comet in early 1997. Unfortunately, our cadence of successfully obtaining photometric data was usually insufficient to tell if there were outbursts due to lack of data on nearby days for comparison. Several reports of an increase in "jetting activity" were noted beginning in May by Birkle (1996) and others, while extracted brightnesses using a 34 arcsec aperture from an extensive set of V-band imaging from a site in Chile revealed at least five outbursts beyond 4 AU during the post-perihelion time frame (Liller 2001). In contrast, based on extensive visual brightness measurements, Womack et al. (2021) found little evidence of outbursts throughout the apparition, though dilution by older material due to the larger field of view for most visible observers precludes the detection of most smaller-sized outbursts. We conclude that larger (2+ magnitude) outbursts did not take place and even moderate outbursts (~0.5-2 magnitude) were relatively rare and apparently only occurred beyond about 4 AU.

*3.5 Water*

As described in Section 2.3, we always compute an equivalent vectorial model water production rate using an empirical computation that starts with the Haser OH production rate, includes factors such as the branching ratio, the $r_H$-dependence of the water velocity, and the distance at which the two models yield equivalent production rates (see Cochran & Schleicher 1993 and Schleicher et al. 1998). Additionally, we have more recently determined a further empirical adjustment for high-production situations that we have encountered during our work on our entire database. Because the size of the adjustment increases with the production rate, Hale-Bopp near perihelion experiences the largest adjustments, reaching a factor of 2.24 on each of two nights within two weeks of each other and on opposite sides of perihelion. This results in a peak water production of nearly $2\times10^{31}$ molecules s$^{-1}$, or about one-half million kg s$^{-1}$, corresponding to about 1000 cubic meters of ice every second assuming a density of 500 kg m$^{-3}$, i.e. one-half that of water. The corresponding mass ratio of released dust to water at perihelion, based on the earlier dust estimates (see Section 3.3) from Jewitt & Matthews (1999) or using Arpigny's empirical relation yields values of 4 to 12 times, respectively; high but not unusually so. An approximate integration of water release for the entire apparition yields $3\times10^{12}$ kg with over ½ of the total released water taking place in a 10-week interval surrounding perihelion. This total is approximately 10 times that lost from Halley during its 1986 apparition (Stewart 1987; Feldman et al. 1987), and corresponds to a total volume of about 6 cubic kilometers of water ice having been vaporized from the surface of Hale-Bopp, again assuming a density of 500 kg m$^{-3}$.

By computing water production rates from our OH measurements, we can make direct comparisons to the ensemble of water values determined and tabulated for H-B by numerous other observers utilizing a wide variety of techniques, as presented in Figure 6. Since in every dataset a single value per night of observation was presented, we have averaged our own measurements as well, thereby reducing the large amount of scatter at large $r_H$ as well as the aperture trends near perihelion. While our dataset, shown in red circles, is quite large with 51 nights having water measurements, the additional datasets are invaluable for both filling in gaps and for providing insights on seasonal behavior. Additional larger datasets plotted in Figure 6 are from Combi et al. (2019), whose data extend to 3.5 AU inbound and to 2.8 AU outbound and where we have computed binned values at 4-day intervals (note that this reference is an update to the original Combi et al. 2000), and Colom et al. (1997) who have numerous measurements inside of a heliocentric distance of 1.85 AU; from Morgenthaler et al. (2001) inside of 1.13 AU; as well as several measurements from Dello Russo et al. (2000) between 1.48 and 2.24 AU, and from Weaver et al. (1997, 1999) between 2.4 and 6.4 AU. A few smaller datasets are also plotted: those from Crovisier et al. (1997,1999) between 2.9 and 4.6 AU; from Stern et al. at 2.3 AU; and from Feldman (1997) and Harris et al. (2002) between 0.9 and 1.0 AU. In all cases we continue to use open symbols for measurements made prior to perihelion and filled symbols for measurements obtained when H-B was receding from the Sun.

**Figure 6 –Water Qs vs log $r_H$**

While details regarding most of these data sets are readily determined from the references, including the specific emissions that were measured such as OH, Lyman-alpha, forbidden oxygen, or water itself, and how water production was computed, in a few instances we provide some clarifications. In particular, the modeling of the 18-cm radio observations depends strongly on the specifics of the inversion curve and on the amount of quenching of the inversion (or anti-inversion) for the OH ground state. The results for OH utilizing a H-B specific model were shown in the "Christmas tree plot" of Biver et al. (1997; 2002) based on values tabulated by Colom et al. (1997), including two lower-limits. However, the latter's tabulated values only include data from 1997 and omit data from 1996 shown in the figures. While the earlier radio measurements are listed in a convenient web site (Crovisier et al. 2002), these do not utilize Biver et al's specific modeling for H-B's unique circumstances and are therefore inappropriate to use as listed. Thus, we do not show the 1996 radio data. Also, note that we have not applied the adjustment for the branching ratio from OH to $H_2O$, that in log space would raise each radio point only by about 0.03 to 0.05, i.e., about the size of each data point. Regarding the Feldman (1997) reference, he summarized several UV data sets that we utilized the original sources, except for a single case for which the data point, by D. G. Kupperman, apparently was never published separately. Finally, the only published result that we did not attempt to plot in Figure 6 was an average for measurements obtained over a 5-week interval near perihelion by Woods et al. (2000); their value of $1.3\times10^{31}$ was consistent with the mid-range of results from the ensemble of other measurements.

Overall, we find a high degree of agreement among most of the water observations. Even when there are differences, these are usually systematic in nature, suggesting that the differences are caused by varying methodologies and/or assumptions, such as gas acceleration or lifetimes. For instance, within the 5-2 AU region in-bound, our values are consistently intermediate between the higher values of Crovisier et al. (1997; 1999) and the lower values by Weaver et al. (1997). We also note that the revised and greatly expanded SOHO/SWAN hydrogen Lyman-alpha data (Combi et al. 2019) are generally higher than most other data sets. The derived pre-perihelion slope for water in log-log space, -3.2, is in excellent agreement with the value of -3.0 from our data alone, but is much steeper than the slope of -2.0 that our OH data yields for the smaller interval between 1.2 and 4 AU used in Section 3.2, due both to the inclusion of the standard adjustment vectorial $r_H^{-0.5}$ term for outflow velocities combined with large increase in $Q$s near perihelion due to collisions (and the correspondingly much higher velocities).

Obviously, our own observations following perihelion were drastically truncated with H-B's move to the southern hemisphere, but as evident from the figure, few measurements were obtained by other groups as well. In fact, except for about a dozen days of data from SOHO/SWAN (Combi et al. 2019), we are only aware of a total of 4 other successful detections and two upper limits beyond 2 AU – all but one space-based. Though sparse, these data points at larger $r_H$ (>2.3 AU) are systematically lower than the corresponding inbound ensemble, thus clearly revealing a pre-/post-perihelion asymmetry – a characteristic we were unable to investigate in Section 3.2 from our OH data alone. Specifically, the slope we derive for this power-law $r_H$ dependence is -4.2, for a difference of -1.0 from inbound measurements at similar distances. As evident from the plotted curve based on a simple water vaporization model (cf. Cowan and A'Hearn 1979), these relatively steeper slopes at large distances are due to ever-greater fractions of solar radiation being needed to first warm the water ice prior to vaporization; note the generally good agreement of the degree of curvature from the model with the water measurements. As mentioned in Section 3.4, we attribute the somewhat higher values inbound just beyond 4 AU to outbursts during this time frame, while the excess measured inside of ~1.2

AU (both pre- and post-perihelion) coincides with the large "northern" source region receiving maximum solar illumination during local summer (see Section 3.6 below).

While we have already discussed the volume of water ice vaporized and released by H-B both during peak activity at perihelion and the total quantity throughout the apparition, we now return to the surface areas associated with this immense outgassing. Based on the updated vaporization model of Cowan & A'Hearn (1979), see A'Hearn (2010; https://pds-smallbodies.astro.umd.edu/tools/ma-evap/index.shtml), we compute an average active area of ~2200 km$^2$ and an area of about 30% greater during the surge at perihelion. Based on the average value, this requires a minimum effective radius for H-B's nucleus of about 13 km, but as is well known, it is thought for a variety of reasons that the size of the nucleus is much greater than this, including the simple fact that the existence of discrete jets necessitates large surface regions with little or no activity. If we assume the adopted value by Lamy et al. (2004) of 37 km for the radius, we can then calculate an approximate active fraction of ~12%. Our narrowband images of H-B show no evidence of a significant icy grain component in the coma (i.e. no tailward OH), unlike many other comets; so the active area computation is likely to be valid. Note that this is one of the lowest active fractions for any of the long-period comets from our database, though only 6 long-period objects have the necessary size determinations to allow an active fractional area to be computed.

### 3.6 Preliminary Nucleus Modeling and Seasonal Effects

Early in this section we discussed the change in slopes of many of the production rates as a function of heliocentric distance (in log-log space) inside of 2 AU, with the greatest changes seen for CN and for dust. We also noted an asymmetry in production rates about perihelion, with CN again exhibiting the strongest effect. Having successfully reproduced such seasonal behavior for several comets in the past based on coma morphology, including 19P/Borrelly (Schleicher et al. 2003), 81P/Wild 2 (Farnham & Schleicher 2005), and 10P/Tempel 2 (Knight et al. 2012), we next incorporated our existing, preliminary model of Hale-Bopp's nucleus to determine to what extent it could explain these and other features associated with the gas and dust production rates.

Our narrowband imaging in early 1997 clearly exhibited outwardly moving spiral morphology, i.e. jets, from which we and other researchers determined H-B's rotational period of ~11.3 hr (Lecacheux et al. 1997; Licandro et al. 1998; Farnham et al. 1998; Jorda & Gutiérrez 2002). The gas jets of CN provided full spirals while the dust exhibited a series of partial spirals on the sunward side presumably due to insufficient gas flow at night to lift the dust grains. In 1996 and again in late 1997 the dust morphology was quite different and instead showed a number of radial features that were interpreted as the edges of side-on corkscrew jets, while gas jets were much more difficult to detect. Our many images and the evolution of the detected features will be the focus of the next paper in this series, while here we simply note that these images, along with images from several other sources, were used in a detailed initial study of H-B's nucleus and associated jets performed and presented in 2004 (Schleicher et al. 2004) and summarized next. In brief, our nucleus/jet model allows us to explore a large parameter space including the pole orientation, and the location and size of multiple source regions across the surface of the nucleus. The jets that are created are then shown as seen from Earth as a function of time as a particular comet orbits the Sun. While we originally developed this model specifically for Hale-Bopp, as mentioned above, we have successfully utilized it for recreating the observed jet morphology of many comets in the past quarter century, adding capabilities along the way as needed for specific targets. In fact, features introduced to yield our initial successful fit of model to observations – for Comet Hyakutake (1996 B2) (Schleicher & Woodney 2003) where we needed to include very large source regions and accelerating dust grains – directly led to our subsequent success at modeling the observed H-B features from 1996 to 1998.

Our preferred solution as presented at the 2004 DPS meeting and which we discuss below, differed only slightly from the values first given in the meeting abstract (Schleicher et al. 2004). A recent reexamination confirms that any additional changes that might arise in our final model (in a future publication) will have only minimal alterations to the source locations and sizes, and that the discussions presented here regarding the model and gas and dust production rates will remain valid. Our preferred solution has the positive pole (conforming to the right-hand rule) located within a few degrees of RA = 18h23m and Dec = -54°, corresponding to an obliquity of about 83°). A near-polar source (latitude of about -78°) dominates the morphology early and late in the apparition, while a high latitude source in the opposite hemisphere (+58°) becomes active at the beginning of 1997 and dominates through the perihelion time frame. Each of these two sources is estimated to have a radius of ~20°, producing relatively broad jets. Additionally, two smaller source regions, one near -35° with a radius of about 15° and the other near the equator with a radius of about 10° are required to reproduce the remaining morphological features. Note that these source sizes are based on the need to match the widths and shapes of the jets during each rotation cycle, and yield a total active fraction of about 9% of the entire nucleus surface, completely consistent with the separately calculated fractional active area based on production rates presented above (~12%). Finally, the central location of each source assumes latitudes on a spherical nucleus because we have no information regarding the actual shape.

To understand the solar illumination of these four source regions throughout the apparition, in Figure 7 we display both the sub-solar and sub-Earth latitudes on the nucleus as a function of time. As can be seen, the sub-solar latitude varies greatly, from about -65° near discovery in 1995, and remaining at relatively large negative values until the start of solar conjunction in late 1996 ($\Delta T \approx$ -118 days), then passing the equator at the end of January (-62 days) and rapidly moving to large positive values by perihelion (+71°) and peaking only 12 days later at +83°. This relatively small offset from perihelion grows significantly as the comet recedes from the Sun, with a sub-solar latitude of 64° on 1997 May 1 (+30 days) but not reaching the equator again until September 8 (+161 days), and slowly dropping to -31° at the end of 1998 (+639 days). When combined with the latitudes of the two largest source regions at -78° and +58°, it becomes evident that which of the two largest sources dominates outgassing switches from the "southern" source through all of 1996 until mid-January 1997 when it becomes nearly inactive (depending on whether or not any residual heat provides some continued vaporization during local winter), just prior to the change in Hale-Bopp's $r_H$-dependent slopes inbound. In contrast, the "northern" source dominates through the first-half of 1997, and produces the major spiral feature seen in imaging during this time frame.

**Figure 7 – Sub-solar & Sub-Earth Latitudes vs ΔT**

We now examine in detail the consequences of our H-B model with regards to the behavior of production rates as a function of time throughout the apparition as discussed earlier in this section. The first possible change in the equilibrium conditions regarding sunlight and source locations takes place as the Sun begins to illuminate the southern-most extent of the large "northern" source region that would not otherwise have experienced any heating since its last perihelion passage ~4200 years ago. This nominally occurred 10 months prior to perihelion, or in early June just at the time of the previously discussed outbursts (see Section 3.4), and even a small amount of heating each day might have been sufficient to trigger outbursts from CO or $CO_2$. Unfortunately, whether this "northern" source actually is beginning to be heated depends on the details of the size and shape of the source region and the nucleus, all effecting local topography, that we will never know. Of course, the more conventional explanation for outbursts is simply the abrupt turning-on of water vaporization, in this case just beyond 4 AU. However, this does not explain the more-shallow slope of CN production also starting in early 1996 June and continuing until 1997 January, which could have been caused by a somewhat different relative abundance of CN vs other species in the "northern" source. In particular, the $C_2$-to-CN ratio steadily increases during this interval, more than would be expected simply due to Haser model inadequacies with $C_2$, and also at a different trend than following perihelion. This latter aspect is clearly associated with differences in abundance ratios between the "northern" and "southern" sources because the length of time the "northern" source is illuminated is very unbalanced about perihelion (see the -180 day and +180 day plots in Figure 8), and thereby explaining the asymmetry in the $C_2$/CN vs time plot in Figure 4.

The modeling can also explain another phenomenon in early 1997 – the unexpectedly strong brightening of the comet in February that continued until H-B was lost to the Sun in early May. The start of the increase coincides with when the Sun's sub-solar latitude was sufficiently positive to permit illumination "around-the-clock" of the major "northern" source, while by perihelion (April 1) the Sun was actually "north" of the source, resulting in extremely intense heating. Finally we note that the sub-solar latitude did not return to the mid-point between the two major sources (-78° and +58°) until 8 months after perihelion, whereas it passed this latitude less than 3 months before perihelion, so it is expected to have a variety of seasonal effects displayed by Hale-Bopp. Many other details regarding the model and our final solution will be presented in a future paper.

## Section 4 – Discussion and Conclusions

It is without question that Comet Hale-Bopp was an exceptional object. Though never closely approaching Earth, nearly anyone old enough to have paid attention during its perihelion passage in early 1997 not only remembers this extraordinary comet, but also reflects on the long duration that it remained visible. H-B was also highly unusual in the amount of gas and dust released, especially for its orbital and observational circumstances. In fact, other comets that have appeared more impressive over the past many decades were either due to closer approaches to either Earth (for instance Halley in 1910) or to the Sun (Ikeya-Seki (1965f = C/1965 S1) in 1965, The Great January Comet of 1910 (C/1910 A1), or McNaught in early 2007 (C/2006 P1)). Perhaps most similar to H-B in history was the Great Comet of 1811, having a perihelion of 1.04 AU, a perigee of 1.22 AU, and an orbital period of several thousand years, it remained visible by eye for over 8 months (Kronk 2003). And while relatively recently discovered Comet

Bernardinelli–Bernstein (C/2014 UN$_{271}$) may be 2-3× larger than H-B, it will never be as active simply given that its closest approach in 2031 will remain beyond Saturn ($q$ = 10.95 AU). Therefore, we repeat, Comet Hale-Bopp was an extremely unusual and truly exceptional object!

The most basic reason for this is quite simple– Hale-Bopp is physically larger than any other comet nucleus that has entered the inner solar system in more than a century. Its estimated radius of 37 km yields a total surface area approximately 300 times that for Halley, and even with a smaller active fraction (about 12% for H-B versus Halley's 28%) still yields ~120× the active area. The average active area of 1800 km$^2$ for H-B not only provides the extremely high water production rates overall but also the much higher gas densities far from the nucleus which, in turn, result in the very extended collisional zone as modeled by Combi et al. (1997). Note that despite its large size, the H-B nucleus appears like a point source by a few thousand kilometers from the nucleus, so that the gas density and the size of the collision zone is completely controlled by the total amount of gas rather than any details regarding the active regions. With significant collisions extending beyond 10$^5$ km, outflow velocities could reach 2-3 km s$^{-1}$, far higher than the canonical ~0.6-1.0 $r_H^{-0.5}$ km s$^{-1}$ at these distances (cf. Combi 1989). One consequence is that the standard Haser scalelengths that we use in converting emission band fluxes to production rates were increasingly inappropriate as H-B approached the Sun and result in significant trends with aperture size. Because each of the observed molecular species has differing parent (and in some cases grandparent) lifetimes, the specific adjustment factors also differ for each species and with heliocentric distance. With insufficient data to disentangle these effects, we ultimately did not make adjustments to the daughter production rates. Fortunately, however, to first approximation the effects were in the same direction and thus largely cancel out when investigating relative abundances. In the case of water, the most abundant volatile in most comets, our studies of several high production rate objects at a variety of distances (including H-B) were used to derive an empirical correction for water production rates as a function of both production and distance. With the appropriate adjustments made, our water results were in excellent agreement with those of most other researchers. As an aside, this same empirical function would successfully correct the unexpectedly low water production rates we computed for Comet Halley near perihelion as compared to other data sets (Schleicher et al. 1998); at that time the effects of high collision rates on aperture trends were masked by similar sized trends due to rotational variability (Millis & Schleicher 1986; Schleicher & Millis 1989).

While the dust grains decouple from the gas well before these large distances, and so do not reach these extreme velocities, the dust grains are entrained with the gas far longer than the few hundred or even a few thousand kilometers measured in other comets, resulting in unusually high dust velocities, especially at peak gas production near perihelion. We defer details regarding both dust and gas velocities to our next paper, focusing on imaging results, but just the unusual trends with aperture size confirms that these outflow velocity effects are caused by continued acceleration in the large collisional regimes.

The extreme heliocentric distances over which H-B was observed is of course also directly due to the very large active area and the resulting high production rates throughout its apparition. With our first photometric measurements taking place only 2 days following discovery at 7.14 AU, and continuing for five and one-half years until 10.58 AU outbound, H-B was the 2$^{nd}$ most extensively studied object in our photometric database, only trailing Halley. These circumstances permitted a higher than normal number of apertures and filters to be utilized throughout the dataset, allowing the exploration of physical phenomena in more detail and over a longer arc of its orbit than is usually possible. A major goal, therefore, was to study the onset and cessation of activity as a function of distance from the Sun for the various gas species. While we successfully measured some species to greater distances than we've achieved ever before, our observational cut-offs were more limited than expected for a variety of reasons. The primary constraint was due to H-B's very high dust-to-gas ratio, which greatly reduced the usual contrast of the gas emission bands to the underlying continuum. Generally poor observing circumstances, including short windows and high airmasses, also dominated the first half-year of the apparition, resulting in our first definitive gas measurements not being obtained until H-B reached 4.7 AU, at which time all 5 of our standard gas species were detected. While observational circumstances were generally better on its outward leg, the low elevation and smaller 0.6-m telescope at Perth meant that OH and NH could not be followed and the S/N was somewhat poorer. Despite these issues, $C_2$ and $C_3$ were easily measured out to 5.1 AU and were probably detected at 6.7 AU, while we successfully measured CN to ~7.7 AU. Note that the maximum distance for gas emission in H-B was a detection of CN at 9.8 AU by Rauer et al. (2003), based on spectra obtained with one of the 8.2-m scopes at the VLT (size sometimes matters). Returning to OH, we note that in only two objects was this daughter of water detected beyond our distance of 4.7 AU, the first being Comet Bowell (1980b = 1980 E1) at 5.2 AU and thought to be an excess due to icy grains in the coma (A'Hearn et al. 1984), and the other being Halley at 5.1 AU. Finally, while there was some hope that the detailed behavior of each gas species with distance might reveal evidence of additional or changing parentage of these species, the practical limitations of our and other investigators data, along with the associated assumptions and modeling issues described throughout this paper, make further investigations along these lines unwarranted and potentially misleading.

The primary driver of activity at large distances was determined to be CO, with relative rates about 4-5× greater than $CO_2$ (Crovisier et al. 1999), similar to that determined for a number of long-period but *not* dynamically new comets (Harrington Pinto et al. 2022). Due to CO's abundance, nearly 30% of water, and very low sublimation temperature, it would have been able to lift the extraordinary amounts of dust that were seen early in the apparition and presumably were a factor late in the apparition as well. It is assumed, as with most comets, that water was the driver when H-B was inside of 3-4 AU, but it remains unclear precisely when the transition occurred. While the huge amount of dust released is directly associated with the unprecedented total active area, we are not convinced that the size of the nucleus is the cause for the very high dust-to-gas ratio even though this seems to be the general assumption. First, unlike the high gas densities away from the nucleus, the amount of dust lifted from the surface is controlled by the gas flow *from the surface* combined with the dirt-to-ice ratio where the ice is vaporizing. The large active area does *not* alter the vaporization rate per unit area, and so cannot be the key factor. Instead, it must be the amount of dust embedded in the ice, and we see no reason why this should be unusual simply due to the size of the nucleus. Thus, we conclude that the observed very high dust-to-gas ratio is coincidental with H-B's enormous size. Our database of comet photometry provides additional credence towards this conclusion because nearly a dozen other comets of various dynamical classes, out of our 200+ objects, display even higher dust-to-gas ratios than H-B (Schleicher & Bair 2016).

A characteristic that is likely associated with the large surface area of the nucleus is simply the number of active source regions. Beginning with Comet Hyakutake in 1996, we have studied gas and dust jet morphology in approximately a dozen objects, and in each case we have found either one or two jets. However, here we measured at least four jets scattered across the surface, but only encompassing about 12% of the entire surface. We suspect the larger number of isolated sources is simply due to the vast areas available.

One consequence of multiple source regions are the relatively small asymmetries in production rates with respect to perihelion for the various species. Some of the most extreme examples of pre-/post-perihelion asymmetries – Comet 19P/Borrelly (Schleicher et al. 2003) and 10P/Tempel 2 (Knight et al. 2012) – were shown to be caused by having a single or dominant source region located near one pole, yielding extreme seasonal variations as the source moved from summer to winter or vice versa. With sources in both hemispheres, seasonal effects are greatly tamped down, whatever the pole orientation might be. With strong evidence for at least 4 jets identified in Hale-Bopp's coma and our preliminary modeling reproducing these features with two major source regions and two smaller ones, it is not surprising that asymmetries with respect to perihelion are indeed generally small. Somewhat unexpectedly is that both CN and water are more asymmetric than other species, not only directly indicating some inhomogeneities in composition among the source regions, but when such variations have been seen in other objects it is usually OH and NH differing from the carbon-bearing species. Our modeling indicates that the proportion of minor species, particularly the parents of $C_2$ and $C_3$, is higher with respect to water in the major northern hemisphere source than the southern sources. These and related issues will be examined in more detail in following papers.

While the production rate asymmetries seen in CN and OH (i.e. water) combined with the lack of asymmetries in the other gaseous species directly yield changing production rate ratios through the apparition, H-B's composition remains within the "Typical" compositional class as defined by A'Hearn et al. (1995) and Schleicher & Bair (2016). This is not surprising, as 86% of all well-determined, non-Jupiter-family comets in our database have this classification, and we conclude that H-B's size itself had no unusual effect on its volatile composition (excepting the artificial effects near perihelion due to the scalelength problems caused by the extreme collisional zone within the coma). Note that, as we see here for H-B, seasonal variations directly associated with differences in composition between source regions are common, but always smaller than the differences between our classes for chemical compositional.

Relatively few outbursts were reported by researchers throughout H-B's apparition. While this might simply be due to its specific surface properties – some comets exhibit frequent activity and breaking off of pieces while others are quite benign – we suspect this is yet another symptom of the very large size of H-B's nucleus. Any unusual activity that might take place, presumably triggered by heating from solar insolation, would not occur simultaneously across large regions on the surface and so would tend to be overwhelmed by the total ongoing activity. In our own case, though our observational set is extensive in both time and distance, it is also generally quite sparse and far from ideal for detecting unusual activity. Even so, we presented evidence of sporadic bursts in the 1996 May/June timeframe, when H-B was just beyond 4 AU. One clear possibility is that this is the approximate distance that water ice is expected to begin to vaporize, and could thus be the onset of water beginning to take over from CO or $CO_2$ vaporization on the hemisphere that has been in sunlight continuously as H-B approached the inner solar system. A more intriguing possibility, however, also exists. As the sub-solar latitude is slowly increasing in mid-1996, the southern-most extent of the large northern hemisphere active region that would just be "seeing" sunlight for the first time in several thousand years and therefore still be at very cold temperatures. In this scenario, a more volatile species than water would be the first to turn-on, and a smaller region would be involved. We would expect that as the northern source was progressively illuminated during the latter half of 1996, hiccups in activity would have continued to occur but eventually be overwhelmed by the total outgassing.

In conclusion, Comet Hale-Bopp was a most unusual object, but this is almost entirely due to its large nucleus size rather than having unique characteristics in its composition or surface properties. Indeed, this large size directly resulted in major differences in the comet's coma, including a much larger collisional zone and much higher outflow velocities of both gas and dust. However, H-B was quite ordinary in many other respects; indeed, if one broke off a 1-2 km piece, such a chunk would behave like the general population of Oort Cloud comets. H-B's brightness made it, along with Comet Halley, one of the two most well-studied comets from Earth in history, and improvements in technology allowed numerous molecular species to be studied for the first time. Even so, a much older technique – photoelectric photometry – provided a wealth of information over a longer baseline than ever before. We continue to wait for the next comet-of-the-century.

## Acknowledgements

We thank R. Millis, S. Lederer, and D. Thompson for acquiring some of these reported observations. We also thank Lowell and Perth Observatories for making facilities available over such a long interval of time, thus permitting many of these studies. This research was originally supported by grants from NASA's Planetary Astronomy Program and most recently by the Solar System Workings Program (Grant Number 80NSSC21K128).

Facilities: Lowell Observatory 42 inch (1.1 m) John S. Hall Telescope; Lowell Observatory 31 inch (0.8 m) telescope (LO:0.8m); Perth Observatory 24 inch (0.6 m) Lowell Telescope.

Software: Data Desk 6.3 by Data Description, Inc.

**Figure Captions:**

Figure 1. The distribution of photometry observations and associated solar phase angles. In addition to the extreme extent of time and the very wide range of heliocentric distances of these narrowband observations, the observing geometries oscillated over a large range of phase angles, for which adjustments were made to dust production rates (see late Section 2.3).

Figure 2: Logarithm of the production rates for the observed molecular species and $A(0°)f\rho$ for the dust as a function of time from perihelion (1997 April 1). Observations obtained from Lowell Observatory are shown as circles while those from Perth Observatory are given as triangles; neither OH nor NH were observed from Perth. Inbound data are plotted with open symbols and outbound as filled. Color-coding is based on aperture size, using the logarithm of the projected radius in km within bins shown at the top of the figure. The large scatter among data taken beyond about 6 AU are generally due to the small fraction of the signal that remains following continuum subtraction, often resulting in negative gas values. Negative results are shown as downward arrows near the bottom of each panel and are also listed as "undefined" in Table 3.

Figure 3: Logarithm of the production rates for the observed molecular species and $A(0°)f\rho$ for the dust as a function of the logarithm of the heliocentric distance. Symbols and colors are identical to Figure 2. However, unlike in Figure 2 where all data are presented, observations at large heliocentric distances that are not considered useful due to large scatter are no longer

shown (see text for details). Linear fits in log-log space were computed and shown as solid lines based on data between 1.2 and 4.0 AU, with pre-perihelion given in red and post-perihelion as blue. These fits are extrapolated to the full range of observations using dotted lines. Due to the gap in observations outbound, the post-perihelion fits actually begin at 2.16 AU, partially explaining the difference between the pre- and post-perihelion slopes, most evident for the dust.

Figure 4: Logarithm of production rate *ratios* plotted as a function of time from perihelion inside of 4.0 AU. Symbols and colors are identical to Figure 2. The trends at larger distances (for instance, NH-to-OH and CN-to-OH) or lack thereof ($C_2$-to-OH and $C_3$-to-CN) are thought to be mostly due to inadequacies with the Haser model and the standard scalelengths when extrapolated to large distances. However, most noteworthy is the large trend seen in $C_2$-to-CN inbound but absent outbound; since model deficiencies cannot be the cause of this asymmetry, this must be indicative of a real seasonal effect due to differences in abundance ratios between source regions.

Figure 5: Detail from Figure 2 for the perihelion time frame for early 1997. Again, the logarithm of the production rates for the observed molecular species and $A(0°)fp$ for the dust as a function of time from perihelion. Aperture trends (see color key) are quite evident, as well as significant rotational variability for gas species, but are nearly nonexistent for the dust.

Figure 6: Logarithm of water production rates plotted as a function of log heliocentric distance as determined by a variety of different types of measurements. Pre-perihelion data are shown as open symbols and post-perihelion are filled. Our data, derived from the OH measurements reported in this paper, are presented as circles, while other data sets are given in the key, including the method of observations (see Section 3.5 for details regarding each data set). The dashed curve represents the prediction of a simple sublimation model scaled to these data at intermediate distances, using an active area of 2200 km$^2$ (see text for details). While the overall behavior of water production is well-matched by the sublimation model, there is a clear asymmetry between inbound and outbound data beyond about 2.3 AU, that largely reverses inside of 2 AU. An excess above this base-line of activity is also evident in the few months surrounding perihelion; these deviations from the simple curve are largely explained by the seasonal effects associated with our preliminary modeling of the major source regions (see Section 3.6).

Figure 7: The sub-solar (solid curve) and sub-Earth (dashed curve) latitudes for Comet Hale-Bopp as a function of time, based on our preliminary, derived pole orientation from our jet modeling (see Section 3.6 for details). Note that the sub-solar latitude only moves significantly northward in the latter half of 1996. Also note the strong asymmetry about perihelion in both curves.

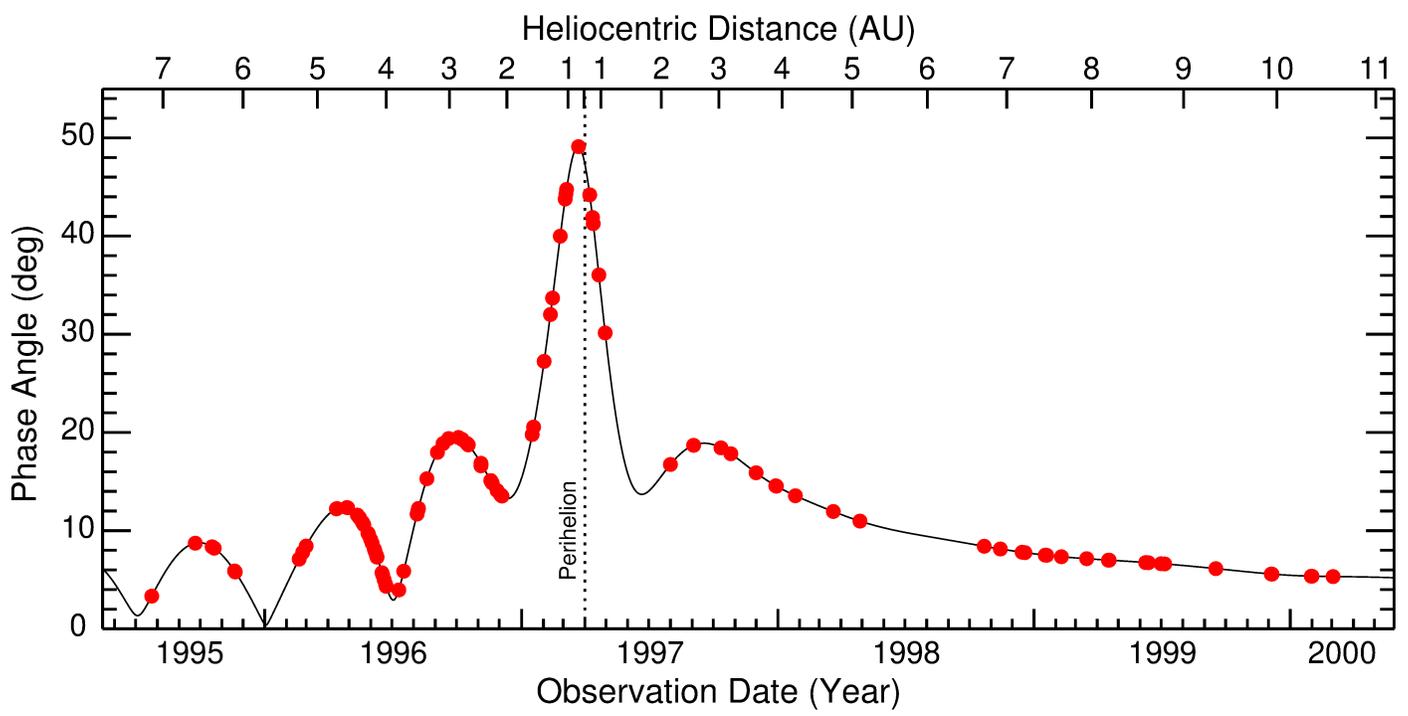

Figure 1

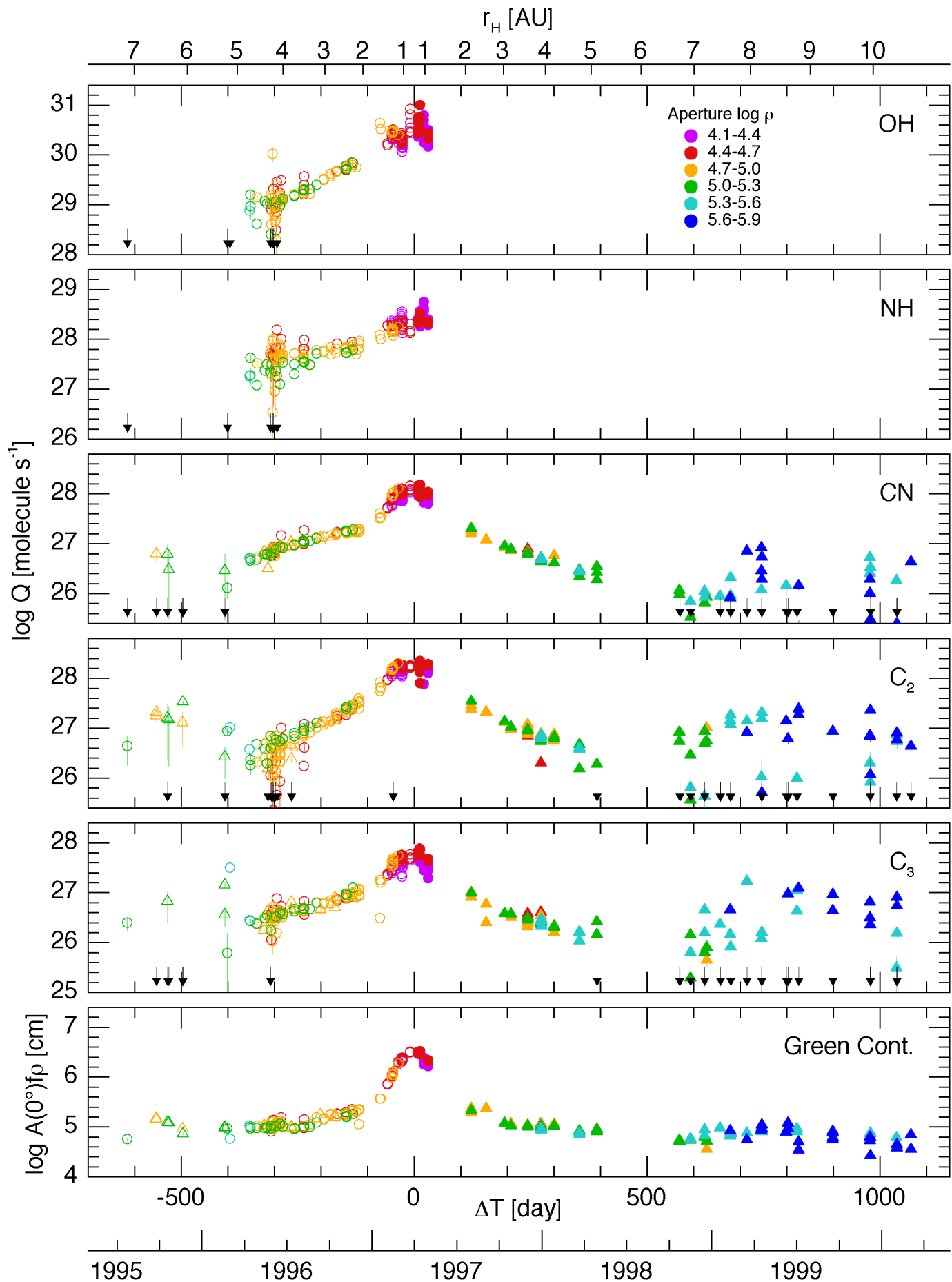

Figure 2

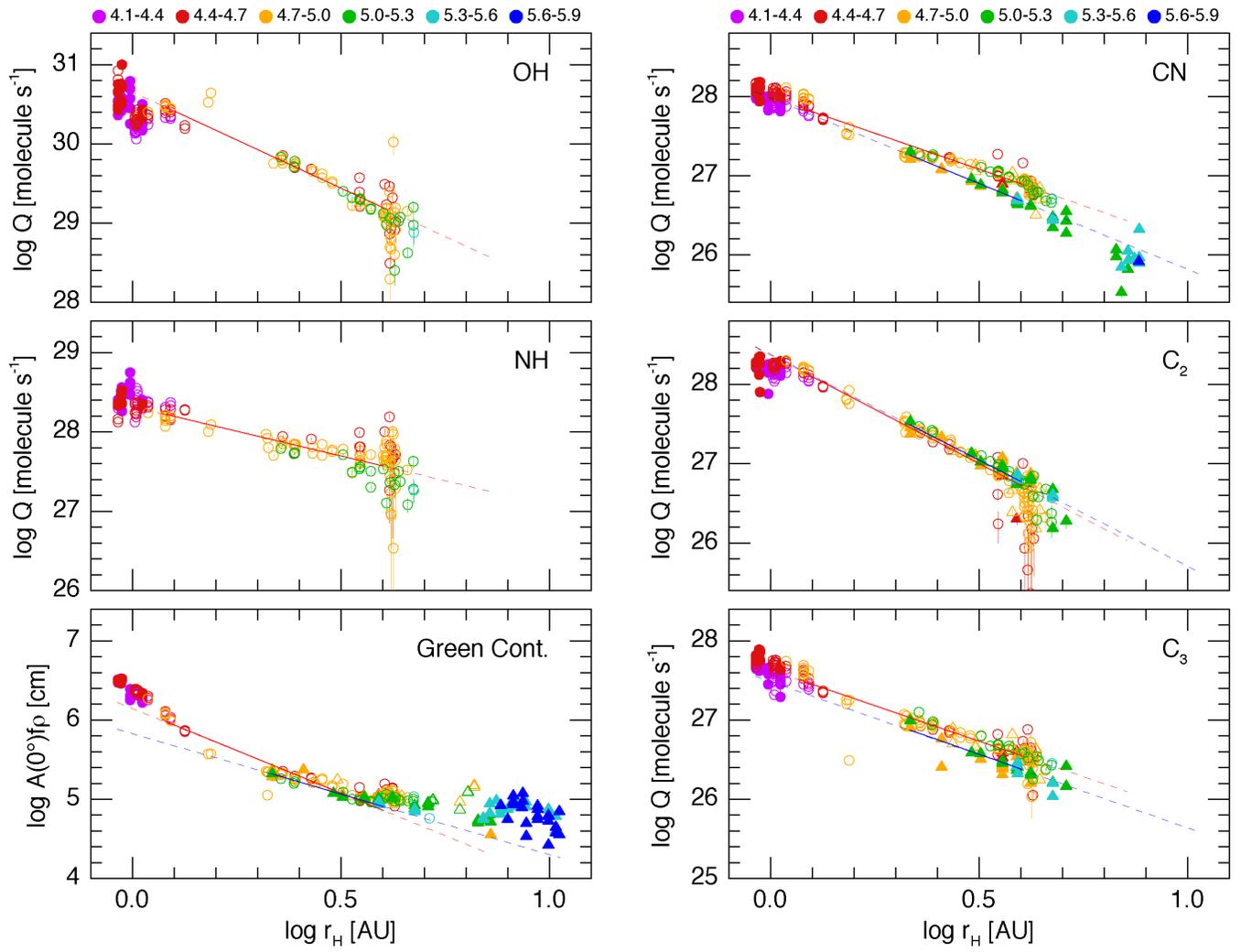

Figure 3

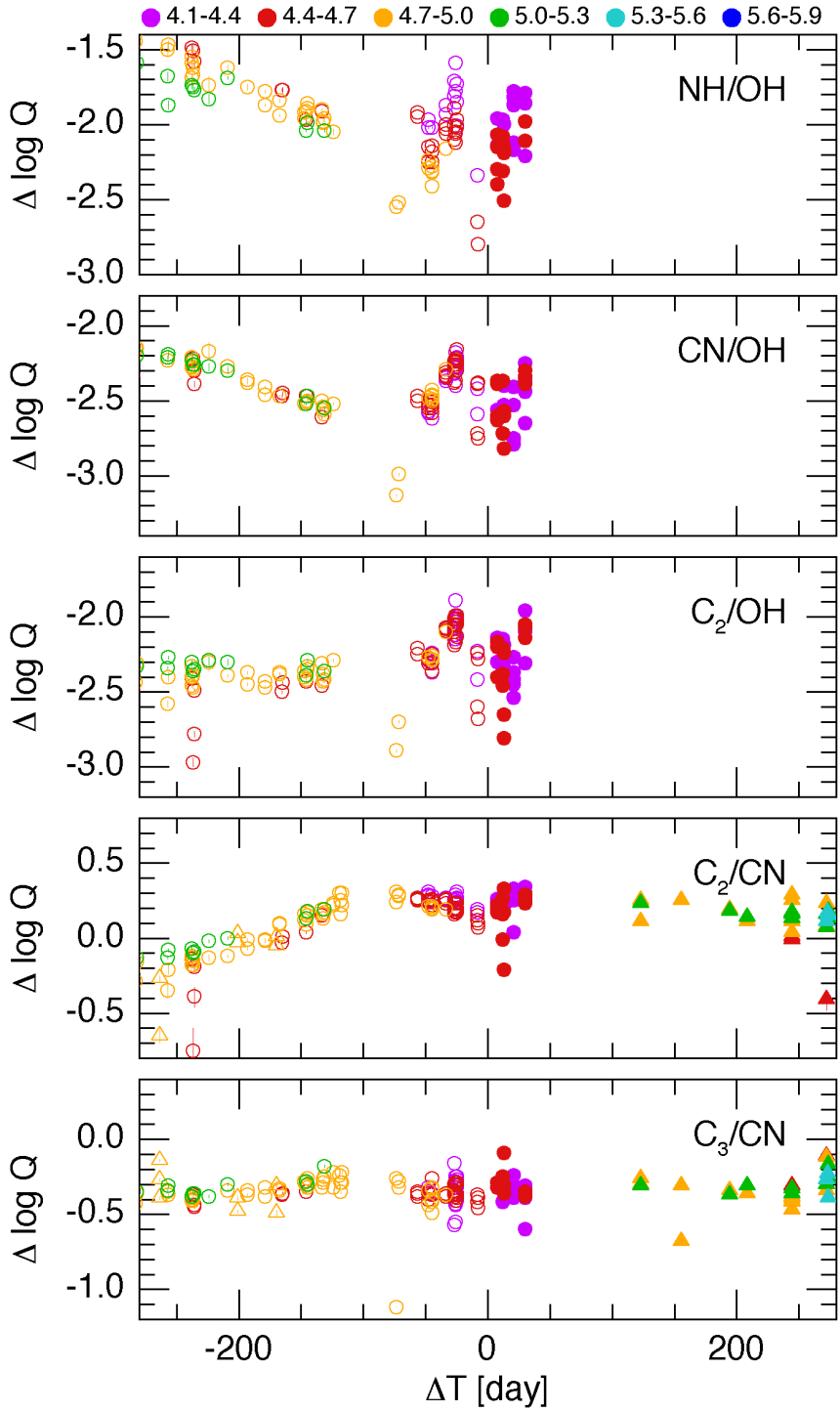

Figure 4

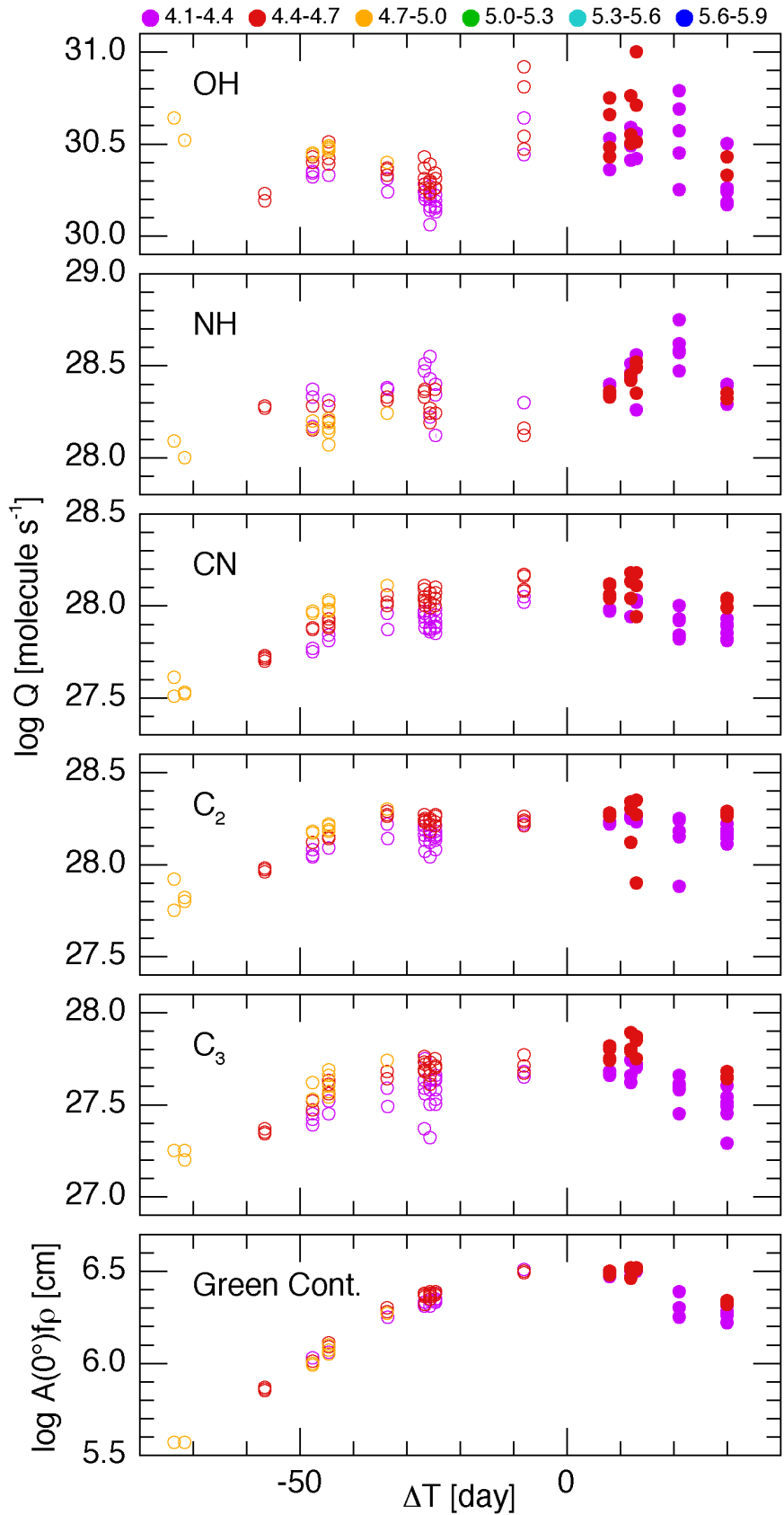

Figure 5

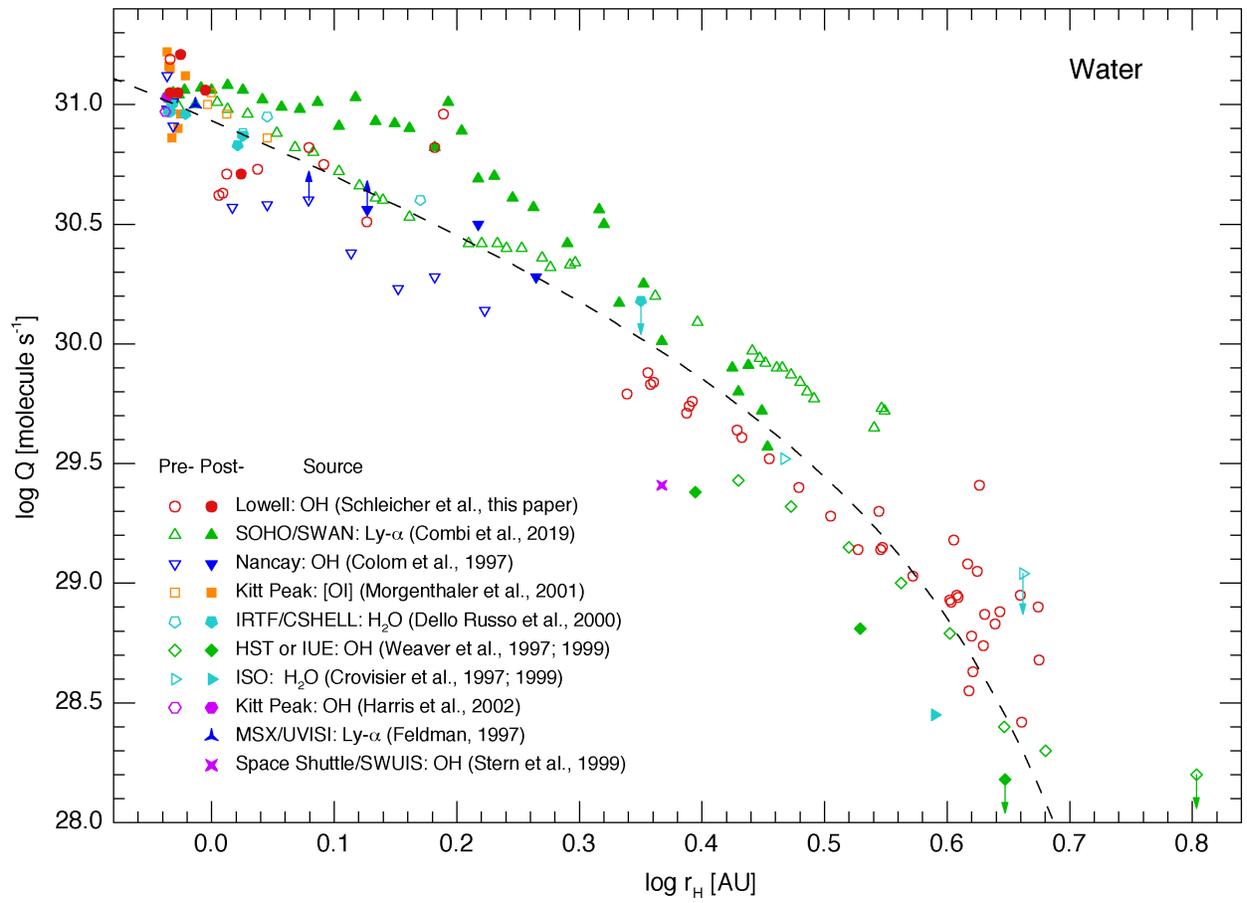

Figure 6

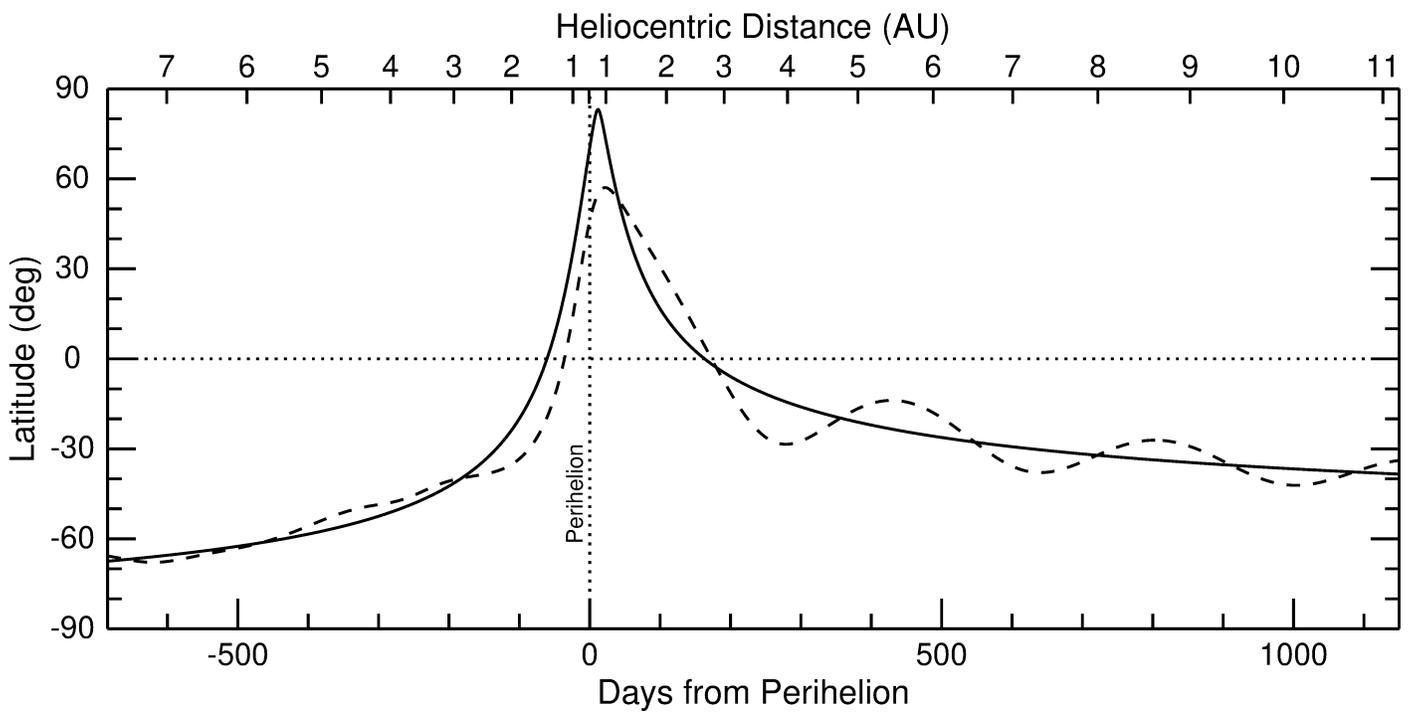

Figure 7

Table 1. Photometry Observing Circumstances and Fluorescence Efficiencies for Comet C/Hale-Bopp.[a]

| UT Date | ΔT (day) | $r_H$ (AU) | Δ (AU) | Phase (°)[b] | Phase Adj.[c] | $\dot{r}_H$ (km s$^{-1}$) | log L/N (erg s$^{-1}$ molecule$^{-1}$) OH | NH | CN | Teles.[d] | Sets | Filt.[e] |
|---|---|---|---|---|---|---|---|---|---|---|---|---|
| 1995 Jul 25.2 | −615.9 | 7.137 | 6.199 | 3.4 | 0.060 | −14.6 | −16.37 | −14.82 | −14.17 | L31 | 1 | I |
| 1995 Sep 25.6 | −553.6 | 6.602 | 6.499 | 8.7 | 0.144 | −15.1 | −16.34 | −14.75 | −14.09 | P24 | 2 | I |
| 1995 Oct 19.6 | −529.6 | 6.391 | 6.691 | 8.3 | 0.138 | −15.3 | −16.33 | −14.72 | −14.06 | P24 | 2 | I |
| 1995 Oct 22.5 | −526.6 | 6.364 | 6.711 | 8.2 | 0.136 | −15.4 | −16.34 | −14.72 | −14.05 | P24 | 1 | I |
| 1995 Nov 20.5 | −497.6 | 6.105 | 6.842 | 5.8 | 0.099 | −15.6 | −16.32 | −14.68 | −14.01 | P24 | 1 | I |
| 1995 Nov 21.5 | −496.6 | 6.096 | 6.844 | 5.7 | 0.097 | −15.6 | −16.32 | −14.68 | −14.01 | P24 | 1 | I |
| 1996 Feb 19.9 | −406.3 | 5.255 | 5.961 | 7.1 | 0.119 | −16.6 | −16.26 | −14.57 | −13.85 | P24 | 2 | I |
| 1996 Feb 25.5 | −400.6 | 5.201 | 5.840 | 7.9 | 0.132 | −16.7 | −16.26 | −14.56 | −13.84 | L42 | 1 | I |
| 1996 Mar 1.5 | −395.6 | 5.153 | 5.729 | 8.5 | 0.141 | −16.7 | −16.25 | −14.55 | −13.83 | L31 | 1 | I |
| 1996 Apr 13.5 | −352.7 | 4.730 | 4.638 | 12.2 | 0.194 | −17.3 | −16.19 | −14.50 | −13.75 | L42 | 1 | I |
| 1996 Apr 14.4 | −351.7 | 4.724 | 4.621 | 12.3 | 0.196 | −17.3 | −16.19 | −14.50 | −13.75 | L42 | 2 | I |
| 1996 Apr 28.5 | −337.7 | 4.579 | 4.237 | 12.3 | 0.196 | −17.5 | −16.17 | −14.48 | −13.72 | L31 | 1 | I |
| 1996 Apr 29.5 | −336.7 | 4.569 | 4.211 | 12.3 | 0.196 | −17.5 | −16.16 | −14.48 | −13.72 | L31 | 1 | I |
| 1996 May 13.8 | −322.3 | 4.423 | 3.841 | 11.5 | 0.184 | −17.7 | −16.14 | −14.46 | −13.69 | P24 | 2 | I |
| 1996 May 16.4 | −319.7 | 4.396 | 3.776 | 11.3 | 0.182 | −17.8 | −16.13 | −14.46 | −13.69 | L31 | 1 | I |
| 1996 May 20.4 | −315.7 | 4.355 | 3.680 | 10.8 | 0.174 | −17.9 | −16.12 | −14.46 | −13.68 | L31 | 1 | I |
| 1996 May 22.8 | −313.4 | 4.331 | 3.625 | 10.5 | 0.170 | −17.9 | −16.12 | −14.45 | −13.68 | P24 | 2 | I |
| 1996 May 28.4 | −307.8 | 4.273 | 3.498 | 9.7 | 0.159 | −18.0 | −16.10 | −14.45 | −13.66 | L42 | 4 | I |
| 1996 May 29.4 | −306.8 | 4.262 | 3.476 | 9.5 | 0.156 | −18.0 | −16.10 | −14.44 | −13.66 | L42 | 5 | I |
| 1996 Jun 1.4 | −303.8 | 4.231 | 3.412 | 9.0 | 0.148 | −18.0 | −16.09 | −14.44 | −13.66 | L31 | 6 | I |
| 1996 Jun 3.4 | −301.8 | 4.211 | 3.370 | 8.6 | 0.142 | −18.1 | −16.09 | −14.44 | −13.65 | L31 | 6 | I |
| 1996 Jun 6.4 | −298.8 | 4.179 | 3.309 | 8.1 | 0.134 | −18.1 | −16.08 | −14.43 | −13.65 | L31 | 3 | I |
| 1996 Jun 7.4 | −297.8 | 4.169 | 3.289 | 7.9 | 0.132 | −18.1 | −16.08 | −14.43 | −13.64 | L31 | 2 | I |
| 1996 Jun 9.3 | −295.8 | 4.148 | 3.252 | 7.4 | 0.124 | −18.2 | −16.07 | −14.43 | −13.64 | L31 | 4 | I |
| 1996 Jun 10.3 | −294.8 | 4.138 | 3.233 | 7.2 | 0.121 | −18.2 | −16.07 | −14.42 | −13.64 | L31 | 5 | I |
| 1996 Jun 17.4 | −287.8 | 4.064 | 3.109 | 5.6 | 0.096 | −18.3 | −16.05 | −14.41 | −13.62 | L42 | 6 | I |
| 1996 Jun 18.3 | −286.8 | 4.054 | 3.094 | 5.4 | 0.092 | −18.3 | −16.05 | −14.41 | −13.62 | L31 | 3 | I |
| 1996 Jun 20.3 | −284.8 | 4.032 | 3.062 | 4.9 | 0.084 | −18.3 | −16.05 | −14.41 | −13.62 | L31 | 2 | I |
| 1996 Jun 22.3 | −282.8 | 4.011 | 3.032 | 4.5 | 0.078 | −18.4 | −16.04 | −14.40 | −13.61 | L31 | 1 | I |
| 1996 Jun 23.3 | −281.8 | 4.001 | 3.018 | 4.3 | 0.074 | −18.4 | −16.04 | −14.40 | −13.61 | L31 | 2 | I |
| 1996 Jul 11.5 | −263.6 | 3.806 | 2.816 | 4.1 | 0.071 | −18.7 | −15.99 | −14.36 | −13.57 | P24 | 3 | I |
| 1996 Jul 18.3 | −256.9 | 3.732 | 2.773 | 6.0 | 0.102 | −18.8 | −15.97 | −14.35 | −13.55 | L42 | 4 | I |
| 1996 Aug 6.3 | −237.9 | 3.524 | 2.734 | 11.8 | 0.189 | −19.2 | −15.91 | −14.30 | −13.50 | L31 | 5 | I |
| 1996 Aug 7.2 | −236.9 | 3.513 | 2.735 | 12.0 | 0.191 | −19.2 | −15.90 | −14.30 | −13.50 | L31 | 4 | I |
| 1996 Aug 8.2 | −235.9 | 3.502 | 2.737 | 12.3 | 0.196 | −19.2 | −15.90 | −14.30 | −13.50 | L31 | 3 | I |
| 1996 Aug 20.3 | −223.9 | 3.367 | 2.769 | 15.4 | 0.236 | −19.5 | −15.86 | −14.26 | −13.46 | L31 | 2 | I |
| 1996 Sep 4.1 | −208.9 | 3.198 | 2.841 | 18.0 | 0.268 | −19.8 | −15.80 | −14.22 | −13.41 | L31 | 2 | I |
| 1996 Sep 12.5 | −200.7 | 3.103 | 2.887 | 18.9 | 0.278 | −20.0 | −15.77 | −14.20 | −13.39 | P24 | 2 | I |
| 1996 Sep 20.1 | −193.0 | 3.014 | 2.929 | 19.4 | 0.284 | −20.2 | −15.74 | −14.18 | −13.36 | L31 | 2 | I |
| 1996 Oct 4.1 | −179.0 | 2.850 | 2.998 | 19.5 | 0.285 | −20.5 | −15.68 | −14.13 | −13.30 | L31 | 2 | I |
| 1996 Oct 13.5 | −169.6 | 2.738 | 3.031 | 19.1 | 0.280 | −20.7 | −15.64 | −14.10 | −13.26 | P24 | 2 | I |
| 1996 Oct 16.1 | −167.2 | 2.708 | 3.038 | 18.9 | 0.278 | −20.7 | −15.63 | −14.09 | −13.25 | L31 | 2 | I |
| 1996 Oct 18.1 | −165.0 | 2.684 | 3.042 | 18.7 | 0.276 | −20.8 | −15.62 | −14.08 | −13.24 | L31 | 2 | I |
| 1996 Nov 5.1 | −147.1 | 2.466 | 3.047 | 16.8 | 0.253 | −21.2 | −15.54 | −14.01 | −13.16 | L31 | 2 | I |
| 1996 Nov 6.1 | −146.1 | 2.453 | 3.046 | 16.7 | 0.252 | −21.2 | −15.53 | −14.00 | −13.16 | L31 | 3 | I |

Table 1—Continued

| UT Date | ΔT (day) | $r_H$ (AU) | Δ (AU) | Phase (°)[b] | Phase Adj.[c] | $\dot{r}_H$ (km s$^{-1}$) | log $L/N$ (erg s$^{-1}$ molecule$^{-1}$) OH | NH | CN | Teles.[d] | Sets | Filt.[e] |
|---|---|---|---|---|---|---|---|---|---|---|---|---|
| 1996 Nov 7.1   | −145.1 | 2.441 | 3.044 | 16.6 | 0.251 | −21.2 | −15.53 | −14.00 | −13.15 | L31 | 3  | I |
| 1996 Nov 19.1  | −133.1 | 2.293 | 3.002 | 15.1 | 0.232 | −21.5 | −15.47 | −13.94 | −13.09 | L31 | 2  | I |
| 1996 Nov 20.1  | −132.1 | 2.281 | 2.997 | 15.0 | 0.231 | −21.5 | −15.46 | −13.94 | −13.09 | L31 | 2  | I |
| 1996 Nov 21.1  | −131.1 | 2.269 | 2.992 | 14.8 | 0.229 | −21.5 | −15.46 | −13.93 | −13.08 | L31 | 2  | I |
| 1996 Nov 28.1  | −124.1 | 2.181 | 2.947 | 14.1 | 0.220 | −21.6 | −15.42 | −13.89 | −13.05 | L31 | 2  | I |
| 1996 Dec 3.1   | −119.1 | 2.119 | 2.908 | 13.6 | 0.213 | −21.7 | −15.39 | −13.87 | −13.02 | L31 | 1  | I |
| 1996 Dec 4.6   | −117.6 | 2.106 | 2.900 | 13.6 | 0.213 | −21.7 | −15.39 | −13.86 | −13.02 | L31 | 4  | B |
| 1997 Jan 17.6  | −73.6  | 1.545 | 2.281 | 20.0 | 0.290 | −21.6 | −15.12 | −13.58 | −12.75 | L42 | 2  | B |
| 1997 Jan 19.6  | −71.6  | 1.520 | 2.244 | 20.8 | 0.299 | −21.5 | −15.11 | −13.56 | −12.73 | L42 | 2  | B |
| 1997 Feb 3.5   | −56.6  | 1.338 | 1.954 | 27.5 | 0.363 | −20.4 | −15.03 | −13.44 | −12.64 | L31 | 4  | B |
| 1997 Feb 12.5  | −47.6  | 1.234 | 1.778 | 32.3 | 0.399 | −19.3 | −14.99 | −13.36 | −12.59 | L42 | 7  | B |
| 1997 Feb 15.5  | −44.6  | 1.201 | 1.720 | 34.0 | 0.410 | −18.7 | −14.99 | −13.33 | −12.58 | L42 | 12 | B |
| 1997 Feb 26.5  | −33.6  | 1.090 | 1.527 | 40.3 | 0.441 | −16.2 | −14.87 | −13.21 | −12.51 | L31 | 6  | I |
| 1997 Mar 5.5   | −26.6  | 1.029 | 1.427 | 44.0 | 0.455 | −13.8 | −14.63 | −13.17 | −12.47 | L31 | 11 | B |
| 1997 Mar 6.5   | −25.6  | 1.021 | 1.415 | 44.5 | 0.456 | −13.5 | −14.60 | −13.17 | −12.46 | L31 | 12 | B |
| 1997 Mar 7.5   | −24.6  | 1.014 | 1.404 | 45.0 | 0.458 | −13.1 | −14.57 | −13.16 | −12.46 | L31 | 10 | B |
| 1997 Mar 24.1  | −8.0   | 0.925 | 1.316 | 49.1 | 0.468 | −4.8  | −14.69 | −13.08 | −12.41 | L31 | 6  | B |
| 1997 Apr 9.1   | +8.0   | 0.925 | 1.431 | 44.1 | 0.455 | +4.8  | −14.53 | −13.01 | −12.27 | L42 | 6  | I |
| 1997 Apr 13.1  | +12.0  | 0.939 | 1.484 | 41.8 | 0.446 | +7.1  | −14.51 | −13.00 | −12.26 | L42 | 6  | I |
| 1997 Apr 14.2  | +13.0  | 0.943 | 1.498 | 41.2 | 0.444 | +7.6  | −14.52 | −13.01 | −12.27 | L42 | 5  | B |
| 1997 Apr 22.2  | +21.0  | 0.988 | 1.621 | 35.9 | 0.421 | +11.5 | −14.38 | −13.09 | −12.33 | L31 | 5  | I |
| 1997 May 1.1   | +30.0  | 1.057 | 1.773 | 30.0 | 0.382 | +15.0 | −14.27 | −13.18 | −12.42 | L31 | 9  | B |
| 1997 Aug 1.9   | +122.8 | 2.165 | 2.874 | 16.7 | 0.252 | +21.7 | −14.88 | −13.80 | −13.13 | P24 | 3  | H |
| 1997 Sep 3.8   | +155.7 | 2.571 | 3.016 | 18.7 | 0.276 | +21.0 | −15.06 | −13.95 | −13.29 | P24 | 2  | H |
| 1997 Oct 12.8  | +194.7 | 3.033 | 3.151 | 18.4 | 0.272 | +20.1 | −15.24 | −14.10 | −13.43 | P24 | 2  | H |
| 1997 Oct 26.7  | +208.6 | 3.193 | 3.213 | 17.8 | 0.266 | +19.8 | −15.29 | −14.15 | −13.48 | P24 | 2  | H |
| 1997 Dec 1.7   | +244.5 | 3.597 | 3.455 | 15.9 | 0.243 | +19.1 | −15.42 | −14.26 | −13.58 | P24 | 10 | H |
| 1997 Dec 29.6  | +272.4 | 3.900 | 3.733 | 14.6 | 0.226 | +18.5 | −15.49 | −14.33 | −13.64 | P24 | 6  | H |
| 1997 Dec 30.6  | +273.4 | 3.911 | 3.745 | 14.6 | 0.226 | +18.5 | −15.50 | −14.33 | −13.65 | P24 | 4  | H |
| 1998 Jan 27.6  | +301.4 | 4.206 | 4.089 | 13.5 | 0.212 | +18.0 | −15.56 | −14.40 | −13.70 | P24 | 4  | H |
| 1998 Mar 22.5  | +355.4 | 4.755 | 4.814 | 11.9 | 0.190 | +17.2 | −15.64 | −14.51 | −13.80 | P24 | 4  | H |
| 1998 Apr 29.5  | +393.3 | 5.127 | 5.286 | 11.0 | 0.177 | +16.7 | −15.69 | −14.57 | −13.85 | P24 | 3  | H |
| 1998 Oct 23.6  | +570.5 | 6.740 | 6.811 | 8.4  | 0.139 | +14.9 | −15.88 | −14.77 | −14.05 | P24 | 3  | H |
| 1998 Nov 15.7  | +593.5 | 6.937 | 6.997 | 8.1  | 0.134 | +14.7 | −15.91 | −14.80 | −14.07 | P24 | 4  | H |
| 1998 Dec 16.6  | +624.5 | 7.199 | 7.256 | 7.8  | 0.130 | +14.5 | −15.94 | −14.82 | −14.11 | P24 | 4  | H |
| 1998 Dec 20.6  | +628.4 | 7.232 | 7.289 | 7.8  | 0.130 | +14.5 | −15.95 | −14.83 | −14.11 | P24 | 2  | H |
| 1999 Jan 18.6  | +657.4 | 7.472 | 7.535 | 7.5  | 0.125 | +14.3 | −15.98 | −14.85 | −14.14 | P24 | 2  | H |
| 1999 Jan 19.6  | +658.4 | 7.481 | 7.543 | 7.5  | 0.125 | +14.3 | −15.98 | −14.85 | −14.14 | P24 | 2  | H |
| 1999 Feb 9.6   | +679.4 | 7.653 | 7.717 | 7.3  | 0.123 | +14.1 | −16.01 | −14.86 | −14.16 | P24 | 2  | H |
| 1999 Feb 10.6  | +680.4 | 7.661 | 7.725 | 7.3  | 0.123 | +14.1 | −16.01 | −14.87 | −14.16 | P24 | 4  | H |
| 1999 Mar 17.5  | +715.4 | 7.943 | 8.002 | 7.1  | 0.119 | +13.9 | −16.04 | −14.89 | −14.19 | P24 | 2  | H |
| 1999 Apr 17.5  | +746.3 | 8.190 | 8.233 | 7.0  | 0.118 | +13.7 | −16.08 | −14.91 | −14.22 | P24 | 4  | H |
| 1999 Apr 18.5  | +747.4 | 8.198 | 8.241 | 7.0  | 0.118 | +13.7 | −16.08 | −14.91 | −14.22 | P24 | 4  | H |
| 1999 Jun 9.4   | +799.3 | 8.604 | 8.613 | 6.8  | 0.115 | +13.4 | −16.13 | −14.94 | −14.27 | P24 | 2  | H |
| 1999 Jun 13.5  | +803.3 | 8.635 | 8.641 | 6.7  | 0.113 | +13.4 | −16.14 | −14.94 | −14.27 | P24 | 2  | H |

Table 1—Continued

| UT Date | ΔT (day) | $r_H$ (AU) | Δ (AU) | Phase (°)[b] | Phase Adj.[c] | $\dot{r}_H$ (km s$^{-1}$) | log $L/N$ (erg s$^{-1}$ molecule$^{-1}$) OH | NH | CN | Teles.[d] | Sets | Filt.[e] |
|---|---|---|---|---|---|---|---|---|---|---|---|---|
| 1999 Jul 2.5 | +822.3 | 8.782 | 8.781 | 6.6 | 0.112 | +13.3 | −16.16 | −14.95 | −14.29 | P24 | 2 | H |
| 1999 Jul 6.5 | +826.3 | 8.812 | 8.811 | 6.6 | 0.112 | +13.3 | −16.16 | −14.95 | −14.29 | P24 | 2 | H |
| 1999 Sep 17.8 | +899.7 | 9.366 | 9.404 | 6.1 | 0.103 | +12.9 | −16.23 | −14.99 | −14.35 | P24 | 4 | H |
| 1999 Dec 6.6 | +979.5 | 9.951 | 10.091 | 5.6 | 0.096 | +12.5 | −16.31 | −15.02 | −14.41 | P24 | 4 | H |
| 1999 Dec 7.6 | +980.4 | 9.958 | 10.098 | 5.6 | 0.096 | +12.5 | −16.31 | −15.02 | −14.41 | P24 | 4 | H |
| 2000 Feb 1.6 | +1036.4 | 10.360 | 10.497 | 5.4 | 0.092 | +12.3 | −16.36 | −15.04 | −14.45 | P24 | 2 | H |
| 2000 Feb 2.6 | +1037.4 | 10.367 | 10.504 | 5.4 | 0.092 | +12.3 | −16.36 | −15.04 | −14.45 | P24 | 4 | H |
| 2000 Mar 3.5 | +1067.4 | 10.579 | 10.674 | 5.3 | 0.091 | +12.2 | −16.38 | −15.05 | −14.47 | P24 | 2 | H |

[a]All parameters are given for the midpoint of each night's observations.

[b]Mean phase angle ($\theta$)

[c]Adjustment to 0° solar phase angle to log($A(\theta)f\rho$) values based on assumed phase function (see Section 2.3).

[d]Telescope ID: L72 = Lowell 72-inch (1.8-m); L42 = Lowell 42-inch (1.1-m); L31 = Lowell 31-inch (0.8-m); P24 = Perth Observatory 24-inch (0.6-m).

[e]Filter ID: I = IHW; H = HB; B = Both IHW and HB.

– 4 –Table 2. Photometric Fluxes and Aperture Abundances for Comet C/Hale-Bopp.

| UT Date | Aperture | | Filt.[a] | log Emission Band Flux[b] | | | | | log Continuum Flux[b] | | | log $M(\rho)$[b] | | | | |
|---|---|---|---|---|---|---|---|---|---|---|---|---|---|---|---|---|
| | Size | log $\rho$ | | (erg cm$^{-2}$ s$^{-1}$) | | | | | (erg cm$^{-2}$ s$^{-1}$ Å$^{-1}$) | | | (molecule) | | | | |
| | (arcsec) | (km) | | OH | NH | CN | $C_3$ | $C_2$ | UV | Blue | Green | OH | NH | CN | $C_3$ | $C_2$ |
| 1995 Jul 25.2 | 81.4 | 5.26 | I | ... | ... | −12.43 | −11.18 | −11.82 | −13.35 | ... | −13.01 | und | 31.12 | 30.78 | 31.56 | 31.27 |
| 1995 Sep 25.5 | 27.7 | 4.81 | I | ... | ... | und | und | −11.82 | −13.24 | ... | −13.12 | ... | ... | und | und | 31.24 |
| 1995 Sep 25.6 | 27.7 | 4.81 | I | ... | ... | −12.23 | und | −11.90 | −13.20 | ... | −13.09 | ... | ... | 30.93 | und | 31.16 |
| 1995 Oct 19.6 | 54.6 | 5.12 | I | ... | ... | und | und | −11.39 | und | ... | −12.86 | ... | ... | und | und | 31.67 |
| 1995 Oct 19.6 | 54.6 | 5.12 | I | ... | ... | −11.70 | −10.89 | und | −13.38 | ... | −12.85 | ... | ... | 31.46 | 31.82 | und |
| 1995 Oct 22.5 | 54.6 | 5.12 | I | ... | ... | −11.99 | und | −11.43 | und | ... | −12.86 | ... | ... | 31.16 | und | 31.63 |
| 1995 Nov 20.5 | 38.8 | 4.98 | I | ... | ... | und | und | −11.67 | −13.23 | ... | −13.07 | ... | ... | und | und | 31.36 |
| 1995 Nov 21.5 | 54.6 | 5.13 | I | ... | ... | und | und | −11.01 | −12.51 | ... | −13.02 | ... | ... | und | und | 32.03 |
| 1996 Feb 19.9 | 54.6 | 5.07 | I | ... | ... | −11.71 | −10.30 | und | −13.87 | ... | −12.73 | ... | ... | 31.15 | 32.14 | und |
| 1996 Feb 19.9 | 77.8 | 5.23 | I | ... | ... | und | −10.72 | −11.63 | −12.82 | ... | −12.57 | ... | ... | und | 31.72 | 31.16 |
| 1996 Feb 25.5 | 70.2 | 5.17 | I | ... | ... | −11.85 | −11.52 | −11.16 | −12.82 | ... | −12.62 | ... | ... | 30.97 | 30.90 | 31.60 |
| 1996 Mar 1.5 | 114.7 | 5.38 | I | ... | ... | −12.65 | −9.55 | −10.74 | und | ... | −12.62 | ... | ... | 30.15 | 32.84 | 32.00 |
| 1996 Apr 13.5 | 146.7 | 5.39 | I | −11.02 | −11.17 | −10.58 | −10.34 | −10.86 | −12.50 | ... | −12.12 | 33.95 | 32.11 | 31.95 | 31.79 | 31.61 |
| 1996 Apr 14.4 | 73.7 | 5.09 | I | −11.41 | −11.30 | −11.05 | −10.66 | −11.63 | −12.84 | ... | −12.48 | 33.56 | 31.97 | 31.48 | 31.47 | 30.84 |
| 1996 Apr 14.4 | 73.7 | 5.09 | I | −11.18 | −11.65 | −11.08 | −10.67 | −11.52 | −12.81 | ... | −12.45 | 33.79 | 31.62 | 31.45 | 31.46 | 30.95 |
| 1996 Apr 28.5 | 114.7 | 5.25 | I | −11.39 | −11.48 | −10.71 | −10.44 | −10.86 | −12.52 | ... | −12.21 | 33.48 | 31.70 | 31.72 | 31.58 | 31.51 |
| 1996 Apr 29.5 | 57.6 | 4.94 | I | −11.34 | −11.54 | −11.14 | −10.80 | −11.68 | −12.83 | ... | −12.47 | 33.52 | 31.64 | 31.28 | 31.21 | 30.68 |
| 1996 May 13.8 | 54.6 | 4.88 | I | ... | ... | −10.99 | −10.66 | −11.30 | −12.73 | ... | −12.45 | ... | ... | 31.32 | 31.25 | 30.96 |
| 1996 May 13.8 | 54.6 | 4.88 | I | ... | ... | −11.14 | −10.88 | −11.57 | −12.68 | ... | −12.41 | ... | ... | 31.17 | 31.03 | 30.69 |
| 1996 May 16.4 | 114.7 | 5.20 | I | −10.86 | −11.13 | −10.52 | −10.25 | −10.87 | −12.47 | ... | −12.11 | 33.87 | 31.93 | 31.77 | 31.64 | 31.37 |
| 1996 May 20.4 | 81.4 | 5.04 | I | −11.13 | −11.24 | −10.74 | −10.34 | −11.38 | −12.62 | ... | −12.23 | 33.57 | 31.80 | 31.52 | 31.52 | 30.82 |
| 1996 May 22.8 | 54.6 | 4.86 | I | ... | ... | −11.00 | −10.45 | −11.76 | −12.70 | ... | −12.35 | ... | ... | 31.24 | 31.39 | 30.42 |
| 1996 May 22.8 | 54.6 | 4.86 | I | ... | ... | −11.30 | −10.47 | und | −12.72 | ... | −12.36 | ... | ... | 30.95 | 31.37 | und |
| 1996 May 28.3 | 51.7 | 4.82 | I | −11.25 | −11.75 | −10.97 | −10.63 | −11.35 | −12.66 | ... | −12.35 | 33.38 | 31.23 | 31.23 | 31.17 | 30.80 |
| 1996 May 28.4 | 146.7 | 5.27 | I | und | und | −10.35 | und | −10.40 | −12.16 | ... | −11.89 | und | und | 31.85 | und | 31.74 |
| 1996 May 28.4 | 36.7 | 4.67 | I | −11.69 | −11.59 | −11.21 | −10.75 | −12.15 | −12.87 | ... | −12.51 | 32.95 | 31.39 | 30.99 | 31.05 | 29.99 |
| 1996 May 28.4 | 51.7 | 4.82 | I | −11.53 | −11.41 | −10.99 | −10.58 | −11.65 | −12.74 | ... | −12.38 | 33.11 | 31.58 | 31.21 | 31.21 | 30.50 |
| 1996 May 29.3 | 51.7 | 4.81 | I | −11.31 | −11.25 | −10.96 | −10.69 | −11.10 | −12.59 | ... | −12.33 | 33.32 | 31.73 | 31.23 | 31.10 | 31.03 |
| 1996 May 29.4 | 104.1 | 5.12 | I | −11.54 | −11.19 | −10.51 | −10.46 | −10.87 | −12.25 | ... | −12.08 | 33.09 | 31.79 | 31.68 | 31.33 | 31.27 |
| 1996 May 29.4 | 51.7 | 4.81 | I | −11.84 | −11.56 | −10.99 | −10.86 | −11.37 | −12.60 | ... | −12.40 | 32.80 | 31.41 | 31.20 | 30.93 | 30.77 |
| 1996 May 29.4 | 36.7 | 4.67 | I | −11.78 | −11.82 | −11.29 | −11.23 | und | −12.79 | ... | −12.59 | 32.85 | 31.16 | 30.91 | 30.56 | und |
| 1996 May 29.4 | 51.7 | 4.81 | I | −11.31 | −11.31 | −10.96 | −10.54 | −11.54 | −12.72 | ... | −12.35 | 33.32 | 31.67 | 31.23 | 31.25 | 30.60 |
| 1996 Jun 1.3 | 57.6 | 4.85 | I | −10.33 | und | −10.87 | −10.48 | und | −12.45 | ... | −12.18 | 34.28 | und | 31.30 | 31.29 | und |
| 1996 Jun 1.3 | 28.7 | 4.55 | I | und | −11.80 | −11.45 | −10.82 | −12.19 | −12.85 | ... | −12.55 | und | 31.16 | 30.72 | 30.95 | 29.92 |
| 1996 Jun 1.4 | 57.6 | 4.85 | I | −11.34 | −10.97 | −10.79 | −10.53 | −11.07 | −12.68 | ... | −12.33 | 33.27 | 31.98 | 31.38 | 31.24 | 31.04 |
| 1996 Jun 1.4 | 40.4 | 4.70 | I | −11.51 | −11.11 | −11.09 | −10.57 | −12.75 | −12.88 | ... | −12.48 | 33.10 | 31.84 | 31.08 | 31.19 | 29.36 |
| 1996 Jun 1.4 | 81.4 | 5.00 | I | −10.74 | −11.17 | −10.68 | −10.31 | −11.06 | −12.39 | ... | −12.12 | 33.86 | 31.78 | 31.49 | 31.46 | 31.06 |
| 1996 Jun 1.4 | 57.6 | 4.85 | I | ... | −12.41 | −10.85 | −10.98 | −11.19 | −12.54 | ... | −12.30 | ... | 30.54 | 31.32 | 30.79 | 30.93 |
| 1996 Jun 3.3 | 57.6 | 4.85 | I | und | und | −10.80 | −10.28 | und | −12.80 | ... | −12.23 | und | und | 31.35 | 31.48 | und |
| 1996 Jun 3.3 | 28.7 | 4.54 | I | −11.54 | −11.81 | −11.36 | −10.77 | und | −12.76 | ... | −12.43 | 33.05 | 31.13 | 30.80 | 30.98 | und |
| 1996 Jun 3.4 | 40.4 | 4.69 | I | und | −11.53 | −11.02 | −10.49 | −12.75 | −12.63 | ... | −12.29 | und | 31.41 | 31.13 | 31.26 | 29.35 |
| 1996 Jun 3.4 | 81.4 | 5.00 | I | −11.02 | −11.74 | −10.61 | −10.34 | −11.23 | −12.24 | ... | −11.97 | 33.57 | 31.20 | 31.54 | 31.41 | 30.87 |
| 1996 Jun 3.4 | 20.5 | 4.40 | I | −11.85 | −12.09 | −11.54 | −10.84 | und | −12.95 | ... | −12.59 | 32.74 | 30.86 | 30.62 | 30.91 | und |
| 1996 Jun 3.4 | 57.6 | 4.85 | I | und | −10.94 | −10.83 | −10.51 | −11.82 | −12.46 | ... | −12.16 | und | 32.01 | 31.32 | 31.24 | 30.28 |

Table 2—Continued

| UT Date | Aperture | | Filt.[a] | log Emission Band Flux[b] | | | | | log Continuum Flux[b] | | | log $M(\rho)$[b] | | | | |
|---|---|---|---|---|---|---|---|---|---|---|---|---|---|---|---|---|
| | Size | log $\rho$ | | (erg cm$^{-2}$ s$^{-1}$) | | | | | (erg cm$^{-2}$ s$^{-1}$ Å$^{-1}$) | | | (molecule) | | | | |
| | (arcsec) | (km) | | OH | NH | CN | $C_3$ | $C_2$ | UV | Blue | Green | OH | NH | CN | $C_3$ | $C_2$ |
| 1996 Jun  6.3 | 57.6 | 4.84 | I | −11.66 | −11.97 | −10.85 | −10.48 | −11.37 | −12.54 | ... | −12.21 | 32.91 | 30.94 | 31.29 | 31.25 | 30.70 |
| 1996 Jun  6.4 | 57.6 | 4.84 | I | −11.55 | −11.53 | −10.89 | −10.50 | −11.57 | −12.54 | ... | −12.21 | 33.02 | 31.39 | 31.25 | 31.23 | 30.51 |
| 1996 Jun  6.4 | 40.4 | 4.69 | I | −11.66 | −11.38 | −11.11 | −10.70 | und | −12.70 | ... | −12.35 | 32.92 | 31.54 | 31.02 | 31.03 | und |
| 1996 Jun  7.3 | 57.6 | 4.84 | I | −11.51 | −11.95 | −10.94 | −10.55 | und | −12.60 | ... | −12.23 | 33.06 | 30.96 | 31.19 | 31.18 | und |
| 1996 Jun  7.4 | 57.6 | 4.84 | I | −11.26 | −11.69 | −10.83 | −10.46 | −11.89 | −12.57 | ... | −12.24 | 33.30 | 31.23 | 31.29 | 31.26 | 30.18 |
| 1996 Jun  9.3 | 57.6 | 4.83 | I | −12.02 | −11.48 | −10.78 | −10.45 | −11.19 | −12.49 | ... | −12.17 | 32.52 | 31.42 | 31.34 | 31.26 | 30.87 |
| 1996 Jun  9.3 | 114.7 | 5.13 | I | −10.76 | −11.05 | −10.32 | −10.10 | −10.58 | −12.23 | ... | −11.91 | 33.78 | 31.85 | 31.79 | 31.61 | 31.48 |
| 1996 Jun  9.3 | 28.7 | 4.53 | I | −12.34 | −11.61 | −11.29 | −10.90 | −11.99 | −12.83 | ... | −12.48 | 32.21 | 31.29 | 30.82 | 30.81 | 30.06 |
| 1996 Jun  9.4 | 57.6 | 4.83 | I | −11.62 | −11.30 | −10.80 | −10.36 | −11.71 | −12.50 | ... | −12.16 | 32.93 | 31.60 | 31.31 | 31.35 | 30.35 |
| 1996 Jun 10.3 | 57.6 | 4.83 | I | −10.92 | −11.32 | −10.79 | −10.38 | −11.41 | −12.50 | ... | −12.17 | 33.62 | 31.57 | 31.31 | 31.32 | 30.64 |
| 1996 Jun 10.3 | 40.4 | 4.68 | I | −11.11 | −10.99 | −11.01 | −10.29 | und | −12.83 | ... | −12.33 | 33.43 | 31.90 | 31.10 | 31.41 | und |
| 1996 Jun 10.3 | 81.4 | 4.98 | I | −11.03 | −11.06 | −10.54 | −10.27 | −11.06 | −12.36 | ... | −12.04 | 33.51 | 31.83 | 31.56 | 31.43 | 30.99 |
| 1996 Jun 10.3 | 28.7 | 4.53 | I | −11.96 | −12.18 | −11.29 | −10.89 | −12.67 | −12.78 | ... | −12.46 | 32.58 | 30.71 | 30.81 | 30.82 | 29.37 |
| 1996 Jun 10.4 | 57.6 | 4.83 | I | und | und | −10.78 | −10.78 | −10.98 | ... | ... | −12.16 | und | und | 31.32 | 30.92 | 31.07 |
| 1996 Jun 17.3 | 51.7 | 4.77 | I | −11.16 | −11.33 | −10.77 | −10.40 | −11.37 | −12.42 | ... | −12.10 | 33.33 | 31.52 | 31.28 | 31.25 | 30.63 |
| 1996 Jun 17.3 | 104.1 | 5.07 | I | −10.87 | −11.34 | −10.31 | −10.04 | −10.63 | −12.18 | ... | −11.84 | 33.62 | 31.50 | 31.75 | 31.61 | 31.36 |
| 1996 Jun 17.4 | 73.7 | 4.92 | I | −11.12 | −11.12 | −10.56 | −10.19 | −10.98 | −12.34 | ... | −12.00 | 33.37 | 31.73 | 31.49 | 31.46 | 31.02 |
| 1996 Jun 17.4 | 36.7 | 4.62 | I | −11.37 | −11.38 | −11.03 | −10.56 | −12.19 | −12.57 | ... | −12.24 | 33.12 | 31.47 | 31.02 | 31.09 | 29.81 |
| 1996 Jun 17.4 | 51.7 | 4.77 | I | −11.16 | −11.08 | −10.77 | −10.37 | −11.55 | −12.47 | ... | −12.12 | 33.33 | 31.76 | 31.28 | 31.28 | 30.45 |
| 1996 Jun 17.4 | 51.7 | 4.77 | I | −11.35 | −11.12 | −10.76 | −10.38 | −11.52 | −12.46 | ... | −12.11 | 33.14 | 31.73 | 31.29 | 31.27 | 30.47 |
| 1996 Jun 18.3 | 57.6 | 4.81 | I | −11.21 | −11.18 | −10.72 | −10.37 | −11.32 | −12.41 | ... | −12.09 | 33.27 | 31.66 | 31.33 | 31.28 | 30.67 |
| 1996 Jun 18.3 | 81.4 | 4.96 | I | −10.91 | −11.00 | −10.49 | −10.14 | −10.91 | −12.30 | ... | −11.97 | 33.58 | 31.84 | 31.56 | 31.51 | 31.09 |
| 1996 Jun 18.3 | 57.6 | 4.81 | I | −11.09 | −11.19 | −10.69 | −10.31 | −11.21 | −12.40 | ... | −12.07 | 33.39 | 31.65 | 31.36 | 31.34 | 30.79 |
| 1996 Jun 20.3 | 40.4 | 4.65 | I | −11.03 | −11.13 | −10.73 | −10.40 | −11.03 | −12.43 | ... | −12.10 | 33.44 | 31.69 | 31.31 | 31.24 | 30.94 |
| 1996 Jun 20.3 | 57.6 | 4.81 | I | −11.13 | −11.07 | −10.71 | −10.39 | −11.08 | −12.40 | ... | −12.09 | 33.34 | 31.75 | 31.32 | 31.24 | 30.90 |
| 1996 Jun 22.3 | 57.6 | 4.80 | I | −11.16 | −11.22 | −10.68 | −10.41 | −11.11 | −12.42 | ... | −12.10 | 33.29 | 31.60 | 31.34 | 31.21 | 30.85 |
| 1996 Jun 23.3 | 114.7 | 5.10 | I | −10.63 | −10.82 | −10.22 | −9.99 | −10.50 | −12.12 | ... | −11.80 | 33.82 | 31.99 | 31.80 | 31.63 | 31.46 |
| 1996 Jun 23.3 | 57.6 | 4.80 | I | −11.17 | −11.15 | −10.68 | −10.34 | −11.00 | −12.42 | ... | −12.11 | 33.28 | 31.66 | 31.33 | 31.28 | 30.96 |
| 1996 Jul 11.5 | 54.6 | 4.75 | I | ... | ... | −10.64 | −10.20 | −11.06 | −12.37 | ... | −12.06 | ... | ... | 31.27 | 31.31 | 30.79 |
| 1996 Jul 11.5 | 54.6 | 4.75 | I | ... | ... | −10.54 | −10.22 | −11.33 | −12.40 | ... | −12.06 | ... | ... | 31.37 | 31.29 | 30.52 |
| 1996 Jul 11.6 | 54.6 | 4.75 | I | ... | ... | −10.63 | −10.05 | und | −12.55 | ... | −12.08 | ... | ... | 31.29 | 31.46 | und |
| 1996 Jul 18.2 | 104.1 | 5.02 | I | −10.52 | −10.82 | −10.13 | −9.88 | −10.40 | −12.16 | ... | −11.82 | 33.78 | 31.86 | 31.76 | 31.60 | 31.42 |
| 1996 Jul 18.2 | 51.7 | 4.72 | I | −10.99 | −11.12 | −10.61 | −10.23 | −11.10 | −12.47 | ... | −12.11 | 33.31 | 31.57 | 31.28 | 31.24 | 30.73 |
| 1996 Jul 18.3 | 146.7 | 5.17 | I | −10.30 | −10.77 | −9.90 | −9.68 | −10.11 | −12.01 | ... | −11.66 | 34.01 | 31.91 | 31.99 | 31.79 | 31.72 |
| 1996 Jul 18.4 | 51.7 | 4.72 | I | −11.02 | −11.12 | −10.59 | −10.25 | −10.95 | −12.44 | ... | −12.11 | 33.28 | 31.57 | 31.29 | 31.23 | 30.88 |
| 1996 Aug  6.2 | 57.6 | 4.76 | I | −10.71 | −10.91 | −10.39 | −10.12 | −10.72 | −12.39 | ... | −12.06 | 33.52 | 31.71 | 31.43 | 31.30 | 31.05 |
| 1996 Aug  6.2 | 114.7 | 5.06 | I | −10.24 | −10.61 | −9.90 | −9.72 | −10.11 | −12.10 | ... | −11.78 | 33.99 | 32.01 | 31.92 | 31.69 | 31.66 |
| 1996 Aug  6.2 | 40.4 | 4.60 | I | −10.98 | −11.12 | −10.64 | −10.30 | −10.93 | −12.55 | ... | −12.21 | 33.25 | 31.50 | 31.18 | 31.12 | 30.83 |
| 1996 Aug  6.3 | 57.6 | 4.76 | I | −10.79 | −11.02 | −10.40 | −10.12 | −10.66 | −12.41 | ... | −12.08 | 33.44 | 31.60 | 31.42 | 31.30 | 31.10 |
| 1996 Aug  6.3 | 81.4 | 4.91 | I | −10.46 | −10.79 | −10.15 | −9.91 | −10.41 | −12.27 | ... | −11.93 | 33.77 | 31.83 | 31.68 | 31.50 | 31.35 |
| 1996 Aug  7.2 | 57.6 | 4.76 | I | −10.71 | −10.97 | −10.40 | −10.09 | −10.72 | −12.42 | ... | −12.07 | 33.52 | 31.65 | 31.42 | 31.32 | 31.04 |
| 1996 Aug  7.2 | 114.7 | 5.06 | I | −10.22 | −10.60 | −9.91 | −9.71 | −10.14 | −12.13 | ... | −11.79 | 34.00 | 32.02 | 31.91 | 31.70 | 31.62 |
| 1996 Aug  7.2 | 28.7 | 4.45 | I | −11.32 | −11.48 | −10.93 | −10.51 | −11.82 | −12.74 | ... | −12.39 | 32.91 | 31.14 | 30.89 | 30.90 | 29.94 |
| 1996 Aug  7.3 | 57.6 | 4.76 | I | −10.74 | −11.06 | −10.39 | −10.11 | −10.68 | −12.41 | ... | −12.08 | 33.48 | 31.56 | 31.43 | 31.30 | 31.08 |

Table 2—Continued

| UT Date | Aperture | | Filt.[a] | log Emission Band Flux[b] | | | | | log Continuum Flux[b] | | | log $M(\rho)$[b] | | | | |
|---|---|---|---|---|---|---|---|---|---|---|---|---|---|---|---|---|
| | Size | log $\rho$ | | (erg cm$^{-2}$ s$^{-1}$) | | | | | (erg cm$^{-2}$ s$^{-1}$ Å$^{-1}$) | | | (molecule) | | | | |
| | (arcsec) | (km) | | OH | NH | CN | $C_3$ | $C_2$ | UV | Blue | Green | OH | NH | CN | $C_3$ | $C_2$ |
| 1996 Aug 8.2 | 40.4 | 4.60 | I | −10.70 | −10.93 | −10.40 | −10.11 | −10.73 | −12.42 | ... | −12.08 | 33.52 | 31.69 | 31.42 | 31.30 | 31.03 |
| 1996 Aug 8.2 | 114.7 | 5.06 | I | −10.23 | −10.62 | −9.91 | −9.71 | −10.14 | −12.13 | ... | −11.78 | 34.00 | 32.00 | 31.91 | 31.70 | 31.62 |
| 1996 Aug 8.2 | 28.7 | 4.45 | I | −11.14 | −11.37 | −10.92 | −10.57 | −11.45 | −12.72 | ... | −12.36 | 33.09 | 31.24 | 30.90 | 30.84 | 30.31 |
| 1996 Aug 20.3 | 114.7 | 5.06 | I | −10.15 | −10.61 | −9.85 | −9.71 | −10.01 | −12.11 | ... | −11.79 | 34.04 | 31.99 | 31.95 | 31.68 | 31.72 |
| 1996 Aug 20.3 | 57.6 | 4.76 | I | −10.73 | −11.13 | −10.31 | −10.02 | −10.59 | −12.38 | ... | −12.03 | 33.46 | 31.47 | 31.48 | 31.37 | 31.15 |
| 1996 Sep 4.2 | 57.6 | 4.77 | I | −10.46 | −10.75 | −10.17 | −9.86 | −10.42 | −12.25 | ... | −11.90 | 33.69 | 31.83 | 31.60 | 31.51 | 31.29 |
| 1996 Sep 4.2 | 114.7 | 5.07 | I | −9.99 | −10.33 | −9.74 | −9.54 | −9.88 | −12.10 | ... | −11.69 | 34.17 | 32.25 | 32.03 | 31.82 | 31.84 |
| 1996 Sep 12.5 | 77.8 | 4.91 | I | ... | ... | −9.93 | −9.84 | −10.09 | −12.00 | ... | −11.71 | ... | ... | 31.83 | 31.51 | 31.61 |
| 1996 Sep 12.5 | 77.8 | 4.91 | I | ... | ... | −10.00 | −9.82 | −10.11 | −12.00 | ... | −11.70 | ... | ... | 31.75 | 31.53 | 31.59 |
| 1996 Sep 20.1 | 57.6 | 4.79 | I | −10.26 | −10.69 | −10.07 | −9.80 | −10.29 | −12.26 | ... | −11.89 | 33.86 | 31.87 | 31.67 | 31.54 | 31.40 |
| 1996 Sep 20.1 | 57.6 | 4.79 | I | −10.27 | −10.71 | −10.07 | −9.83 | −10.24 | −12.25 | ... | −11.91 | 33.85 | 31.85 | 31.67 | 31.51 | 31.45 |
| 1996 Oct 4.1 | 57.6 | 4.80 | I | −10.07 | −10.55 | −9.97 | −9.72 | −10.14 | −12.15 | ... | −11.77 | 34.01 | 31.99 | 31.73 | 31.59 | 31.52 |
| 1996 Oct 4.1 | 57.6 | 4.80 | I | −10.12 | −10.69 | −9.98 | −9.78 | −10.15 | −12.13 | ... | −11.78 | 33.96 | 31.85 | 31.73 | 31.53 | 31.51 |
| 1996 Oct 13.5 | 77.8 | 4.93 | I | ... | ... | −9.69 | −9.51 | −9.90 | −12.07 | ... | −11.67 | ... | ... | 31.99 | 31.77 | 31.74 |
| 1996 Oct 13.5 | 54.6 | 4.78 | I | ... | ... | −9.93 | −9.86 | −10.08 | −12.06 | ... | −11.78 | ... | ... | 31.75 | 31.43 | 31.56 |
| 1996 Oct 16.1 | 57.6 | 4.80 | I | −9.97 | −10.50 | −9.89 | −9.67 | −9.95 | −12.12 | ... | −11.77 | 34.07 | 32.00 | 31.78 | 31.61 | 31.68 |
| 1996 Oct 16.1 | 81.4 | 4.95 | I | −9.72 | −10.34 | −9.65 | −9.52 | −9.70 | −11.97 | ... | −11.62 | 34.32 | 32.16 | 32.02 | 31.76 | 31.93 |
| 1996 Oct 18.1 | 40.4 | 4.65 | I | −10.18 | −10.65 | −10.08 | −9.83 | −10.27 | −12.20 | ... | −11.86 | 33.86 | 31.85 | 31.58 | 31.44 | 31.34 |
| 1996 Oct 18.1 | 40.4 | 4.65 | I | −10.17 | −10.64 | −10.07 | −9.80 | −10.21 | −12.21 | ... | −11.85 | 33.86 | 31.85 | 31.60 | 31.47 | 31.41 |
| 1996 Nov 5.1 | 57.6 | 4.80 | I | −9.73 | −10.34 | −9.70 | −9.49 | −9.74 | −11.95 | ... | −11.60 | 34.23 | 32.09 | 31.88 | 31.71 | 31.81 |
| 1996 Nov 5.1 | 81.4 | 4.95 | I | −9.50 | −10.14 | −9.48 | −9.36 | −9.49 | −11.82 | ... | −11.48 | 34.46 | 32.28 | 32.10 | 31.84 | 32.06 |
| 1996 Nov 6.1 | 57.6 | 4.80 | I | −9.74 | −10.34 | −9.70 | −9.52 | −9.78 | −11.97 | ... | −11.61 | 34.21 | 32.07 | 31.88 | 31.68 | 31.77 |
| 1996 Nov 6.1 | 40.4 | 4.65 | I | −10.02 | −10.69 | −9.93 | −9.71 | −10.06 | −12.09 | ... | −11.73 | 33.93 | 31.72 | 31.64 | 31.48 | 31.49 |
| 1996 Nov 6.1 | 114.7 | 5.10 | I | −9.26 | −9.96 | −9.26 | −9.23 | −9.29 | −11.69 | ... | −11.34 | 34.69 | 32.46 | 32.31 | 31.97 | 32.25 |
| 1996 Nov 7.1 | 57.6 | 4.80 | I | −9.76 | −10.32 | −9.71 | −9.49 | −9.73 | −11.96 | ... | −11.61 | 34.19 | 32.10 | 31.85 | 31.71 | 31.81 |
| 1996 Nov 7.1 | 57.6 | 4.80 | I | −9.75 | −10.33 | −9.71 | −9.53 | −9.70 | −11.95 | ... | −11.61 | 34.19 | 32.08 | 31.86 | 31.66 | 31.84 |
| 1996 Nov 7.1 | 114.7 | 5.10 | I | −9.31 | −9.98 | −9.28 | −9.26 | −9.25 | −11.70 | ... | −11.36 | 34.63 | 32.44 | 32.29 | 31.93 | 32.28 |
| 1996 Nov 19.1 | 57.6 | 4.80 | I | −9.63 | −10.22 | −9.59 | −9.45 | −9.57 | −11.80 | ... | −11.46 | 34.24 | 32.12 | 31.91 | 31.68 | 31.91 |
| 1996 Nov 19.1 | 40.4 | 4.64 | I | −9.76 | −10.38 | −9.83 | −9.61 | −9.84 | −11.94 | ... | −11.58 | 34.11 | 31.96 | 31.67 | 31.52 | 31.63 |
| 1996 Nov 20.1 | 57.6 | 4.80 | I | −9.57 | −10.25 | −9.59 | −9.42 | −9.59 | −11.80 | ... | −11.46 | 34.29 | 32.09 | 31.91 | 31.70 | 31.88 |
| 1996 Nov 20.1 | 57.6 | 4.80 | I | −9.56 | −10.26 | −9.59 | −9.40 | −9.63 | −11.80 | ... | −11.44 | 34.30 | 32.08 | 31.90 | 31.72 | 31.84 |
| 1996 Nov 21.1 | 57.6 | 4.80 | I | −9.52 | −10.17 | −9.57 | −9.37 | −9.57 | −11.81 | ... | −11.46 | 34.33 | 32.16 | 31.91 | 31.74 | 31.89 |
| 1996 Nov 21.1 | 114.7 | 5.09 | I | −9.10 | −9.80 | −9.14 | −9.05 | −9.12 | −11.60 | ... | −11.21 | 34.76 | 32.54 | 32.34 | 32.06 | 32.34 |
| 1996 Nov 28.1 | 57.6 | 4.79 | I | −9.55 | −10.29 | −9.54 | −9.42 | −9.48 | −11.75 | ... | −11.42 | 34.26 | 31.99 | 31.90 | 31.64 | 31.93 |
| 1996 Nov 28.1 | 57.6 | 4.79 | I | ... | ... | −9.53 | −9.32 | −9.48 | −11.78 | ... | −11.42 | ... | ... | 31.90 | 31.75 | 31.93 |
| 1996 Dec 3.1 | 57.6 | 4.78 | I | ... | −10.15 | −9.44 | −9.26 | −9.31 | −11.68 | ... | −11.34 | ... | 32.10 | 31.96 | 31.77 | 32.06 |
| 1996 Dec 4.1 | 57.6 | 4.78 | H | ... | ... | −9.49 | −9.36 | −9.50 | ... | −11.49 | −11.64 | ... | ... | 31.90 | 31.66 | 31.86 |
| 1996 Dec 4.1 | 57.6 | 4.78 | I | ... | −10.02 | −9.44 | −9.37 | −9.36 | −11.66 | ... | −11.33 | ... | 32.21 | 31.95 | 31.65 | 32.01 |
| 1996 Dec 5.1 | 57.6 | 4.78 | H | ... | ... | −9.47 | −9.34 | −9.34 | ... | −11.33 | −11.33 | ... | ... | 31.91 | 31.67 | 32.02 |
| 1996 Dec 5.1 | 57.6 | 4.78 | I | ... | −9.96 | −9.42 | −9.23 | −9.37 | −11.67 | ... | −11.32 | ... | 32.26 | 31.96 | 31.79 | 31.99 |
| 1997 Jan 17.6 | 73.7 | 4.79 | H | ... | ... | −8.54 | −9.41 | −8.40 | ... | −11.63 | −10.72 | ... | ... | 32.37 | 31.13 | 32.49 |
| 1997 Jan 17.6 | 73.7 | 4.79 | I | −8.02 | −9.21 | −8.64 | −8.66 | −8.58 | −11.02 | ... | −10.70 | 35.27 | 32.53 | 32.27 | 31.89 | 32.31 |
| 1997 Jan 19.6 | 73.7 | 4.78 | H | ... | ... | −8.61 | −8.68 | −8.50 | ... | −10.69 | −10.71 | ... | ... | 32.27 | 31.83 | 32.36 |
| 1997 Jan 19.6 | 73.7 | 4.78 | I | −8.12 | −9.28 | −8.60 | −8.64 | −8.48 | −10.99 | ... | −10.69 | 35.14 | 32.44 | 32.29 | 31.88 | 32.38 |

Table 2—Continued

| UT Date | Aperture | | Filt.[a] | log Emission Band Flux[b] | | | | | log Continuum Flux[b] | | | log $M(\rho)$[b] | | | | |
|---|---|---|---|---|---|---|---|---|---|---|---|---|---|---|---|---|
| | Size | log $\rho$ | | (erg cm$^{-2}$ s$^{-1}$) | | | | | (erg cm$^{-2}$ s$^{-1}$ Å$^{-1}$) | | | (molecule) | | | | |
| | (arcsec) | (km) | | OH | NH | CN | $C_3$ | $C_2$ | UV | Blue | Green | OH | NH | CN | $C_3$ | $C_2$ |
| 1997 Feb 3.5 | 57.6 | 4.61 | H | ... | ... | −8.39 | −8.45 | −8.28 | ... | −10.41 | −10.41 | ... | ... | 32.28 | 31.84 | 32.35 |
| 1997 Feb 3.5 | 57.6 | 4.61 | I | −8.39 | −8.95 | −8.37 | −8.44 | −8.27 | −10.70 | ... | −10.40 | 34.67 | 32.52 | 32.30 | 31.84 | 32.36 |
| 1997 Feb 3.5 | 57.6 | 4.61 | H | ... | ... | −8.40 | −8.46 | −8.28 | ... | −10.39 | −10.41 | ... | ... | 32.27 | 31.82 | 32.35 |
| 1997 Feb 3.5 | 57.6 | 4.61 | I | −8.43 | −8.95 | −8.38 | −8.43 | −8.26 | −10.71 | ... | −10.41 | 34.62 | 32.52 | 32.29 | 31.86 | 32.37 |
| 1997 Feb 12.5 | 73.7 | 4.68 | H | −7.99 | −8.79 | −8.01 | −8.17 | −7.88 | −10.43 | −10.06 | −10.08 | 34.95 | 32.52 | 32.53 | 31.96 | 32.60 |
| 1997 Feb 12.5 | 73.7 | 4.68 | I | −7.96 | −8.65 | −7.99 | −8.12 | −7.88 | −10.36 | ... | −10.07 | 34.98 | 32.66 | 32.55 | 32.01 | 32.60 |
| 1997 Feb 12.5 | 146.7 | 4.98 | H | −7.60 | −8.38 | −7.60 | −8.02 | −7.53 | −10.13 | −9.78 | −9.79 | 35.34 | 32.92 | 32.94 | 32.11 | 32.95 |
| 1997 Feb 12.5 | 146.7 | 4.98 | I | −7.61 | −8.34 | −7.60 | −7.93 | −7.54 | −10.07 | ... | −9.79 | 35.33 | 32.97 | 32.94 | 32.20 | 32.93 |
| 1997 Feb 12.6 | 36.7 | 4.37 | H | −8.49 | −9.22 | −8.51 | −8.44 | −8.34 | −10.74 | −10.35 | −10.37 | 34.45 | 32.09 | 32.03 | 31.69 | 32.13 |
| 1997 Feb 12.6 | 36.7 | 4.37 | I | −8.46 | −9.06 | −8.49 | −8.38 | −8.31 | −10.67 | ... | −10.38 | 34.48 | 32.25 | 32.05 | 31.75 | 32.17 |
| 1997 Feb 12.6 | 36.7 | 4.37 | I | −8.47 | −9.02 | −8.48 | −8.41 | −8.34 | −10.65 | ... | −10.35 | 34.47 | 32.29 | 32.06 | 31.72 | 32.14 |
| 1997 Feb 15.5 | 73.7 | 4.66 | H | −7.91 | −8.68 | −7.94 | −8.07 | −7.82 | −10.34 | −9.98 | −10.00 | 35.00 | 32.57 | 32.56 | 32.01 | 32.61 |
| 1997 Feb 15.5 | 73.7 | 4.66 | I | −7.86 | −8.70 | −7.92 | −8.00 | −7.81 | −10.27 | ... | −9.97 | 35.05 | 32.56 | 32.58 | 32.08 | 32.62 |
| 1997 Feb 15.5 | 36.7 | 4.36 | H | −8.39 | −9.14 | −8.41 | −8.34 | −8.26 | −10.65 | −10.29 | −10.30 | 34.51 | 32.12 | 32.09 | 31.73 | 32.17 |
| 1997 Feb 15.5 | 36.7 | 4.36 | I | −8.33 | −9.06 | −8.39 | −8.28 | −8.25 | −10.58 | ... | −10.29 | 34.58 | 32.19 | 32.11 | 31.80 | 32.17 |
| 1997 Feb 15.5 | 104.1 | 4.81 | H | −7.71 | −8.49 | −7.71 | −7.98 | −7.62 | −10.19 | −9.82 | −9.84 | 35.19 | 32.76 | 32.79 | 32.10 | 32.80 |
| 1997 Feb 15.5 | 104.1 | 4.81 | I | −7.72 | −8.53 | −7.70 | −7.90 | −7.62 | −10.12 | ... | −9.82 | 35.19 | 32.72 | 32.80 | 32.18 | 32.81 |
| 1997 Feb 15.5 | 146.7 | 4.96 | H | −7.54 | −8.29 | −7.52 | −7.92 | −7.47 | −10.04 | −9.68 | −9.69 | 35.37 | 32.96 | 32.98 | 32.16 | 32.95 |
| 1997 Feb 15.5 | 146.7 | 4.96 | I | −7.55 | −8.44 | −7.52 | −7.83 | −7.46 | −9.97 | ... | −9.69 | 35.36 | 32.82 | 32.98 | 32.25 | 32.97 |
| 1997 Feb 15.6 | 51.7 | 4.51 | H | −8.18 | −8.90 | −8.15 | −8.14 | −8.01 | −10.46 | −10.11 | −10.13 | 34.72 | 32.35 | 32.35 | 31.94 | 32.42 |
| 1997 Feb 15.6 | 51.7 | 4.51 | I | −8.16 | −8.84 | −8.15 | −8.07 | −8.01 | −10.41 | ... | −10.09 | 34.75 | 32.41 | 32.35 | 32.01 | 32.41 |
| 1997 Feb 15.6 | 146.7 | 4.96 | H | −7.57 | −8.36 | −7.51 | −7.97 | und | −9.95 | −9.65 | −9.67 | 35.34 | 32.90 | 32.99 | 32.10 | und |
| 1997 Feb 15.6 | 36.7 | 4.36 | I | −8.46 | −9.03 | −8.39 | −8.28 | −8.26 | −10.60 | ... | −10.29 | 34.45 | 32.22 | 32.11 | 31.80 | 32.17 |
| 1997 Feb 26.5 | 57.6 | 4.50 | I | −7.97 | −8.53 | −7.83 | −7.90 | −7.68 | −10.04 | ... | −9.75 | 34.71 | 32.50 | 32.50 | 31.99 | 32.56 |
| 1997 Feb 26.5 | 81.4 | 4.65 | I | −7.77 | −8.34 | −7.61 | −7.80 | −7.49 | −9.90 | ... | −9.62 | 34.92 | 32.69 | 32.72 | 32.10 | 32.74 |
| 1997 Feb 26.5 | 114.7 | 4.80 | I | −7.57 | −8.22 | −7.41 | −7.69 | −7.33 | −9.76 | ... | −9.48 | 35.12 | 32.81 | 32.92 | 32.21 | 32.90 |
| 1997 Feb 26.5 | 40.4 | 4.35 | I | −8.23 | −8.71 | −8.09 | −8.05 | −7.92 | −10.21 | ... | −9.93 | 34.46 | 32.32 | 32.25 | 31.84 | 32.32 |
| 1997 Feb 26.5 | 57.6 | 4.50 | I | −8.01 | −8.53 | −7.85 | −7.91 | −7.68 | −10.06 | ... | −9.78 | 34.68 | 32.50 | 32.48 | 31.98 | 32.55 |
| 1997 Feb 26.6 | 28.7 | 4.20 | I | −8.52 | −8.95 | −8.37 | −8.26 | −8.21 | −10.41 | ... | −10.10 | 34.16 | 32.08 | 31.96 | 31.63 | 32.03 |
| 1997 Mar 5.5 | 28.7 | 4.17 | I | ... | ... | −8.22 | −8.30 | −8.07 | ... | ... | −9.97 | ... | ... | 32.01 | 31.48 | 32.06 |
| 1997 Mar 5.5 | 57.6 | 4.47 | H | −7.74 | ... | −7.74 | −7.79 | −7.63 | −9.98 | −9.61 | −9.63 | 34.65 | ... | 32.48 | 31.99 | 32.50 |
| 1997 Mar 5.5 | 57.6 | 4.47 | I | −7.63 | −8.45 | −7.72 | −7.80 | −7.62 | −9.89 | ... | −9.61 | 34.76 | 32.48 | 32.51 | 31.99 | 32.51 |
| 1997 Mar 5.5 | 81.4 | 4.62 | H | −7.50 | ... | −7.51 | −7.68 | −7.43 | −9.82 | −9.47 | −9.49 | 34.89 | ... | 32.72 | 32.11 | 32.70 |
| 1997 Mar 5.5 | 81.4 | 4.62 | I | −7.49 | −8.21 | −7.49 | −7.65 | −7.43 | −9.75 | ... | −9.46 | 34.89 | 32.72 | 32.73 | 32.14 | 32.70 |
| 1997 Mar 5.5 | 40.4 | 4.32 | H | −8.02 | ... | −8.00 | −7.97 | −7.85 | −10.20 | −9.80 | −9.84 | 34.36 | ... | 32.22 | 31.81 | 32.28 |
| 1997 Mar 5.5 | 40.4 | 4.32 | I | −7.98 | −8.54 | −7.99 | −7.94 | −7.89 | −10.12 | ... | −9.81 | 34.40 | 32.39 | 32.23 | 31.84 | 32.24 |
| 1997 Mar 5.5 | 57.6 | 4.47 | H | −7.77 | ... | −7.77 | −7.80 | −7.64 | −10.02 | −9.65 | −9.64 | 34.61 | ... | 32.46 | 31.98 | 32.49 |
| 1997 Mar 5.5 | 57.6 | 4.47 | I | −7.79 | −8.41 | −7.75 | −7.76 | −7.61 | −9.94 | ... | −9.67 | 34.59 | 32.52 | 32.48 | 32.02 | 32.52 |
| 1997 Mar 5.5 | 28.7 | 4.17 | H | −8.28 | ... | −8.28 | −8.11 | −8.12 | −10.37 | −9.97 | −9.97 | 34.11 | ... | 31.94 | 31.67 | 32.01 |
| 1997 Mar 5.5 | 28.7 | 4.17 | I | −8.26 | −8.74 | −8.25 | −7.92 | −8.19 | −10.31 | ... | −9.96 | 34.12 | 32.20 | 31.97 | 31.86 | 31.94 |
| 1997 Mar 6.5 | 57.6 | 4.47 | H | −7.79 | ... | −7.77 | −7.85 | −7.67 | −10.02 | −9.64 | −9.63 | 34.56 | ... | 32.44 | 31.91 | 32.44 |
| 1997 Mar 6.5 | 57.6 | 4.47 | I | −7.63 | −8.50 | −7.74 | −7.74 | −7.64 | −9.95 | ... | −9.63 | 34.72 | 32.42 | 32.47 | 32.03 | 32.48 |
| 1997 Mar 6.5 | 28.7 | 4.17 | H | −8.38 | ... | −8.28 | −8.16 | −8.08 | −10.37 | −10.00 | −9.99 | 33.97 | ... | 31.94 | 31.61 | 32.04 |
| 1997 Mar 6.5 | 28.7 | 4.17 | I | −8.18 | −9.00 | −8.29 | −8.35 | −8.13 | −10.24 | ... | −9.92 | 34.17 | 31.91 | 31.93 | 31.42 | 31.99 |

Table 2—Continued

| UT Date | Aperture | | Filt.[a] | log Emission Band Flux[b] | | | | | log Continuum Flux[b] | | | log $M(\rho)$[b] | | | | |
| | Size | log $\rho$ | | (erg cm$^{-2}$ s$^{-1}$) | | | | | (erg cm$^{-2}$ s$^{-1}$ Å$^{-1}$) | | | (molecule) | | | | |
| | (arcsec) | (km) | | OH | NH | CN | $C_3$ | $C_2$ | UV | Blue | Green | OH | NH | CN | $C_3$ | $C_2$ |
|---|---|---|---|---|---|---|---|---|---|---|---|---|---|---|---|---|
| 1997 Mar 6.5 | 81.4 | 4.62 | H | −7.54 | ... | −7.52 | −7.74 | −7.44 | −9.82 | −9.44 | −9.45 | 34.81 | ... | 32.69 | 32.03 | 32.67 |
| 1997 Mar 6.5 | 81.4 | 4.62 | I | −7.53 | −8.32 | −7.52 | −7.71 | −7.45 | −9.75 | ... | −9.45 | 34.82 | 32.59 | 32.70 | 32.06 | 32.66 |
| 1997 Mar 6.5 | 57.6 | 4.47 | H | −7.79 | ... | −7.77 | −7.86 | −7.63 | −9.98 | −9.61 | −9.63 | 34.56 | ... | 32.44 | 31.91 | 32.48 |
| 1997 Mar 6.5 | 57.6 | 4.47 | I | −7.79 | −8.58 | −7.74 | −7.81 | −7.60 | −9.89 | ... | −9.61 | 34.57 | 32.34 | 32.47 | 31.96 | 32.52 |
| 1997 Mar 6.5 | 40.4 | 4.32 | H | −8.03 | ... | −8.02 | −8.05 | −7.88 | −10.15 | −9.79 | −9.80 | 34.32 | ... | 32.20 | 31.72 | 32.24 |
| 1997 Mar 6.5 | 40.4 | 4.32 | I | −8.02 | −8.57 | −8.02 | −8.06 | −7.89 | −10.09 | ... | −9.77 | 34.33 | 32.35 | 32.20 | 31.71 | 32.23 |
| 1997 Mar 6.5 | 28.7 | 4.17 | H | −8.29 | ... | −8.29 | −8.09 | −8.21 | −10.39 | −9.94 | −9.95 | 34.06 | ... | 31.92 | 31.68 | 31.90 |
| 1997 Mar 6.5 | 28.7 | 4.17 | I | −8.31 | −8.68 | −8.28 | −8.04 | −8.13 | −10.31 | ... | −9.97 | 34.04 | 32.24 | 31.94 | 31.73 | 31.99 |
| 1997 Mar 7.5 | 57.6 | 4.47 | H | −7.73 | ... | −7.75 | −7.76 | −7.61 | −9.98 | −9.60 | −9.61 | 34.59 | ... | 32.45 | 32.00 | 32.49 |
| 1997 Mar 7.5 | 57.6 | 4.47 | I | −7.65 | −8.39 | −7.71 | −7.70 | −7.63 | −9.90 | ... | −9.59 | 34.67 | 32.51 | 32.49 | 32.05 | 32.47 |
| 1997 Mar 7.5 | 28.7 | 4.16 | H | −8.26 | ... | −8.29 | −8.07 | −8.07 | −10.36 | −9.92 | −9.94 | 34.06 | ... | 31.91 | 31.68 | 32.03 |
| 1997 Mar 7.5 | 28.7 | 4.16 | I | −8.28 | −9.10 | −8.23 | −8.15 | −8.16 | −10.23 | ... | −9.94 | 34.03 | 31.81 | 31.97 | 31.60 | 31.94 |
| 1997 Mar 7.5 | 81.4 | 4.62 | H | −7.49 | ... | −7.51 | −7.70 | −7.42 | −9.80 | −9.44 | −9.44 | 34.83 | ... | 32.69 | 32.06 | 32.68 |
| 1997 Mar 7.5 | 81.4 | 4.62 | I | −7.55 | −8.31 | −7.48 | −7.68 | −7.41 | −9.72 | ... | −9.43 | 34.77 | 32.60 | 32.72 | 32.08 | 32.69 |
| 1997 Mar 7.5 | 40.4 | 4.31 | H | −7.98 | ... | −8.00 | −7.91 | −7.83 | −10.15 | −9.77 | −9.79 | 34.33 | ... | 32.21 | 31.85 | 32.28 |
| 1997 Mar 7.5 | 40.4 | 4.31 | I | −8.01 | −8.65 | −7.96 | −7.88 | −7.86 | −10.07 | ... | −9.75 | 34.31 | 32.25 | 32.24 | 31.87 | 32.25 |
| 1997 Mar 7.5 | 28.7 | 4.16 | H | −8.25 | ... | −8.27 | −8.02 | −8.09 | −10.34 | −9.93 | −9.95 | 34.07 | ... | 31.93 | 31.73 | 32.02 |
| 1997 Mar 7.5 | 28.7 | 4.16 | I | −8.28 | −8.82 | −8.25 | −8.13 | −8.11 | −10.26 | ... | −9.94 | 34.04 | 32.09 | 31.95 | 31.63 | 31.99 |
| 1997 Mar 24.1 | 40.4 | 4.29 | H | −7.83 | ... | −7.83 | −7.81 | −7.70 | −9.86 | −9.51 | −9.53 | 34.55 | ... | 32.27 | 31.81 | 32.26 |
| 1997 Mar 24.1 | 40.4 | 4.29 | I | −7.63 | −8.55 | −7.80 | −7.78 | −7.69 | −9.81 | ... | −9.52 | 34.75 | 32.22 | 32.29 | 31.84 | 32.27 |
| 1997 Mar 24.1 | 57.6 | 4.44 | H | −7.59 | ... | −7.58 | −7.67 | −7.50 | −9.72 | −9.36 | −9.39 | 34.79 | ... | 32.51 | 31.95 | 32.47 |
| 1997 Mar 24.1 | 57.6 | 4.44 | I | −7.26 | −8.46 | −7.57 | −7.71 | −7.52 | −9.65 | ... | −9.37 | 35.12 | 32.30 | 32.53 | 31.91 | 32.45 |
| 1997 Mar 24.1 | 81.4 | 4.59 | H | −7.34 | ... | −7.34 | −7.55 | −7.31 | −9.57 | −9.21 | −9.23 | 35.04 | ... | 32.75 | 32.07 | 32.65 |
| 1997 Mar 24.1 | 81.4 | 4.59 | I | −6.97 | −8.30 | −7.33 | −7.61 | −7.33 | −9.50 | ... | −9.22 | 35.41 | 32.47 | 32.76 | 32.01 | 32.64 |
| 1997 Apr 9.1 | 36.7 | 4.28 | I | −7.82 | −8.47 | −7.81 | −7.88 | −7.77 | −9.91 | ... | −9.62 | 34.47 | 32.30 | 32.22 | 31.82 | 32.27 |
| 1997 Apr 9.1 | 51.7 | 4.43 | I | −7.56 | −8.29 | −7.57 | −7.71 | −7.56 | −9.76 | ... | −9.46 | 34.73 | 32.49 | 32.46 | 31.99 | 32.48 |
| 1997 Apr 9.1 | 73.7 | 4.58 | I | −7.32 | −8.11 | −7.33 | −7.57 | −7.37 | −9.61 | ... | −9.30 | 34.97 | 32.66 | 32.69 | 32.12 | 32.66 |
| 1997 Apr 9.2 | 36.7 | 4.28 | I | −7.66 | −8.48 | −7.82 | −7.85 | −7.76 | −9.91 | ... | −9.62 | 34.63 | 32.30 | 32.21 | 31.84 | 32.27 |
| 1997 Apr 9.2 | 51.7 | 4.43 | I | −7.33 | −8.29 | −7.55 | −7.72 | −7.55 | −9.74 | ... | −9.44 | 34.96 | 32.49 | 32.48 | 31.97 | 32.48 |
| 1997 Apr 9.2 | 73.7 | 4.58 | I | −7.06 | −8.09 | −7.32 | −7.60 | ... | −9.60 | ... | −9.28 | 35.24 | 32.68 | 32.70 | 32.10 | ... |
| 1997 Apr 13.1 | 36.7 | 4.30 | I | −7.77 | −8.44 | −7.76 | −7.82 | −7.76 | −9.90 | ... | −9.61 | 34.53 | 32.36 | 32.29 | 31.92 | 32.32 |
| 1997 Apr 13.1 | 51.7 | 4.44 | I | −7.49 | −8.23 | −7.50 | −7.68 | −7.72 | −9.73 | ... | −9.44 | 34.81 | 32.56 | 32.56 | 32.06 | 32.36 |
| 1997 Apr 13.1 | 73.7 | 4.60 | I | −7.25 | −8.02 | −7.27 | −7.54 | −7.34 | −9.58 | ... | −9.30 | 35.05 | 32.77 | 32.79 | 32.20 | 32.74 |
| 1997 Apr 13.1 | 26.2 | 4.15 | I | −7.91 | −8.58 | −8.05 | −8.01 | −8.09 | −10.06 | ... | −9.79 | 34.40 | 32.21 | 32.00 | 31.73 | 31.99 |
| 1997 Apr 13.2 | 36.7 | 4.30 | I | −7.59 | −8.41 | −7.76 | −7.95 | −7.77 | −9.89 | ... | −9.61 | 34.71 | 32.39 | 32.29 | 31.79 | 32.31 |
| 1997 Apr 13.2 | 51.7 | 4.44 | I | −7.23 | −8.20 | −7.58 | −7.69 | −7.54 | −9.82 | ... | −9.50 | 35.08 | 32.59 | 32.47 | 32.05 | 32.54 |
| 1997 Apr 14.1 | 36.7 | 4.30 | H | −7.78 | −8.62 | −7.79 | −7.87 | −7.80 | −9.98 | −9.62 | −9.62 | 34.54 | 32.19 | 32.28 | 31.88 | 32.30 |
| 1997 Apr 14.1 | 36.7 | 4.30 | I | −7.64 | −8.31 | −7.78 | −7.85 | −7.76 | −9.91 | ... | −9.62 | 34.68 | 32.49 | 32.29 | 31.90 | 32.33 |
| 1997 Apr 14.2 | 51.7 | 4.45 | H | −7.48 | −8.32 | −7.69 | −7.64 | −7.58 | −9.85 | −9.46 | −9.46 | 34.83 | 32.49 | 32.38 | 32.11 | 32.52 |
| 1997 Apr 14.2 | 51.7 | 4.45 | I | −7.29 | −8.14 | −7.52 | −7.74 | −7.95 | −9.75 | ... | −9.46 | 35.03 | 32.67 | 32.54 | 32.01 | 32.15 |
| 1997 Apr 14.2 | 73.7 | 4.60 | I | −6.80 | −7.96 | −7.28 | −7.57 | −7.34 | −9.60 | ... | −9.31 | 35.51 | 32.85 | 32.78 | 32.18 | 32.76 |
| 1997 Apr 22.1 | 28.7 | 4.23 | I | −7.99 | −8.69 | −8.20 | −8.27 | −8.36 | −10.29 | ... | −10.02 | 34.25 | 32.27 | 32.00 | 31.59 | 31.85 |
| 1997 Apr 22.2 | 28.7 | 4.23 | I | −7.79 | −8.58 | −8.13 | −8.11 | −8.06 | −10.19 | ... | −9.89 | 34.46 | 32.38 | 32.07 | 31.75 | 32.14 |
| 1997 Apr 22.2 | 28.7 | 4.23 | I | −7.68 | −8.41 | −8.23 | −8.14 | −8.09 | −10.28 | ... | −9.97 | 34.57 | 32.54 | 31.97 | 31.72 | 32.11 |

Table 2—Continued

| UT Date | Aperture | | Filt.[a] | log Emission Band Flux[b] | | | | | log Continuum Flux[b] | | | log $M(\rho)$[b] | | | | |
|---|---|---|---|---|---|---|---|---|---|---|---|---|---|---|---|---|
| | Size | log $\rho$ | | (erg cm$^{-2}$ s$^{-1}$) | | | | | (erg cm$^{-2}$ s$^{-1}$ Å$^{-1}$) | | | (molecule) | | | | |
| | (arcsec) | (km) | | OH | NH | CN | $C_3$ | $C_2$ | UV | Blue | Green | OH | NH | CN | $C_3$ | $C_2$ |
| 1997 Apr 22.2 | 40.4 | 4.38 | I | −7.34 | −8.36 | −7.93 | −8.02 | −7.81 | −10.05 | ... | −9.74 | 34.90 | 32.59 | 32.28 | 31.84 | 32.40 |
| 1997 Apr 22.2 | 40.4 | 4.38 | I | −7.24 | −8.31 | −7.86 | −7.96 | −7.80 | −10.03 | ... | −9.73 | 35.01 | 32.64 | 32.34 | 31.90 | 32.40 |
| 1997 May 1.1 | 28.7 | 4.27 | H | −7.93 | ... | −8.28 | −8.30 | −8.17 | −10.35 | −10.00 | −10.01 | 34.29 | ... | 32.08 | 31.69 | 32.17 |
| 1997 May 1.1 | 28.7 | 4.27 | I | −8.01 | −8.91 | −8.26 | −8.21 | −8.12 | −10.31 | ... | −10.03 | 34.21 | 32.21 | 32.11 | 31.79 | 32.22 |
| 1997 May 1.1 | 20.5 | 4.12 | H | −8.24 | ... | −8.57 | −8.48 | −8.45 | −10.66 | −10.25 | −10.27 | 33.99 | ... | 31.79 | 31.52 | 31.89 |
| 1997 May 1.1 | 20.5 | 4.12 | I | −8.15 | −9.13 | −8.56 | −8.43 | −8.39 | −10.51 | ... | −10.22 | 34.07 | 31.99 | 31.80 | 31.56 | 31.95 |
| 1997 May 1.1 | 40.4 | 4.41 | I | −7.65 | −8.73 | −8.00 | −8.07 | −7.87 | −10.14 | ... | −9.86 | 34.57 | 32.40 | 32.37 | 31.92 | 32.47 |
| 1997 May 1.1 | 57.6 | 4.57 | H | −7.44 | ... | −7.77 | −7.98 | −7.71 | −10.05 | −9.71 | −9.71 | 34.78 | ... | 32.60 | 32.02 | 32.63 |
| 1997 May 1.1 | 57.6 | 4.57 | I | −7.34 | −8.53 | −7.76 | −7.95 | −7.68 | −9.99 | ... | −9.71 | 34.88 | 32.59 | 32.60 | 32.04 | 32.66 |
| 1997 May 1.2 | 28.7 | 4.27 | H | −7.95 | ... | −8.30 | −8.51 | −8.21 | −10.38 | −10.04 | −10.07 | 34.27 | ... | 32.07 | 31.48 | 32.14 |
| 1997 May 1.2 | 28.7 | 4.27 | I | −7.68 | −9.01 | −8.33 | −8.27 | −8.16 | −10.38 | ... | −10.07 | 34.54 | 32.12 | 32.03 | 31.73 | 32.18 |
| 1997 Aug 1.9 | 154.9 | 5.21 | H | ... | ... | −9.00 | −9.06 | −8.84 | ... | −10.99 | −11.00 | ... | ... | 32.50 | 31.98 | 32.54 |
| 1997 Aug 1.9 | 54.6 | 4.76 | H | ... | ... | −9.62 | −9.33 | −9.59 | ... | −11.43 | −11.40 | ... | ... | 31.88 | 31.71 | 31.80 |
| 1997 Aug 1.9 | 54.6 | 4.76 | H | ... | ... | −9.68 | −9.44 | −9.51 | ... | −11.47 | −11.49 | ... | ... | 31.82 | 31.60 | 31.88 |
| 1997 Sep 3.8 | 77.8 | 4.93 | H | ... | ... | −9.82 | −9.95 | −9.63 | ... | −11.39 | −11.44 | ... | ... | 31.88 | 31.28 | 31.94 |
| 1997 Sep 3.8 | 54.6 | 4.78 | H | ... | ... | −10.04 | −9.74 | −9.87 | ... | −11.59 | −11.60 | ... | ... | 31.65 | 31.49 | 31.70 |
| 1997 Oct 12.8 | 109.9 | 5.10 | H | ... | ... | −9.94 | −9.79 | −9.84 | ... | −11.75 | −11.75 | ... | ... | 31.94 | 31.62 | 31.92 |
| 1997 Oct 12.8 | 54.6 | 4.80 | H | ... | ... | −10.42 | −10.09 | −10.31 | ... | −12.07 | −12.06 | ... | ... | 31.46 | 31.32 | 31.45 |
| 1997 Oct 26.7 | 109.9 | 5.11 | H | ... | ... | −10.08 | −9.85 | −10.02 | ... | −11.84 | −11.84 | ... | ... | 31.86 | 31.62 | 31.80 |
| 1997 Oct 26.7 | 54.6 | 4.80 | H | ... | ... | −10.55 | −10.22 | −10.53 | ... | −12.11 | −12.12 | ... | ... | 31.39 | 31.25 | 31.29 |
| 1997 Dec 1.6 | 54.6 | 4.84 | H | ... | ... | −10.76 | −10.45 | −10.80 | ... | −12.31 | −12.26 | ... | ... | 31.34 | 31.19 | 31.18 |
| 1997 Dec 1.6 | 109.9 | 5.14 | H | ... | ... | −10.35 | −10.10 | −10.25 | ... | −11.98 | −11.98 | ... | ... | 31.76 | 31.54 | 31.74 |
| 1997 Dec 1.6 | 77.8 | 4.99 | H | ... | ... | −10.57 | −10.31 | −10.36 | ... | −12.10 | −12.11 | ... | ... | 31.54 | 31.33 | 31.62 |
| 1997 Dec 1.6 | 154.9 | 5.29 | H | ... | ... | −10.10 | −9.94 | −10.01 | ... | −11.84 | −11.84 | ... | ... | 32.00 | 31.70 | 31.97 |
| 1997 Dec 1.6 | 109.9 | 5.14 | H | ... | ... | −10.31 | −10.09 | −10.26 | ... | −11.95 | −11.95 | ... | ... | 31.79 | 31.54 | 31.72 |
| 1997 Dec 1.6 | 77.8 | 4.99 | H | ... | ... | −10.57 | −10.30 | −10.40 | ... | −12.11 | −12.11 | ... | ... | 31.54 | 31.34 | 31.59 |
| 1997 Dec 1.7 | 54.6 | 4.84 | H | ... | ... | −10.79 | −10.48 | −10.76 | ... | −12.24 | −12.24 | ... | ... | 31.32 | 31.15 | 31.22 |
| 1997 Dec 1.7 | 38.8 | 4.69 | H | ... | ... | −10.94 | −10.47 | −11.04 | ... | −12.40 | −12.40 | ... | ... | 31.16 | 31.16 | 30.95 |
| 1997 Dec 1.8 | 54.6 | 4.84 | H | ... | ... | −10.81 | −10.56 | −10.73 | ... | −12.24 | −12.26 | ... | ... | 31.29 | 31.08 | 31.26 |
| 1997 Dec 1.8 | 38.8 | 4.69 | H | ... | ... | −11.00 | −10.53 | −11.08 | ... | −12.42 | −12.42 | ... | ... | 31.10 | 31.10 | 30.91 |
| 1997 Dec 29.6 | 154.9 | 5.32 | H | ... | ... | −10.35 | −10.11 | −10.31 | ... | −11.97 | −11.96 | ... | ... | 31.89 | 31.67 | 31.81 |
| 1997 Dec 29.6 | 109.9 | 5.17 | H | ... | ... | −10.57 | −10.24 | −10.49 | ... | −12.08 | −12.08 | ... | ... | 31.67 | 31.53 | 31.63 |
| 1997 Dec 29.6 | 77.8 | 5.02 | H | ... | ... | −10.80 | −10.44 | −10.82 | ... | −12.22 | −12.21 | ... | ... | 31.43 | 31.34 | 31.30 |
| 1997 Dec 29.6 | 54.6 | 4.87 | H | ... | ... | −11.01 | −10.61 | −11.02 | ... | −12.35 | −12.34 | ... | ... | 31.23 | 31.16 | 31.11 |
| 1997 Dec 29.6 | 38.8 | 4.72 | H | ... | ... | −11.30 | −10.64 | −11.17 | ... | −12.50 | −12.50 | ... | ... | 30.93 | 31.14 | 30.96 |
| 1997 Dec 29.6 | 27.7 | 4.57 | H | ... | ... | −11.49 | −10.77 | −11.99 | ... | −12.71 | −12.61 | ... | ... | 30.75 | 31.01 | 30.13 |
| 1997 Dec 30.6 | 154.9 | 5.32 | H | ... | ... | −10.34 | −10.06 | −10.25 | ... | −11.97 | −11.97 | ... | ... | 31.90 | 31.72 | 31.88 |
| 1997 Dec 30.6 | 77.8 | 5.02 | H | ... | ... | −10.84 | −10.34 | −10.79 | ... | −12.23 | −12.23 | ... | ... | 31.40 | 31.44 | 31.33 |
| 1997 Dec 30.6 | 154.9 | 5.32 | H | ... | ... | −10.32 | −10.20 | −10.24 | ... | −11.97 | −11.99 | ... | ... | 31.92 | 31.59 | 31.88 |
| 1997 Dec 30.6 | 77.8 | 5.02 | H | ... | ... | −10.80 | −10.40 | −10.70 | ... | −12.23 | −12.23 | ... | ... | 31.44 | 31.38 | 31.42 |
| 1998 Jan 27.6 | 77.8 | 5.06 | H | ... | ... | −10.97 | −10.58 | −10.88 | ... | −12.28 | −12.28 | ... | ... | 31.40 | 31.34 | 31.38 |
| 1998 Jan 27.6 | 54.6 | 4.91 | H | ... | ... | −11.06 | −10.77 | −11.18 | ... | −12.42 | −12.42 | ... | ... | 31.31 | 31.15 | 31.09 |
| 1998 Jan 27.6 | 54.6 | 4.91 | H | ... | ... | −11.07 | −10.89 | −11.05 | ... | −12.40 | −12.41 | ... | ... | 31.31 | 31.03 | 31.22 |
| 1998 Jan 27.6 | 77.8 | 5.06 | H | ... | ... | −10.96 | −10.61 | −10.85 | ... | −12.30 | −12.29 | ... | ... | 31.41 | 31.31 | 31.41 |

Table 2—Continued

| UT Date | Aperture | | Filt.[a] | log Emission Band Flux[b] | | | | | log Continuum Flux[b] | | | log $M(\rho)$[b] | | | | |
|---|---|---|---|---|---|---|---|---|---|---|---|---|---|---|---|---|
| | Size | log $\rho$ | | (erg cm$^{-2}$ s$^{-1}$) | | | | | (erg cm$^{-2}$ s$^{-1}$ Å$^{-1}$) | | | (molecule) | | | | |
| | (arcsec) | (km) | | OH | NH | CN | $C_3$ | $C_2$ | UV | Blue | Green | OH | NH | CN | $C_3$ | $C_2$ |
| 1998 Mar 22.5 | 154.9 | 5.43 | H | ... | ... | −10.87 | −10.76 | −10.84 | ... | −12.34 | −12.33 | ... | ... | 31.74 | 31.41 | 31.67 |
| 1998 Mar 22.5 | 77.8 | 5.13 | H | ... | ... | −11.43 | −10.88 | −11.20 | ... | −12.58 | −12.59 | ... | ... | 31.18 | 31.29 | 31.32 |
| 1998 Mar 22.5 | 154.9 | 5.43 | H | ... | ... | −10.88 | −10.59 | −10.81 | ... | −12.30 | −12.29 | ... | ... | 31.73 | 31.58 | 31.71 |
| 1998 Mar 22.6 | 77.8 | 5.13 | H | ... | ... | −11.29 | −10.89 | −11.69 | ... | −12.57 | −12.54 | ... | ... | 31.32 | 31.28 | 30.82 |
| 1998 Apr 29.5 | 77.8 | 5.17 | H | ... | ... | −11.33 | und | und | ... | −12.69 | −12.61 | ... | ... | 31.41 | und | und |
| 1998 Apr 29.5 | 54.6 | 5.02 | H | ... | ... | −11.85 | −10.97 | −11.97 | ... | −12.82 | −12.81 | ... | ... | 30.89 | 31.34 | 30.70 |
| 1998 Apr 29.5 | 77.8 | 5.17 | H | ... | ... | −11.45 | −11.04 | und | ... | −12.68 | −12.61 | ... | ... | 31.29 | 31.27 | und |
| 1998 Oct 23.6 | 77.8 | 5.28 | H | ... | ... | −12.20 | und | −11.50 | ... | −13.16 | −13.15 | ... | ... | 30.96 | und | 31.62 |
| 1998 Oct 23.6 | 77.8 | 5.28 | H | ... | ... | −12.29 | und | −11.69 | ... | −13.16 | −13.14 | ... | ... | 30.87 | und | 31.43 |
| 1998 Oct 23.7 | 77.8 | 5.28 | H | ... | ... | und | und | und | ... | −13.16 | −13.17 | ... | ... | und | und | und |
| 1998 Nov 15.7 | 104.1 | 5.42 | H | ... | ... | und | und | und | ... | −12.76 | −13.01 | ... | ... | und | und | und |
| 1998 Nov 15.7 | 51.7 | 5.12 | H | ... | ... | −13.08 | −11.70 | −12.30 | ... | −13.36 | −13.33 | ... | ... | 30.14 | 31.12 | 30.86 |
| 1998 Nov 15.7 | 104.1 | 5.42 | H | ... | ... | −12.26 | −11.66 | −12.46 | ... | −13.05 | −13.03 | ... | ... | 30.95 | 31.16 | 30.71 |
| 1998 Nov 15.7 | 51.7 | 5.12 | H | ... | ... | und | −12.58 | −13.20 | ... | −13.32 | −13.35 | ... | ... | und | 30.25 | 29.96 |
| 1998 Dec 16.6 | 146.7 | 5.59 | H | ... | ... | −11.89 | −10.70 | −12.45 | ... | −13.09 | −12.84 | ... | ... | 31.39 | 32.19 | 30.78 |
| 1998 Dec 16.6 | 73.7 | 5.29 | H | ... | ... | −12.59 | −11.90 | −11.87 | ... | −13.18 | −13.17 | ... | ... | 30.68 | 30.99 | 31.37 |
| 1998 Dec 16.7 | 146.7 | 5.59 | H | ... | ... | −12.01 | −11.15 | und | ... | −12.86 | −12.73 | ... | ... | 31.27 | 31.73 | und |
| 1998 Dec 16.7 | 73.7 | 5.29 | H | ... | ... | −12.59 | und | −11.63 | ... | −13.14 | −13.16 | ... | ... | 30.69 | und | 31.60 |
| 1998 Dec 20.6 | 51.7 | 5.14 | H | ... | ... | −12.74 | −12.01 | −12.08 | ... | −13.45 | −13.41 | ... | ... | 30.54 | 30.88 | 31.16 |
| 1998 Dec 20.6 | 36.7 | 4.99 | H | ... | ... | −13.96 | −12.49 | −12.08 | ... | −13.64 | −13.73 | ... | ... | 29.33 | 30.41 | 31.16 |
| 1999 Jan 18.6 | 104.1 | 5.45 | H | ... | ... | −12.27 | −11.21 | und | ... | −12.96 | −12.88 | ... | ... | 31.07 | 31.74 | und |
| 1999 Jan 18.6 | 73.7 | 5.30 | H | ... | ... | und | −10.95 | −11.86 | ... | −13.14 | −13.14 | ... | ... | und | 32.00 | 31.44 |
| 1999 Jan 19.6 | 146.7 | 5.60 | H | ... | ... | −12.03 | und | und | ... | −12.77 | −12.84 | ... | ... | 31.31 | und | und |
| 1999 Jan 19.6 | 73.7 | 5.30 | H | ... | ... | −11.89 | und | und | ... | −13.01 | −12.92 | ... | ... | 31.45 | und | und |
| 1999 Feb 9.6 | 146.7 | 5.61 | H | ... | ... | −12.11 | −10.77 | und | ... | −13.11 | −12.82 | ... | ... | 31.27 | 32.22 | und |
| 1999 Feb 9.6 | 73.7 | 5.31 | H | ... | ... | und | −11.63 | −11.51 | ... | −13.18 | −13.23 | ... | ... | und | 31.36 | 31.83 |
| 1999 Feb 10.6 | 104.1 | 5.46 | H | ... | ... | −12.29 | −11.69 | und | ... | −13.03 | −12.98 | ... | ... | 31.09 | 31.30 | und |
| 1999 Feb 10.6 | 73.7 | 5.31 | H | ... | ... | −12.18 | und | −11.59 | ... | −13.22 | −13.17 | ... | ... | 31.20 | und | 31.75 |
| 1999 Feb 10.6 | 104.1 | 5.46 | H | ... | ... | −12.37 | und | und | ... | −13.00 | −12.99 | ... | ... | 31.02 | und | und |
| 1999 Feb 10.6 | 73.7 | 5.31 | H | ... | ... | und | und | −11.39 | ... | −13.16 | −13.18 | ... | ... | und | und | 31.95 |
| 1999 Mar 17.5 | 146.7 | 5.63 | H | ... | ... | −11.23 | und | −11.32 | ... | −12.88 | −13.04 | ... | ... | 32.22 | und | 32.08 |
| 1999 Mar 17.5 | 73.7 | 5.33 | H | ... | ... | und | −10.61 | −11.57 | ... | −13.24 | −13.21 | ... | ... | und | 32.44 | 31.83 |
| 1999 Apr 17.5 | 146.7 | 5.64 | H | ... | ... | −11.68 | und | und | ... | −12.89 | −12.89 | ... | ... | 31.83 | und | und |
| 1999 Apr 17.5 | 73.7 | 5.34 | H | ... | ... | −12.54 | und | −12.74 | ... | −13.12 | −13.20 | ... | ... | 30.97 | und | 30.71 |
| 1999 Apr 17.5 | 146.7 | 5.64 | H | ... | ... | −11.21 | −12.70 | und | ... | −12.97 | −12.87 | ... | ... | 32.29 | 30.41 | und |
| 1999 Apr 17.5 | 73.7 | 5.34 | H | ... | ... | und | −11.80 | und | ... | −13.09 | −13.13 | ... | ... | und | 31.31 | und |
| 1999 Apr 18.5 | 146.7 | 5.64 | H | ... | ... | −11.85 | und | und | ... | −12.97 | −12.78 | ... | ... | 31.66 | und | und |
| 1999 Apr 18.5 | 73.7 | 5.34 | H | ... | ... | und | und | −11.46 | ... | −13.12 | −13.22 | ... | ... | und | und | 32.00 |
| 1999 Apr 18.6 | 146.7 | 5.64 | H | ... | ... | −11.40 | und | −12.58 | ... | −13.01 | −12.86 | ... | ... | 32.10 | und | 30.88 |
| 1999 Apr 18.6 | 73.7 | 5.34 | H | ... | ... | und | −11.68 | −11.58 | ... | −13.13 | −13.16 | ... | ... | und | 31.42 | 31.88 |
| 1999 Jun 9.4 | 146.7 | 5.66 | H | ... | ... | und | und | −11.21 | ... | −12.43 | −12.99 | ... | ... | und | und | 32.32 |
| 1999 Jun 9.4 | 73.7 | 5.36 | H | ... | ... | −12.54 | und | und | ... | −13.47 | −13.27 | ... | ... | 31.05 | und | und |
| 1999 Jun 13.4 | 146.7 | 5.66 | H | ... | ... | und | und | −11.58 | ... | −12.51 | −12.92 | ... | ... | und | und | 31.96 |
| 1999 Jun 13.5 | 146.7 | 5.66 | H | ... | ... | −13.14 | −10.62 | und | ... | −12.94 | −12.81 | ... | ... | 30.45 | 32.58 | und |

Table 2—Continued

| UT Date | Aperture | | Filt.[a] | log Emission Band Flux[b] | | | | | log Continuum Flux[b] | | | log $M(\rho)$[b] | | | | |
|---|---|---|---|---|---|---|---|---|---|---|---|---|---|---|---|---|
| | Size | log $\rho$ | | (erg cm$^{-2}$ s$^{-1}$) | | | | | (erg cm$^{-2}$ s$^{-1}$ Å$^{-1}$) | | | (molecule) | | | | |
| | (arcsec) | (km) | | OH | NH | CN | $C_3$ | $C_2$ | UV | Blue | Green | OH | NH | CN | $C_3$ | $C_2$ |
| 1999 Jul 2.5 | 77.8 | 5.39 | H | ... | ... | *und* | −11.32 | *und* | ... | −13.24 | −13.22 | ... | ... | *und* | 31.90 | *und* |
| 1999 Jul 2.5 | 77.8 | 5.39 | H | ... | ... | *und* | −10.91 | −12.84 | ... | −13.34 | −13.28 | ... | ... | *und* | 32.32 | 30.73 |
| 1999 Jul 6.5 | 154.9 | 5.69 | H | ... | ... | −13.17 | *und* | −11.09 | ... | −12.89 | −13.35 | ... | ... | 30.46 | *und* | 32.49 |
| 1999 Jul 6.5 | 154.9 | 5.69 | H | ... | ... | −12.06 | −10.50 | −10.98 | ... | −13.01 | −13.19 | ... | ... | 31.57 | 32.72 | 32.60 |
| 1999 Sep 17.8 | 146.7 | 5.70 | H | ... | ... | *und* | *und* | *und* | ... | −12.91 | −13.20 | ... | ... | *und* | *und* | *und* |
| 1999 Sep 17.8 | 146.7 | 5.70 | H | ... | ... | *und* | *und* | −11.56 | ... | −12.97 | −13.24 | ... | ... | *und* | *und* | 32.13 |
| 1999 Sep 17.8 | 146.7 | 5.70 | H | ... | ... | *und* | −10.74 | *und* | ... | −12.98 | −13.11 | ... | ... | *und* | 32.60 | *und* |
| 1999 Sep 17.8 | 146.7 | 5.70 | H | ... | ... | *und* | −11.05 | *und* | ... | −12.91 | −13.07 | ... | ... | *und* | 32.28 | *und* |
| 1999 Dec 6.6 | 146.7 | 5.73 | H | ... | ... | −12.45 | −10.97 | −11.77 | ... | −13.62 | −13.33 | ... | ... | 31.41 | 32.48 | 32.03 |
| 1999 Dec 6.6 | 73.7 | 5.43 | H | ... | ... | −12.22 | *und* | *und* | ... | −13.57 | −13.50 | ... | ... | 31.64 | *und* | *und* |
| 1999 Dec 6.6 | 146.7 | 5.73 | H | ... | ... | −12.17 | −11.30 | −11.73 | ... | −13.58 | −13.26 | ... | ... | 31.70 | 32.16 | 32.07 |
| 1999 Dec 6.6 | 73.7 | 5.43 | H | ... | ... | −12.53 | *und* | −12.78 | ... | −13.46 | −13.50 | ... | ... | 31.33 | *und* | 31.02 |
| 1999 Dec 7.5 | 146.7 | 5.73 | H | ... | ... | −12.97 | −11.44 | −11.23 | ... | −13.36 | −13.64 | ... | ... | 30.89 | 32.02 | 32.57 |
| 1999 Dec 7.6 | 73.7 | 5.43 | H | ... | ... | −12.42 | *und* | −13.97 | ... | −13.57 | −13.49 | ... | ... | 31.45 | *und* | 29.83 |
| 1999 Dec 7.6 | 146.7 | 5.73 | H | ... | ... | *und* | *und* | −12.52 | ... | −13.21 | −13.28 | ... | ... | *und* | *und* | 31.28 |
| 1999 Dec 7.6 | 73.7 | 5.43 | H | ... | ... | *und* | *und* | −13.16 | ... | −13.51 | −13.49 | ... | ... | *und* | *und* | 30.64 |
| 2000 Feb 1.6 | 77.8 | 5.47 | H | ... | ... | −12.71 | *und* | *und* | ... | −13.54 | −13.60 | ... | ... | 31.23 | *und* | *und* |
| 2000 Feb 1.6 | 77.8 | 5.47 | H | ... | ... | *und* | −12.71 | *und* | ... | −13.63 | −13.61 | ... | ... | *und* | 30.81 | *und* |
| 2000 Feb 2.5 | 154.9 | 5.77 | H | ... | ... | −13.10 | −11.09 | −11.71 | ... | −13.58 | −13.52 | ... | ... | 30.85 | 32.44 | 32.16 |
| 2000 Feb 2.5 | 77.8 | 5.47 | H | ... | ... | *und* | *und* | −12.37 | ... | −13.65 | −13.79 | ... | ... | *und* | *und* | 31.50 |
| 2000 Feb 2.6 | 154.9 | 5.77 | H | ... | ... | *und* | −10.91 | −11.85 | ... | −13.71 | −13.44 | ... | ... | *und* | 32.61 | 32.02 |
| 2000 Feb 2.6 | 77.8 | 5.47 | H | ... | ... | *und* | −12.02 | −12.25 | ... | −13.62 | −13.78 | ... | ... | *und* | 31.51 | 31.62 |
| 2000 Mar 3.5 | 154.9 | 5.78 | H | ... | ... | −11.89 | ... | *und* | ... | −13.47 | −13.27 | ... | ... | 32.09 | ... | *und* |
| 2000 Mar 3.5 | 154.9 | 5.78 | H | ... | ... | −11.89 | ... | −12.01 | ... | −13.47 | −13.56 | ... | ... | 32.09 | ... | 31.89 |

[a]Filter ID: I = IHW; H = HB.

[b] *"und"* stands for "undefined". For the gases, this means that the emission flux was measured but was less than zero following sky and continuum removal. For the continuum, this means the continuum flux was measured but following sky subtraction it was less than zero.

Table 3.  Photometric Production Rates for Comet C/Hale-Bopp.

| UT Date | $\Delta T$ | $\log r_H$ | $\log \rho$ | Filt.[a] | $\log Q$[b,c] (molecules s$^{-1}$) | | | | | $\log A(\theta)f\rho$[b,c] (cm) | | | $\log Q$[b] |
|---|---|---|---|---|---|---|---|---|---|---|---|---|---|
| | (day) | (AU) | (km) | | OH | NH | CN | $C_3$ | $C_2$ | UV | Blue | Green | $H_2O$ |
| 1995 Jul  25.2 | −615.9 | 0.854 | 5.26 | I | ... | ... | 25.94 .24 | 26.40 .12 | 26.64 .20 | 4.58 .06 | ... | 4.69 .01 | ... |
| 1995 Sep  25.5 | −553.6 | 0.820 | 4.81 | I | ... | ... | und | und | 27.32 .07 | 5.11 .04 | ... | 5.00 .00 | ... |
| 1995 Sep  25.6 | −553.6 | 0.820 | 4.81 | I | ... | ... | 26.80 .11 | und | 27.24 .09 | 5.16 .04 | ... | 5.03 .00 | ... |
| 1995 Oct  19.6 | −529.6 | 0.806 | 5.12 | I | ... | ... | und | und | 27.20 .27 | und | ... | 4.95 .00 | ... |
| 1995 Oct  19.6 | −529.5 | 0.806 | 5.12 | I | ... | ... | 26.79 .20 | 26.82 .21 | und | 4.66 .17 | ... | 4.96 .00 | ... |
| 1995 Oct  22.5 | −526.6 | 0.804 | 5.12 | I | ... | ... | 26.48 .40 | und | 27.16 .27 | und | ... | 4.95 .00 | ... |
| 1995 Nov  20.5 | −497.6 | 0.786 | 4.98 | I | ... | ... | und | und | 27.10 .22 | 4.93 .16 | ... | 4.86 .01 | ... |
| 1995 Nov  21.5 | −496.6 | 0.785 | 5.13 | I | ... | ... | und | und | 27.52 .02 | 5.50 .04 | ... | 4.76 .01 | ... |
| 1996 Feb  19.9 | −406.3 | 0.721 | 5.07 | I | ... | ... | 26.45 .35 | 27.15 .05 | und | 3.96 .19 | ... | 4.86 .00 | ... |
| 1996 Feb  19.9 | −406.3 | 0.721 | 5.23 | I | ... | ... | und | 26.55 .16 | 26.42 .21 | 4.85 .05 | ... | 4.87 .00 | ... |
| 1996 Feb  25.5 | −400.6 | 0.716 | 5.17 | I | ... | ... | 26.11 .14 | 25.79 .38 | 26.94 .07 | 4.88 .04 | ... | 4.84 .01 | ... |
| 1996 Mar   1.5 | −395.6 | 0.712 | 5.38 | I | ... | ... | 24.97 .99 | 27.50 .03 | 27.01 .09 | und | ... | 4.62 .01 | ... |
| 1996 Apr  13.5 | −352.7 | 0.675 | 5.39 | I | 28.88 .11 | 27.26 .12 | 26.71 .01 | 26.45 .04 | 26.57 .06 | 4.70 .02 | ... | 4.84 .00 | 28.50 |
| 1996 Apr  14.4 | −351.7 | 0.674 | 5.09 | I | 28.97 .17 | 27.63 .07 | 26.70 .02 | 26.43 .04 | 26.26 .13 | 4.65 .02 | ... | 4.78 .00 | 28.60 |
| 1996 Apr  14.4 | −351.7 | 0.674 | 5.09 | I | 29.20 .08 | 27.28 .13 | 26.66 .02 | 26.43 .04 | 26.37 .10 | 4.69 .01 | ... | 4.81 .00 | 28.87 |
| 1996 Apr  28.5 | −337.7 | 0.661 | 5.25 | I | 28.62 .05 | 27.08 .08 | 26.68 .01 | 26.38 .02 | 26.68 .03 | 4.72 .01 | ... | 4.79 .00 | 28.21 |
| 1996 Apr  29.5 | −336.7 | 0.660 | 4.94 | I | 29.15 .05 | 27.52 .06 | 26.72 .01 | 26.36 .03 | 26.32 .10 | 4.70 .01 | ... | 4.83 .00 | 28.83 |
| 1996 May  13.8 | −322.3 | 0.646 | 4.88 | I | ... | ... | 26.84 .02 | 26.46 .02 | 26.68 .03 | 4.75 .01 | ... | 4.80 .00 | ... |
| 1996 May  13.8 | −322.3 | 0.646 | 4.88 | I | ... | ... | 26.69 .02 | 26.24 .03 | 26.41 .05 | 4.80 .01 | ... | 4.84 .00 | ... |
| 1996 May  16.4 | −319.7 | 0.643 | 5.20 | I | 29.07 .02 | 27.37 .05 | 26.79 .01 | 26.49 .02 | 26.58 .04 | 4.68 .01 | ... | 4.80 .00 | 28.74 |
| 1996 May  20.4 | −315.7 | 0.639 | 5.04 | I | 29.02 .05 | 27.50 .07 | 26.79 .01 | 26.54 .02 | 26.29 .12 | 4.66 .01 | ... | 4.82 .00 | 28.69 |
| 1996 May  22.8 | −313.4 | 0.637 | 4.86 | I | ... | ... | 26.79 .01 | 26.63 .01 | 26.18 .08 | 4.74 .01 | ... | 4.86 .00 | ... |
| 1996 May  22.8 | −313.4 | 0.637 | 4.86 | I | ... | ... | 26.50 .02 | 26.62 .01 | und | 4.72 .01 | ... | 4.85 .00 | ... |
| 1996 May  28.3 | −307.8 | 0.631 | 4.82 | I | 29.19 .06 | 27.29 .20 | 26.84 .01 | 26.46 .04 | 26.61 .08 | 4.78 .01 | ... | 4.86 .00 | 28.90 |
| 1996 May  28.4 | −307.8 | 0.631 | 5.27 | I | und | und | 26.75 .01 | und | 26.84 .02 | 4.82 .01 | ... | 4.86 .00 | und |
| 1996 May  28.4 | −307.7 | 0.631 | 4.67 | I | 29.01 .05 | 27.71 .07 | 26.85 .01 | 26.54 .02 | 26.05 .22 | 4.71 .01 | ... | 4.84 .00 | 28.69 |
| 1996 May  28.4 | −307.7 | 0.631 | 4.82 | I | 28.91 .04 | 27.64 .05 | 26.82 .01 | 26.50 .02 | 26.31 .09 | 4.70 .01 | ... | 4.83 .00 | 28.58 |
| 1996 May  29.3 | −306.8 | 0.630 | 4.81 | I | 29.13 .09 | 27.79 .10 | 26.84 .02 | 26.39 .06 | 26.85 .05 | 4.84 .02 | ... | 4.87 .00 | 28.83 |
| 1996 May  29.4 | −306.8 | 0.630 | 5.12 | I | 28.40 .13 | 27.33 .12 | 26.81 .01 | 26.25 .05 | 26.59 .04 | 4.88 .01 | ... | 4.81 .00 | 28.22 |
| 1996 May  29.4 | −306.8 | 0.630 | 4.81 | I | 28.60 .12 | 27.47 .13 | 26.81 .01 | 26.22 .06 | 26.58 .06 | 4.83 .01 | ... | 4.80 .00 | 28.21 |
| 1996 May  29.4 | −306.8 | 0.630 | 4.67 | I | 28.91 .07 | 27.48 .14 | 26.76 .01 | 26.05 .08 | und | 4.79 .01 | ... | 4.75 .00 | 28.58 |
| 1996 May  29.4 | −306.7 | 0.630 | 4.81 | I | 29.13 .03 | 27.73 .04 | 26.84 .01 | 26.54 .02 | 26.41 .07 | 4.71 .01 | ... | 4.84 .00 | 28.83 |
| 1996 Jun   1.3 | −303.8 | 0.626 | 4.85 | I | 30.02 .11 | und | 26.84 .05 | 26.53 .12 | und | 4.92 .04 | ... | 4.95 .01 | 29.87 |
| 1996 Jun   1.3 | −303.8 | 0.626 | 4.55 | I | und | 27.68 .26 | 26.77 .05 | 26.61 .08 | 26.18 .46 | 4.82 .03 | ... | 4.88 .01 | und |
| 1996 Jun   1.4 | −303.8 | 0.626 | 4.85 | I | 29.01 .15 | 27.97 .10 | 26.92 .03 | 26.48 .08 | 26.79 .12 | 4.69 .04 | ... | 4.80 .01 | 28.69 |
| 1996 Jun   1.4 | −303.8 | 0.626 | 4.70 | I | 29.10 .14 | 28.10 .09 | 26.88 .03 | 26.64 .06 | 25.37 .99 | 4.65 .04 | ... | 4.81 .01 | 28.79 |
| 1996 Jun   1.4 | −303.7 | 0.626 | 5.00 | I | 29.35 .05 | 27.52 .17 | 26.79 .02 | 26.51 .06 | 26.56 .13 | 4.83 .02 | ... | 4.87 .01 | 29.09 |
| 1996 Jun   1.4 | −303.7 | 0.626 | 4.85 | I | ... | 26.53 .84 | 26.86 .02 | 26.03 .16 | 26.67 .12 | 4.83 .02 | ... | 4.84 .01 | ... |
| 1996 Jun   3.3 | −301.8 | 0.624 | 4.85 | I | und | und | 26.90 .05 | 26.72 .09 | und | 4.56 .07 | ... | 4.89 .01 | und |
| 1996 Jun   3.3 | −301.8 | 0.624 | 4.54 | I | 29.32 .17 | 27.65 .41 | 26.85 .07 | 26.65 .12 | 0.00 .68 | 4.91 .04 | ... | 5.00 .01 | 29.05 |
| 1996 Jun   3.4 | −301.8 | 0.624 | 4.69 | I | und | 27.67 .24 | 26.93 .03 | 26.72 .06 | 25.36 .99 | 4.89 .03 | ... | 4.99 .01 | und |
| 1996 Jun   3.4 | −301.8 | 0.624 | 5.00 | I | 29.07 .10 | 26.94 .52 | 26.85 .02 | 26.47 .08 | 26.38 .22 | 4.98 .02 | ... | 5.01 .01 | 28.76 |
| 1996 Jun   3.4 | −301.8 | 0.624 | 4.40 | I | 29.26 .13 | 27.64 .31 | 26.92 .04 | 26.80 .06 | und | 4.86 .02 | ... | 4.99 .01 | 28.99 |
| 1996 Jun   3.4 | −301.7 | 0.624 | 4.85 | I | und | 28.00 .09 | 26.87 .03 | 26.49 .08 | 26.03 .44 | 4.90 .02 | ... | 4.97 .01 | und |
| 1996 Jun   6.3 | −298.8 | 0.621 | 4.84 | I | 28.67 .13 | 26.95 .38 | 26.85 .02 | 26.50 .04 | 26.47 .13 | 4.81 .01 | ... | 4.90 .00 | 28.29 |

Table 3—Continued

| UT Date | ΔT (day) | log $r_H$ (AU) | log $\rho$ (km) | Filt.[a] | log $Q$[b,c] (molecules s$^{-1}$) OH | NH | CN | $C_3$ | $C_2$ | log $A(\theta)f\rho$[b,c] (cm) UV | Blue | Green | log $Q$[b] $H_2O$ |
|---|---|---|---|---|---|---|---|---|---|---|---|---|---|
| 1996 Jun  6.4 −298.8 | 0.621 | 4.84 | I | 28.78 .08 | 27.40 .13 | 26.80 .01 | 26.49 .03 | 26.27 .15 | 4.80 .01 | ... | 4.90 .00 | 28.42 |
| 1996 Jun  6.4 −298.7 | 0.621 | 4.69 | I | 28.93 .07 | 27.81 .07 | 26.83 .01 | 26.49 .04 | und | 4.80 .01 | ... | 4.92 .00 | 28.60 |
| 1996 Jun  7.3 −297.8 | 0.620 | 4.84 | I | 28.81 .06 | 26.97 .21 | 26.75 .01 | 26.43 .03 | und | 4.75 .01 | ... | 4.88 .00 | 28.47 |
| 1996 Jun  7.4 −297.7 | 0.620 | 4.84 | I | 29.06 .04 | 27.23 .12 | 26.85 .01 | 26.52 .02 | 25.94 .19 | 4.77 .01 | ... | 4.87 .00 | 28.75 |
| 1996 Jun  9.3 −295.9 | 0.618 | 4.83 | I | 28.29 .16 | 27.44 .11 | 26.90 .01 | 26.52 .02 | 26.64 .06 | 4.85 .01 | ... | 4.93 .00 | 28.11 |
| 1996 Jun  9.3 −295.8 | 0.618 | 5.13 | I | 29.06 .04 | 27.36 .07 | 26.88 .01 | 26.52 .02 | 26.77 .02 | 4.80 .01 | ... | 4.89 .00 | 28.75 |
| 1996 Jun  9.3 −295.8 | 0.618 | 4.53 | I | 28.49 .17 | 27.83 .08 | 26.89 .01 | 26.49 .04 | 26.34 .19 | 4.80 .01 | ... | 4.92 .00 | 28.31 |
| 1996 Jun  9.4 −295.8 | 0.618 | 4.83 | I | 28.69 .07 | 27.61 .06 | 26.88 .01 | 26.61 .02 | 26.12 .16 | 4.83 .01 | ... | 4.94 .00 | 28.32 |
| 1996 Jun 10.3 −294.9 | 0.617 | 4.83 | I | 29.39 .06 | 27.59 .08 | 26.88 .01 | 26.59 .02 | 26.42 .10 | 4.83 .01 | ... | 4.93 .00 | 29.14 |
| 1996 Jun 10.3 −294.9 | 0.617 | 4.68 | I | 29.46 .06 | 28.19 .03 | 26.92 .01 | 26.88 .01 | und | 4.66 .01 | ... | 4.92 .00 | 29.22 |
| 1996 Jun 10.3 −294.8 | 0.617 | 4.98 | I | 29.02 .04 | 27.59 .05 | 26.89 .01 | 26.51 .02 | 26.52 .05 | 4.82 .01 | ... | 4.91 .00 | 28.71 |
| 1996 Jun 10.3 −294.8 | 0.617 | 4.53 | I | 28.86 .08 | 27.26 .21 | 26.89 .01 | 26.50 .03 | 25.66 .52 | 4.85 .01 | ... | 4.94 .00 | 28.52 |
| 1996 Jun 10.4 −294.8 | 0.617 | 4.83 | I | und | und | 26.89 .00 | 26.19 .01 | 26.84 .03 | ... | ... | 4.94 .00 | und |
| 1996 Jun 17.3 −287.9 | 0.609 | 4.77 | I | 29.19 .02 | 27.64 .05 | 26.95 .00 | 26.59 .01 | 26.50 .06 | 4.92 .00 | ... | 5.01 .00 | 28.92 |
| 1996 Jun 17.3 −287.8 | 0.609 | 5.07 | I | 28.98 .02 | 27.10 .07 | 26.93 .00 | 26.59 .01 | 26.74 .02 | 4.86 .00 | ... | 4.96 .00 | 28.67 |
| 1996 Jun 17.4 −287.8 | 0.609 | 4.92 | I | 28.97 .02 | 27.58 .04 | 26.91 .00 | 26.61 .01 | 26.64 .03 | 4.85 .00 | ... | 4.96 .00 | 28.66 |
| 1996 Jun 17.4 −287.8 | 0.609 | 4.62 | I | 29.23 .03 | 27.84 .04 | 26.93 .01 | 26.64 .02 | 25.93 .23 | 4.93 .00 | ... | 5.02 .00 | 28.96 |
| 1996 Jun 17.4 −287.7 | 0.609 | 4.77 | I | 29.19 .02 | 27.88 .03 | 26.95 .01 | 26.63 .01 | 26.32 .08 | 4.88 .00 | ... | 4.99 .00 | 28.92 |
| 1996 Jun 17.4 −287.7 | 0.609 | 4.77 | I | 29.00 .03 | 27.84 .03 | 26.96 .01 | 26.62 .01 | 26.34 .09 | 4.89 .00 | ... | 5.00 .00 | 28.70 |
| 1996 Jun 18.3 −286.9 | 0.608 | 4.81 | I | 29.06 .03 | 27.70 .04 | 26.92 .01 | 26.56 .02 | 26.46 .07 | 4.88 .01 | ... | 4.97 .00 | 28.76 |
| 1996 Jun 18.3 −286.8 | 0.608 | 4.96 | I | 29.11 .02 | 27.62 .04 | 26.91 .00 | 26.61 .01 | 26.63 .04 | 4.85 .00 | ... | 4.94 .00 | 28.83 |
| 1996 Jun 18.3 −286.8 | 0.608 | 4.81 | I | 29.18 .02 | 27.68 .04 | 26.95 .01 | 26.62 .01 | 26.58 .06 | 4.89 .00 | ... | 4.99 .00 | 28.90 |
| 1996 Jun 20.3 −284.9 | 0.606 | 4.65 | I | 29.49 .02 | 28.00 .04 | 27.16 .01 | 26.73 .02 | 27.00 .04 | 5.01 .01 | ... | 5.10 .00 | 29.27 |
| 1996 Jun 20.3 −284.8 | 0.606 | 4.81 | I | 29.13 .03 | 27.80 .04 | 26.91 .01 | 26.53 .02 | 26.69 .04 | 4.88 .01 | ... | 4.96 .00 | 28.84 |
| 1996 Jun 22.3 −282.8 | 0.603 | 4.80 | I | 29.09 .02 | 27.64 .04 | 26.94 .01 | 26.51 .02 | 26.65 .04 | 4.85 .00 | ... | 4.94 .00 | 28.80 |
| 1996 Jun 23.3 −281.9 | 0.602 | 5.10 | I | 29.12 .02 | 27.53 .03 | 26.92 .00 | 26.57 .01 | 26.79 .02 | 4.85 .00 | ... | 4.93 .00 | 28.84 |
| 1996 Jun 23.3 −281.8 | 0.602 | 4.80 | I | 29.08 .03 | 27.70 .04 | 26.93 .01 | 26.57 .01 | 26.77 .03 | 4.85 .00 | ... | 4.92 .00 | 28.79 |
| 1996 Jul 11.5 −263.6 | 0.580 | 4.75 | I | ... | ... | 26.93 .01 | 26.66 .01 | 26.66 .02 | 4.84 .00 | ... | 4.93 .00 | ... |
| 1996 Jul 11.5 −263.6 | 0.580 | 4.75 | I | ... | ... | 27.03 .01 | 26.64 .01 | 26.38 .04 | 4.82 .00 | ... | 4.92 .00 | ... |
| 1996 Jul 11.6 −263.6 | 0.580 | 4.75 | I | ... | ... | 26.95 .01 | 26.81 .01 | und | 4.67 .00 | ... | 4.91 .00 | ... |
| 1996 Jul 18.2 −257.0 | 0.572 | 5.02 | I | 29.18 .01 | 27.50 .03 | 26.97 .00 | 26.63 .01 | 26.84 .01 | 4.75 .00 | ... | 4.86 .00 | 28.93 |
| 1996 Jul 18.2 −257.0 | 0.572 | 4.72 | I | 29.21 .02 | 27.71 .04 | 26.98 .01 | 26.63 .01 | 26.63 .05 | 4.75 .01 | ... | 4.87 .00 | 28.96 |
| 1996 Jul 18.3 −256.8 | 0.572 | 5.17 | I | 29.17 .01 | 27.30 .03 | 26.98 .00 | 26.67 .01 | 26.90 .01 | 4.76 .00 | ... | 4.87 .00 | 28.92 |
| 1996 Jul 18.4 −256.8 | 0.572 | 4.72 | I | 29.18 .03 | 27.71 .03 | 26.99 .00 | 26.62 .01 | 26.78 .03 | 4.78 .01 | ... | 4.88 .00 | 28.93 |
| 1996 Aug  6.2 −237.9 | 0.547 | 4.76 | I | 29.31 .01 | 27.76 .02 | 27.03 .00 | 26.62 .01 | 26.85 .02 | 4.72 .00 | ... | 4.82 .00 | 29.11 |
| 1996 Aug  6.2 −237.9 | 0.547 | 5.06 | I | 29.29 .01 | 27.55 .02 | 27.06 .00 | 26.68 .00 | 26.99 .01 | 4.72 .00 | ... | 4.81 .00 | 29.09 |
| 1996 Aug  6.2 −237.9 | 0.547 | 4.60 | I | 29.30 .02 | 27.81 .03 | 27.03 .00 | 26.63 .01 | 26.89 .02 | 4.72 .00 | ... | 4.83 .00 | 29.10 |
| 1996 Aug  6.3 −237.9 | 0.547 | 4.76 | I | 29.23 .02 | 27.64 .03 | 27.02 .00 | 26.62 .01 | 26.90 .02 | 4.71 .01 | ... | 4.80 .00 | 29.02 |
| 1996 Aug  6.3 −237.8 | 0.547 | 4.91 | I | 29.31 .02 | 27.62 .03 | 27.04 .00 | 26.65 .01 | 26.91 .01 | 4.69 .01 | ... | 4.80 .00 | 29.11 |
| 1996 Aug  7.2 −237.0 | 0.546 | 4.76 | I | 29.31 .01 | 27.69 .02 | 27.02 .00 | 26.64 .01 | 26.84 .02 | 4.69 .00 | ... | 4.81 .00 | 29.11 |
| 1996 Aug  7.2 −236.9 | 0.546 | 5.06 | I | 29.31 .01 | 27.56 .02 | 27.05 .00 | 26.69 .01 | 26.95 .01 | 4.68 .00 | ... | 4.79 .00 | 29.11 |
| 1996 Aug  7.2 −236.9 | 0.546 | 4.45 | I | 29.21 .03 | 27.70 .05 | 26.99 .01 | 26.62 .01 | 26.24 .15 | 4.67 .01 | ... | 4.79 .00 | 28.99 |
| 1996 Aug  7.3 −236.9 | 0.546 | 4.76 | I | 29.27 .02 | 27.60 .03 | 27.03 .00 | 26.62 .01 | 26.88 .02 | 4.71 .00 | ... | 4.80 .00 | 29.06 |
| 1996 Aug  8.2 −236.0 | 0.544 | 4.60 | I | 29.57 .01 | 27.99 .02 | 27.27 .00 | 26.82 .01 | 27.08 .02 | 4.85 .00 | ... | 4.96 .00 | 29.42 |
| 1996 Aug  8.2 −235.9 | 0.544 | 5.06 | I | 29.30 .02 | 27.53 .02 | 27.04 .00 | 26.68 .01 | 26.95 .01 | 4.69 .00 | ... | 4.80 .00 | 29.10 |

Table 3—Continued

| UT Date | ΔT (day) | log $r_H$ (AU) | log $\rho$ (km) | Filt.[a] | log $Q^{b,c}$ (molecules s$^{-1}$) | | | | | log $A(\theta)f\rho^{b,c}$ (cm) | | | log $Q^b$ |
|---|---|---|---|---|---|---|---|---|---|---|---|---|---|
| | | | | | OH | NH | CN | $C_3$ | $C_2$ | UV | Blue | Green | $H_2O$ |
| 1996 Aug  8.2 | −235.9 | 0.544 | 4.45 | I | 29.39 .02 | 27.81 .04 | 27.00 .01 | 26.56 .02 | 26.61 .06 | 4.69 .01 | ... | 4.81 .00 | 29.20 |
| 1996 Aug 20.3 | −223.9 | 0.527 | 5.06 | I | 29.32 .02 | 27.49 .03 | 27.05 .00 | 26.67 .01 | 27.03 .01 | 4.68 .01 | ... | 4.76 .00 | 29.14 |
| 1996 Aug 20.3 | −223.9 | 0.527 | 4.76 | I | 29.22 .06 | 27.48 .06 | 27.05 .01 | 26.67 .01 | 26.92 .02 | 4.70 .01 | ... | 4.82 .00 | 29.02 |
| 1996 Sep  4.2 | −209.0 | 0.505 | 4.77 | I | 29.40 .01 | 27.78 .02 | 27.13 .00 | 26.79 .01 | 27.01 .01 | 4.80 .00 | ... | 4.92 .00 | 29.25 |
| 1996 Sep  4.2 | −208.9 | 0.505 | 5.07 | I | 29.40 .02 | 27.71 .01 | 27.10 .00 | 26.80 .01 | 27.10 .01 | 4.65 .01 | ... | 4.82 .00 | 29.25 |
| 1996 Sep 12.5 | −200.7 | 0.492 | 4.91 | I | ... | ... | 27.13 .00 | 26.65 .01 | 27.10 .00 | 4.90 .00 | ... | 4.96 .00 | ... |
| 1996 Sep 12.5 | −200.6 | 0.492 | 4.91 | I | ... | ... | 27.05 .00 | 26.66 .01 | 27.08 .00 | 4.90 .00 | ... | 4.97 .00 | ... |
| 1996 Sep 20.1 | −193.0 | 0.479 | 4.79 | I | 29.52 .01 | 27.77 .02 | 27.14 .00 | 26.80 .00 | 27.07 .01 | 4.75 .00 | ... | 4.89 .00 | 29.40 |
| 1996 Sep 20.1 | −193.0 | 0.479 | 4.79 | I | 29.50 .01 | 27.75 .02 | 27.14 .00 | 26.77 .01 | 27.13 .01 | 4.76 .00 | ... | 4.87 .00 | 29.39 |
| 1996 Oct  4.1 | −179.0 | 0.455 | 4.80 | I | 29.62 .01 | 27.84 .01 | 27.16 .00 | 26.84 .00 | 27.15 .01 | 4.82 .00 | ... | 4.97 .00 | 29.55 |
| 1996 Oct  4.1 | −179.0 | 0.455 | 4.80 | I | 29.57 .01 | 27.70 .01 | 27.16 .00 | 26.78 .00 | 27.14 .01 | 4.84 .00 | ... | 4.96 .00 | 29.49 |
| 1996 Oct 13.5 | −169.6 | 0.437 | 4.93 | I | ... | ... | 27.20 .01 | 26.89 .01 | 27.15 .01 | 4.75 .01 | ... | 4.91 .00 | ... |
| 1996 Oct 13.5 | −169.6 | 0.437 | 4.78 | I | ... | ... | 27.19 .01 | 26.70 .02 | 27.20 .01 | 4.90 .01 | ... | 4.95 .00 | ... |
| 1996 Oct 16.1 | −167.0 | 0.433 | 4.80 | I | 29.65 .01 | 27.81 .01 | 27.18 .00 | 26.85 .00 | 27.27 .00 | 4.81 .00 | ... | 4.93 .00 | 29.60 |
| 1996 Oct 16.1 | −167.0 | 0.433 | 4.95 | I | 29.67 .00 | 27.73 .01 | 27.20 .00 | 26.86 .00 | 27.30 .00 | 4.81 .00 | ... | 4.93 .00 | 29.62 |
| 1996 Oct 18.1 | −165.0 | 0.429 | 4.65 | I | 29.68 .01 | 27.91 .01 | 27.21 .00 | 26.84 .00 | 27.18 .01 | 4.88 .00 | ... | 4.99 .00 | 29.63 |
| 1996 Oct 18.1 | −165.0 | 0.429 | 4.65 | I | 29.68 .01 | 27.91 .02 | 27.23 .00 | 26.87 .01 | 27.24 .01 | 4.87 .00 | ... | 4.99 .00 | 29.64 |
| 1996 Nov  5.1 | −147.1 | 0.392 | 4.80 | I | 29.76 .01 | 27.84 .01 | 27.24 .00 | 26.95 .00 | 27.36 .00 | 4.90 .00 | ... | 5.02 .00 | 29.76 |
| 1996 Nov  5.1 | −147.0 | 0.392 | 4.95 | I | 29.76 .01 | 27.80 .01 | 27.24 .00 | 26.96 .00 | 27.40 .00 | 4.89 .00 | ... | 4.99 .00 | 29.76 |
| 1996 Nov  6.1 | −146.1 | 0.390 | 4.80 | I | 29.74 .01 | 27.83 .01 | 27.23 .00 | 26.92 .00 | 27.32 .00 | 4.88 .00 | ... | 5.00 .00 | 29.73 |
| 1996 Nov  6.1 | −146.1 | 0.390 | 4.65 | I | 29.70 .01 | 27.73 .02 | 27.23 .00 | 26.88 .01 | 27.27 .01 | 4.91 .00 | ... | 5.04 .00 | 29.69 |
| 1996 Nov  6.1 | −146.0 | 0.390 | 5.10 | I | 29.78 .01 | 27.74 .01 | 27.26 .00 | 26.98 .00 | 27.39 .00 | 4.86 .00 | ... | 4.97 .00 | 29.78 |
| 1996 Nov  7.1 | −145.1 | 0.388 | 4.80 | I | 29.71 .01 | 27.85 .01 | 27.21 .00 | 26.95 .00 | 27.36 .00 | 4.88 .00 | ... | 5.00 .00 | 29.71 |
| 1996 Nov  7.1 | −145.1 | 0.388 | 4.80 | I | 29.72 .01 | 27.83 .01 | 27.22 .00 | 26.90 .00 | 27.39 .00 | 4.89 .00 | ... | 5.00 .00 | 29.71 |
| 1996 Nov  7.1 | −145.0 | 0.388 | 5.10 | I | 29.71 .01 | 27.72 .01 | 27.24 .00 | 26.94 .00 | 27.42 .00 | 4.84 .00 | ... | 4.95 .00 | 29.71 |
| 1996 Nov 19.1 | −133.1 | 0.360 | 4.80 | I | 29.75 .01 | 27.85 .01 | 27.25 .00 | 26.93 .01 | 27.44 .00 | 4.99 .00 | ... | 5.09 .00 | 29.77 |
| 1996 Nov 19.1 | −133.1 | 0.360 | 4.64 | I | 29.85 .01 | 27.94 .02 | 27.24 .01 | 26.92 .01 | 27.39 .01 | 5.00 .00 | ... | 5.12 .00 | 29.89 |
| 1996 Nov 20.1 | −132.1 | 0.358 | 4.80 | I | 29.79 .01 | 27.81 .02 | 27.25 .00 | 26.96 .01 | 27.41 .00 | 4.97 .00 | ... | 5.09 .00 | 29.83 |
| 1996 Nov 20.1 | −132.1 | 0.358 | 4.80 | I | 29.80 .01 | 27.81 .02 | 27.24 .00 | 26.97 .01 | 27.37 .00 | 4.98 .00 | ... | 5.10 .00 | 29.84 |
| 1996 Nov 21.1 | −131.1 | 0.356 | 4.80 | I | 29.84 .02 | 27.88 .02 | 27.25 .00 | 27.00 .01 | 27.42 .00 | 4.96 .01 | ... | 5.08 .00 | 29.88 |
| 1996 Nov 21.1 | −131.1 | 0.356 | 5.09 | I | 29.83 .01 | 27.79 .01 | 27.28 .00 | 27.10 .01 | 27.47 .00 | 4.87 .01 | ... | 5.03 .00 | 29.87 |
| 1996 Nov 28.1 | −124.1 | 0.339 | 4.79 | I | 29.75 .01 | 27.70 .02 | 27.23 .00 | 26.91 .00 | 27.46 .00 | 4.99 .00 | ... | 5.08 .00 | 29.80 |
| 1996 Nov 28.1 | −124.1 | 0.339 | 4.79 | I | ... | ... | 27.24 .00 | 27.02 .00 | 27.46 .00 | 4.95 .00 | ... | 5.08 .00 | ... |
| 1996 Dec  3.1 | −119.1 | 0.326 | 4.78 | I | ... | 27.80 .02 | 27.29 .00 | 27.05 .00 | 27.59 .00 | 5.02 .00 | ... | 5.13 .00 | ... |
| 1996 Dec  4.1 | −118.1 | 0.323 | 4.78 | H | ... | ... | 27.23 .00 | 26.94 .00 | 27.39 .00 | ... | 4.97 .00 | 4.83 .00 | ... |
| 1996 Dec  4.1 | −118.1 | 0.323 | 4.78 | I | ... | 27.92 .01 | 27.28 .00 | 26.93 .01 | 27.53 .00 | 5.04 .00 | ... | 5.13 .00 | ... |
| 1996 Dec  5.1 | −117.1 | 0.321 | 4.78 | H | ... | ... | 27.24 .00 | 26.95 .00 | 27.54 .00 | ... | 5.12 .00 | 5.14 .00 | ... |
| 1996 Dec  5.1 | −117.1 | 0.321 | 4.78 | I | ... | 27.97 .01 | 27.29 .00 | 27.07 .01 | 27.51 .00 | 5.02 .00 | ... | 5.13 .00 | ... |
| 1997 Jan 17.6 | −73.6 | 0.189 | 4.79 | H | ... | ... | 27.61 .00 | 26.49 .00 | 27.92 .00 | ... | 4.34 .00 | 5.28 .00 | ... |
| 1997 Jan 17.6 | −73.6 | 0.189 | 4.79 | I | 30.64 .01 | 28.09 .01 | 27.51 .00 | 27.25 .00 | 27.75 .00 | 5.19 .00 | ... | 5.28 .00 | 30.96 |
| 1997 Jan 19.6 | −71.6 | 0.182 | 4.78 | H | ... | ... | 27.52 .00 | 27.20 .00 | 27.80 .00 | ... | 5.26 .00 | 5.27 .00 | ... |
| 1997 Jan 19.6 | −71.6 | 0.182 | 4.78 | I | 30.52 .01 | 28.00 .01 | 27.53 .00 | 27.25 .00 | 27.82 .00 | 5.20 .00 | ... | 5.27 .00 | 30.82 |
| 1997 Feb  3.5 | −56.6 | 0.126 | 4.61 | H | ... | ... | 27.71 .00 | 27.35 .00 | 27.96 .00 | ... | 5.48 .00 | 5.50 .00 | ... |
| 1997 Feb  3.5 | −56.6 | 0.126 | 4.61 | I | 30.23 .00 | 28.28 .00 | 27.73 .00 | 27.35 .00 | 27.98 .00 | 5.43 .00 | ... | 5.49 .00 | 30.53 |
| 1997 Feb  3.5 | −56.6 | 0.126 | 4.61 | H | ... | ... | 27.70 .00 | 27.34 .00 | 27.97 .00 | ... | 5.50 .00 | 5.50 .00 | ... |

Table 3—Continued

| UT Date | ΔT (day) | log $r_H$ (AU) | log $\rho$ (km) | Filt.[a] | log $Q$[b,c] (molecules s$^{-1}$) | | | | | log $A(\theta)f\rho$[b,c] (cm) | | | log $Q$[b] |
|---|---|---|---|---|---|---|---|---|---|---|---|---|---|
| | | | | | OH | NH | CN | $C_3$ | $C_2$ | UV | Blue | Green | $H_2O$ |
| 1997 Feb 3.5 | −56.6 | 0.126 | 4.61 | I | 30.19 .00 | 28.27 .00 | 27.72 .00 | 27.37 .00 | 27.98 .00 | 5.42 .00 | ... | 5.48 .00 | 30.48 |
| 1997 Feb 12.5 | −47.6 | und | 4.68 | H | 30.40 .00 | 28.15 .00 | 27.87 .00 | 27.47 .00 | 28.12 .00 | 5.57 .00 | 5.61 .00 | 5.61 .00 | 30.76 |
| 1997 Feb 12.5 | −47.6 | 0.091 | 4.68 | I | 30.43 .00 | 28.28 .00 | 27.88 .00 | 27.52 .00 | 28.12 .00 | 5.55 .00 | ... | 5.61 .00 | 30.79 |
| 1997 Feb 12.5 | −47.6 | 0.091 | 4.98 | H | 30.45 .00 | 28.16 .00 | 27.97 .00 | 27.53 .00 | 28.18 .00 | 5.57 .00 | 5.59 .00 | 5.60 .00 | 30.81 |
| 1997 Feb 12.5 | −47.6 | 0.091 | 4.98 | I | 30.44 .00 | 28.20 .00 | 27.96 .00 | 27.62 .00 | 28.17 .00 | 5.54 .00 | ... | 5.59 .00 | 30.81 |
| 1997 Feb 12.6 | −47.6 | 0.091 | 4.37 | H | 30.32 .00 | 28.17 .00 | 27.75 .00 | 27.39 .00 | 28.04 .00 | 5.57 .00 | 5.62 .00 | 5.63 .00 | 30.67 |
| 1997 Feb 12.6 | −47.6 | 0.091 | 4.37 | I | 30.35 .00 | 28.33 .01 | 27.77 .00 | 27.45 .00 | 28.08 .00 | 5.54 .00 | ... | 5.60 .00 | 30.69 |
| 1997 Feb 12.6 | −47.6 | 0.091 | 4.37 | I | 30.34 .00 | 28.37 .00 | 27.77 .00 | 27.42 .00 | 28.05 .00 | 5.56 .00 | ... | 5.63 .00 | 30.68 |
| 1997 Feb 15.5 | −44.6 | 0.080 | 4.66 | H | 30.46 .00 | 28.21 .00 | 27.91 .00 | 27.54 .00 | 28.15 .00 | 5.62 .00 | 5.65 .00 | 5.66 .00 | 30.84 |
| 1997 Feb 15.5 | −44.6 | 0.080 | 4.66 | I | 30.51 .00 | 28.20 .00 | 27.93 .00 | 27.61 .00 | 28.15 .00 | 5.61 .00 | ... | 5.66 .00 | 30.90 |
| 1997 Feb 15.5 | −44.6 | 0.080 | 4.36 | H | 30.39 .00 | 28.21 .00 | 27.81 .00 | 27.45 .00 | 28.09 .00 | 5.61 .00 | 5.65 .00 | 5.66 .00 | 30.76 |
| 1997 Feb 15.5 | −44.6 | 0.080 | 4.36 | I | 30.46 .00 | 28.28 .00 | 27.84 .00 | 27.52 .00 | 28.09 .00 | 5.60 .00 | ... | 5.65 .00 | 30.84 |
| 1997 Feb 15.5 | −44.6 | 0.080 | 4.81 | H | 30.47 .00 | 28.19 .00 | 27.97 .00 | 27.57 .00 | 28.18 .00 | 5.63 .00 | 5.66 .00 | 5.66 .00 | 30.85 |
| 1997 Feb 15.5 | −44.6 | 0.080 | 4.81 | I | 30.47 .00 | 28.16 .00 | 27.98 .00 | 27.66 .00 | 28.19 .00 | 5.60 .00 | ... | 5.66 .00 | 30.85 |
| 1997 Feb 15.5 | −44.6 | 0.080 | 4.96 | H | 30.49 .00 | 28.21 .00 | 28.02 .00 | 27.60 .00 | 28.21 .00 | 5.63 .00 | 5.65 .00 | 5.66 .00 | 30.87 |
| 1997 Feb 15.5 | −44.6 | 0.080 | 4.96 | I | 30.48 .00 | 28.07 .00 | 28.02 .00 | 27.69 .00 | 28.22 .00 | 5.60 .00 | ... | 5.64 .00 | 30.86 |
| 1997 Feb 15.6 | −44.6 | 0.080 | 4.51 | H | 30.39 .00 | 28.21 .00 | 27.88 .00 | 27.56 .00 | 28.14 .00 | 5.65 .00 | 5.68 .00 | 5.68 .00 | 30.76 |
| 1997 Feb 15.6 | −44.6 | 0.080 | 4.51 | I | 30.42 .00 | 28.28 .00 | 27.89 .00 | 27.63 .00 | 28.14 .00 | 5.62 .00 | ... | 5.70 .00 | 30.79 |
| 1997 Feb 15.6 | −44.6 | 0.080 | 4.96 | H | 30.46 .00 | 28.14 .00 | 28.03 .00 | 27.54 .00 | und | 5.72 .00 | 5.68 .00 | 5.69 .00 | 30.84 |
| 1997 Feb 15.6 | −44.6 | 0.080 | 4.36 | I | 30.33 .00 | 28.31 .01 | 27.84 .00 | 27.52 .00 | 28.09 .00 | 5.58 .00 | ... | 5.65 .00 | 30.68 |
| 1997 Feb 26.5 | −33.6 | 0.037 | 4.50 | I | 30.36 .00 | 28.33 .00 | 28.02 .00 | 27.64 .00 | 28.27 .00 | 5.81 .00 | ... | 5.86 .00 | 30.75 |
| 1997 Feb 26.5 | −33.6 | 0.037 | 4.65 | I | 30.37 .00 | 28.31 .00 | 28.06 .00 | 27.68 .00 | 28.29 .00 | 5.79 .00 | ... | 5.84 .00 | 30.77 |
| 1997 Feb 26.5 | −33.6 | 0.037 | 4.80 | I | 30.40 .00 | 28.24 .00 | 28.11 .00 | 27.74 .00 | 28.30 .00 | 5.79 .00 | ... | 5.83 .00 | 30.80 |
| 1997 Feb 26.5 | −33.6 | 0.037 | 4.35 | I | 30.31 .00 | 28.38 .00 | 27.96 .00 | 27.59 .00 | 28.22 .00 | 5.79 .00 | ... | 5.83 .00 | 30.70 |
| 1997 Feb 26.5 | −33.6 | 0.037 | 4.50 | I | 30.33 .00 | 28.33 .00 | 28.00 .00 | 27.64 .00 | 28.26 .00 | 5.78 .00 | ... | 5.83 .00 | 30.72 |
| 1997 Feb 26.6 | −33.6 | 0.037 | 4.20 | I | 30.24 .00 | 28.37 .00 | 27.87 .00 | 27.49 .00 | 28.14 .00 | 5.73 .00 | ... | 5.81 .00 | 30.61 |
| 1997 Mar 5.5 | −26.6 | 0.012 | 4.17 | I | ... | ... | 27.94 .00 | 27.37 .00 | 28.19 .00 | ... | ... | 5.86 .00 | ... |
| 1997 Mar 5.5 | −26.6 | 0.012 | 4.47 | H | 30.31 .00 | ... | 28.03 .00 | 27.69 .00 | 28.23 .00 | 5.87 .00 | 5.91 .00 | 5.92 .00 | 30.72 |
| 1997 Mar 5.5 | −26.6 | 0.012 | 4.47 | I | 30.43 .00 | 28.33 .00 | 28.05 .00 | 27.68 .00 | 28.24 .00 | 5.88 .00 | ... | 5.92 .00 | 30.85 |
| 1997 Mar 5.5 | −26.6 | 0.012 | 4.62 | H | 30.37 .00 | ... | 28.09 .00 | 27.73 .00 | 28.27 .00 | 5.88 .00 | 5.91 .00 | 5.91 .00 | 30.78 |
| 1997 Mar 5.5 | −26.6 | 0.012 | 4.62 | I | 30.37 .00 | 28.36 .00 | 28.11 .00 | 27.76 .00 | 28.27 .00 | 5.86 .00 | ... | 5.92 .00 | 30.79 |
| 1997 Mar 5.5 | −26.6 | 0.012 | 4.32 | H | 30.24 .00 | ... | 27.96 .00 | 27.59 .00 | 28.21 .00 | 5.80 .00 | 5.88 .00 | 5.86 .00 | 30.63 |
| 1997 Mar 5.5 | −26.6 | 0.012 | 4.32 | I | 30.28 .00 | 28.47 .00 | 27.97 .00 | 27.63 .00 | 28.16 .00 | 5.80 .00 | ... | 5.87 .00 | 30.68 |
| 1997 Mar 5.5 | −26.6 | 0.012 | 4.47 | H | 30.28 .00 | ... | 28.00 .00 | 27.68 .00 | 28.23 .00 | 5.84 .00 | 5.88 .00 | 5.90 .00 | 30.68 |
| 1997 Mar 5.5 | −26.6 | 0.012 | 4.47 | I | 30.26 .00 | 28.37 .00 | 28.02 .00 | 27.72 .00 | 28.25 .00 | 5.82 .00 | ... | 5.86 .00 | 30.66 |
| 1997 Mar 5.5 | −26.6 | 0.012 | 4.17 | H | 30.20 .00 | ... | 27.88 .00 | 27.56 .00 | 28.13 .00 | 5.79 .00 | 5.86 .00 | 5.88 .00 | 30.59 |
| 1997 Mar 5.5 | −26.6 | 0.012 | 4.17 | I | 30.22 .00 | 28.51 .00 | 27.91 .00 | 27.75 .00 | 28.07 .00 | 5.76 .00 | ... | 5.87 .00 | 30.61 |
| 1997 Mar 6.5 | −25.6 | 0.009 | 4.47 | H | 30.23 .00 | ... | 27.99 .00 | 27.61 .00 | 28.18 .00 | 5.82 .00 | 5.88 .00 | 5.90 .00 | 30.63 |
| 1997 Mar 6.5 | −25.6 | 0.009 | 4.47 | I | 30.39 .00 | 28.27 .00 | 28.02 .00 | 27.73 .00 | 28.22 .00 | 5.81 .00 | ... | 5.89 .00 | 30.81 |
| 1997 Mar 6.5 | −25.6 | 0.009 | 4.17 | H | 30.06 .00 | ... | 27.88 .00 | 27.50 .00 | 28.17 .00 | 5.77 .00 | 5.81 .00 | 5.85 .00 | 30.43 |
| 1997 Mar 6.5 | −25.6 | 0.009 | 4.17 | I | 30.27 .00 | 28.22 .01 | 27.87 .00 | 27.32 .00 | 28.12 .00 | 5.82 .00 | ... | 5.90 .00 | 30.67 |
| 1997 Mar 6.5 | −25.6 | 0.009 | 4.62 | H | 30.29 .00 | ... | 28.07 .00 | 27.66 .00 | 28.24 .00 | 5.87 .00 | 5.92 .00 | 5.93 .00 | 30.70 |
| 1997 Mar 6.5 | −25.6 | 0.009 | 4.62 | I | 30.30 .00 | 28.24 .00 | 28.07 .00 | 27.69 .00 | 28.24 .00 | 5.85 .00 | ... | 5.92 .00 | 30.71 |
| 1997 Mar 6.5 | −25.6 | 0.009 | 4.47 | H | 30.23 .00 | ... | 27.99 .00 | 27.61 .00 | 28.22 .00 | 5.86 .00 | 5.91 .00 | 5.91 .00 | 30.63 |
| 1997 Mar 6.5 | −25.6 | 0.009 | 4.47 | I | 30.24 .00 | 28.19 .00 | 28.02 .00 | 27.66 .00 | 28.25 .00 | 5.86 .00 | ... | 5.91 .00 | 30.63 |

Table 3—Continued

| UT Date | ΔT (day) | log $r_H$ (AU) | log $\rho$ (km) | Filt.[a] | log $Q^{b,c}$ OH | NH | CN | $C_3$ | $C_2$ (molecules s$^{-1}$) | log $A(\theta)f\rho^{b,c}$ UV | Blue | Green (cm) | log $Q^b$ $H_2O$ |
|---|---|---|---|---|---|---|---|---|---|---|---|---|---|
| 1997 Mar 6.5 | −25.6 | 0.009 | 4.32 | H | 30.20 .00 | ... | 27.94 .00 | 27.50 .00 | 28.17 .00 | 5.85 .00 | 5.88 .00 | 5.89 .00 | 30.59 |
| 1997 Mar 6.5 | −25.6 | 0.009 | 4.32 | I | 30.21 .00 | 28.43 .00 | 27.93 .00 | 27.50 .00 | 28.16 .00 | 5.82 .00 | ... | 5.90 .00 | 30.60 |
| 1997 Mar 6.5 | −25.6 | 0.009 | 4.17 | H | 30.16 .00 | ... | 27.86 .00 | 27.58 .00 | 28.04 .00 | 5.76 .00 | 5.87 .00 | 5.89 .00 | 30.54 |
| 1997 Mar 6.5 | −25.6 | 0.009 | 4.17 | I | 30.14 .00 | 28.55 .00 | 27.88 .00 | 27.63 .00 | 28.12 .00 | 5.75 .00 | ... | 5.86 .00 | 30.52 |
| 1997 Mar 7.5 | −24.6 | 0.006 | 4.47 | H | 30.26 .00 | ... | 28.00 .00 | 27.70 .00 | 28.23 .00 | 5.85 .00 | 5.91 .00 | 5.92 .00 | 30.67 |
| 1997 Mar 7.5 | −24.6 | 0.006 | 4.47 | I | 30.34 .00 | 28.37 .00 | 28.04 .00 | 27.75 .00 | 28.21 .00 | 5.84 .00 | ... | 5.93 .00 | 30.76 |
| 1997 Mar 7.5 | −24.6 | 0.006 | 4.16 | H | 30.15 .00 | ... | 27.85 .00 | 27.58 .00 | 28.16 .00 | 5.78 .00 | 5.88 .00 | 5.89 .00 | 30.54 |
| 1997 Mar 7.5 | −24.6 | 0.006 | 4.16 | I | 30.13 .00 | 28.12 .01 | 27.91 .00 | 27.50 .00 | 28.08 .00 | 5.81 .00 | ... | 5.88 .00 | 30.52 |
| 1997 Mar 7.5 | −24.6 | 0.006 | 4.62 | H | 30.31 .00 | ... | 28.07 .00 | 27.70 .00 | 28.26 .00 | 5.88 .00 | 5.92 .00 | 5.93 .00 | 30.73 |
| 1997 Mar 7.5 | −24.6 | 0.006 | 4.62 | I | 30.26 .00 | 28.24 .00 | 28.10 .00 | 27.71 .00 | 28.27 .00 | 5.87 .00 | ... | 5.93 .00 | 30.66 |
| 1997 Mar 7.5 | −24.6 | 0.006 | 4.31 | H | 30.22 .00 | ... | 27.95 .00 | 27.64 .00 | 28.21 .00 | 5.84 .00 | 5.89 .00 | 5.90 .00 | 30.62 |
| 1997 Mar 7.5 | −24.6 | 0.006 | 4.31 | I | 30.19 .00 | 28.34 .00 | 27.98 .00 | 27.66 .00 | 28.18 .00 | 5.83 .00 | ... | 5.92 .00 | 30.58 |
| 1997 Mar 7.5 | −24.6 | 0.006 | 4.16 | H | 30.16 .00 | ... | 27.88 .00 | 27.63 .00 | 28.15 .00 | 5.79 .00 | 5.88 .00 | 5.87 .00 | 30.55 |
| 1997 Mar 7.5 | −24.6 | 0.006 | 4.16 | I | 30.13 .00 | 28.40 .00 | 27.89 .00 | 27.53 .00 | 28.13 .00 | 5.79 .00 | ... | 5.87 .00 | 30.52 |
| 1997 Mar 24.1 | −8.0 | −0.034 | 4.29 | H | 30.44 .00 | ... | 28.02 .00 | 27.65 .00 | 28.21 .00 | 6.02 .00 | 6.04 .00 | 6.04 .00 | 30.91 |
| 1997 Mar 24.1 | −8.0 | −0.034 | 4.29 | I | 30.64 .00 | 28.30 .00 | 28.05 .00 | 27.68 .00 | 28.22 .00 | 5.98 .00 | ... | 6.03 .00 | 31.14 |
| 1997 Mar 24.1 | −8.0 | −0.034 | 4.44 | H | 30.47 .00 | ... | 28.08 .00 | 27.71 .00 | 28.23 .00 | 6.00 .00 | 6.03 .00 | 6.03 .00 | 30.95 |
| 1997 Mar 24.1 | −8.0 | −0.034 | 4.44 | I | 30.81 .00 | 28.16 .00 | 28.09 .00 | 27.67 .00 | 28.21 .00 | 5.99 .00 | ... | 6.04 .00 | 31.34 |
| 1997 Mar 24.1 | −8.0 | −0.034 | 4.59 | H | 30.54 .00 | ... | 28.16 .00 | 27.77 .00 | 28.26 .00 | 6.01 .00 | 6.03 .00 | 6.03 .00 | 31.03 |
| 1997 Mar 24.1 | −8.0 | −0.034 | 4.59 | I | 30.92 .00 | 28.12 .00 | 28.17 .00 | 27.71 .00 | 28.24 .00 | 5.99 .00 | ... | 6.04 .00 | 31.46 |
| 1997 Apr 9.1 | +8.0 | −0.034 | 4.28 | I | 30.36 .00 | 28.40 .00 | 27.98 .00 | 27.66 .00 | 28.22 .00 | 5.96 .00 | ... | 6.02 .00 | 30.82 |
| 1997 Apr 9.1 | +8.0 | −0.034 | 4.43 | I | 30.43 .00 | 28.36 .00 | 28.04 .00 | 27.75 .00 | 28.26 .00 | 5.96 .00 | ... | 6.02 .00 | 30.90 |
| 1997 Apr 9.1 | +8.0 | −0.034 | 4.58 | I | 30.48 .00 | 28.33 .00 | 28.11 .00 | 27.82 .00 | 28.28 .00 | 5.96 .00 | ... | 6.03 .00 | 30.96 |
| 1997 Apr 9.2 | +8.0 | −0.034 | 4.28 | I | 30.53 .00 | 28.39 .00 | 27.97 .00 | 27.68 .00 | 28.23 .00 | 5.96 .00 | ... | 6.02 .00 | 31.01 |
| 1997 Apr 9.2 | +8.0 | −0.034 | 4.43 | I | 30.66 .00 | 28.36 .00 | 28.06 .00 | 27.74 .00 | 28.26 .00 | 5.98 .00 | ... | 6.05 .00 | 31.16 |
| 1997 Apr 9.2 | +8.0 | −0.034 | 4.58 | I | 30.75 .00 | 28.35 .00 | 28.12 .00 | 27.80 .00 | ... | 5.97 .00 | ... | 6.05 .00 | 31.27 |
| 1997 Apr 13.1 | +12.0 | −0.027 | 4.30 | I | 30.41 .00 | 28.44 .00 | 28.04 .00 | 27.74 .00 | 28.26 .00 | 6.00 .00 | ... | 6.06 .00 | 30.87 |
| 1997 Apr 13.1 | +12.0 | −0.027 | 4.44 | I | 30.50 .00 | 28.42 .00 | 28.13 .00 | 27.80 .00 | 28.12 .00 | 6.02 .00 | ... | 6.07 .00 | 30.97 |
| 1997 Apr 13.1 | +12.0 | −0.027 | 4.60 | I | 30.55 .00 | 28.42 .00 | 28.18 .00 | 27.89 .00 | 28.34 .00 | 6.01 .00 | ... | 6.06 .00 | 31.03 |
| 1997 Apr 13.1 | +12.0 | −0.027 | 4.15 | I | 30.49 .00 | 28.51 .00 | 27.94 .00 | 27.66 .00 | 28.12 .00 | 5.99 .00 | ... | 6.02 .00 | 30.96 |
| 1997 Apr 13.2 | +12.0 | −0.027 | 4.30 | I | 30.59 .00 | 28.46 .00 | 28.04 .00 | 27.62 .00 | 28.25 .00 | 6.01 .00 | ... | 6.05 .00 | 31.08 |
| 1997 Apr 13.2 | +12.0 | −0.027 | 4.44 | I | 30.76 .00 | 28.45 .00 | 28.04 .00 | 27.79 .00 | 28.30 .00 | 5.93 .00 | ... | 6.01 .00 | 31.28 |
| 1997 Apr 14.1 | +13.0 | −0.025 | 4.30 | H | 30.42 .00 | 28.26 .00 | 28.02 .00 | 27.70 .00 | 28.23 .00 | 6.01 .00 | 6.05 .00 | 6.06 .00 | 30.87 |
| 1997 Apr 14.1 | +13.0 | −0.025 | 4.30 | I | 30.56 .00 | 28.56 .00 | 28.03 .00 | 27.72 .00 | 28.26 .00 | 5.99 .00 | ... | 6.05 .00 | 31.04 |
| 1997 Apr 14.2 | +13.0 | −0.025 | 4.45 | H | 30.51 .00 | 28.35 .00 | 27.94 .00 | 27.85 .00 | 28.27 .00 | 5.99 .00 | 6.06 .00 | 6.08 .00 | 30.98 |
| 1997 Apr 14.2 | +13.0 | −0.025 | 4.45 | I | 30.71 .00 | 28.52 .00 | 28.11 .00 | 27.75 .00 | 27.90 .00 | 6.01 .00 | ... | 6.06 .00 | 31.21 |
| 1997 Apr 14.2 | +13.0 | −0.025 | 4.60 | I | 31.00 .00 | 28.49 .00 | 28.18 .00 | 27.87 .00 | 28.35 .00 | 6.01 .00 | ... | 6.06 .00 | 31.56 |
| 1997 Apr 22.1 | +21.0 | −0.005 | 4.23 | I | 30.25 .00 | 28.47 .00 | 27.84 .00 | 27.45 .00 | 27.88 .00 | 5.80 .00 | ... | 5.83 .00 | 30.66 |
| 1997 Apr 22.2 | +21.0 | −0.005 | 4.23 | I | 30.45 .00 | 28.58 .00 | 27.92 .00 | 27.61 .00 | 28.18 .00 | 5.90 .00 | ... | 5.97 .00 | 30.90 |
| 1997 Apr 22.2 | +21.0 | −0.005 | 4.23 | I | 30.57 .00 | 28.75 .00 | 27.82 .00 | 27.58 .00 | 28.15 .00 | 5.81 .00 | ... | 5.88 .00 | 31.03 |
| 1997 Apr 22.2 | +21.0 | −0.005 | 4.38 | I | 30.69 .00 | 28.57 .00 | 27.93 .00 | 27.60 .00 | 28.24 .00 | 5.89 .00 | ... | 5.96 .00 | 31.18 |
| 1997 Apr 22.2 | +21.0 | −0.005 | 4.38 | I | 30.79 .00 | 28.62 .00 | 28.00 .00 | 27.66 .00 | 28.25 .00 | 5.91 .00 | ... | 5.97 .00 | 31.30 |
| 1997 May 1.1 | +30.0 | 0.024 | 4.27 | H | 30.26 .00 | ... | 27.90 .00 | 27.51 .00 | 28.18 .00 | 5.93 .00 | 5.94 .00 | 5.95 .00 | 30.65 |
| 1997 May 1.1 | +30.0 | 0.024 | 4.27 | I | 30.18 .00 | 28.39 .01 | 27.93 .00 | 27.60 .00 | 28.22 .00 | 5.88 .00 | ... | 5.92 .00 | 30.55 |
| 1997 May 1.1 | +30.0 | 0.024 | 4.12 | H | 30.17 .01 | ... | 27.81 .00 | 27.45 .00 | 28.11 .00 | 5.76 .00 | 5.84 .00 | 5.84 .00 | 30.55 |

Table 3—Continued

| UT Date | ΔT (day) | log $r_H$ (AU) | log $\rho$ (km) | Filt.[a] | log $Q$[b,c] (molecules s$^{-1}$) OH | NH | CN | $C_3$ | $C_2$ | log $A(\theta)f\rho$[b,c] (cm) UV | Blue | Green | log $Q$[b] $H_2O$ |
|---|---|---|---|---|---|---|---|---|---|---|---|---|---|
| 1997 May 1.1 | +30.0 | 0.024 | 4.12 | I | 30.26 .01 | 28.40 .01 | 27.82 .00 | 27.49 .00 | 28.16 .00 | 5.83 .00 | ... | 5.88 .00 | 30.65 |
| 1997 May 1.1 | +30.0 | 0.024 | 4.41 | I | 30.33 .00 | 28.35 .01 | 27.99 .00 | 27.64 .00 | 28.28 .00 | 5.90 .00 | ... | 5.94 .00 | 30.73 |
| 1997 May 1.1 | +30.0 | 0.024 | 4.57 | H | 30.33 .01 | ... | 28.03 .00 | 27.65 .00 | 28.26 .00 | 5.93 .00 | 5.93 .00 | 5.96 .00 | 30.74 |
| 1997 May 1.1 | +30.0 | 0.024 | 4.57 | I | 30.43 .00 | 28.32 .00 | 28.04 .00 | 27.68 .00 | 28.29 .00 | 5.90 .00 | ... | 5.94 .00 | 30.85 |
| 1997 May 1.2 | +30.0 | 0.024 | 4.27 | H | 30.24 .02 | ... | 27.89 .00 | 27.29 .00 | 28.14 .00 | 5.89 .00 | 5.91 .00 | 5.89 .00 | 30.62 |
| 1997 May 1.2 | +30.0 | 0.024 | 4.27 | I | 30.50 .01 | 28.29 .01 | 27.85 .00 | 27.54 .00 | 28.19 .00 | 5.81 .00 | ... | 5.88 .00 | 30.94 |
| 1997 Aug 1.9 | +122.8 | 0.335 | 5.21 | H | ... | ... | 27.30 .00 | 26.99 .00 | 27.53 .00 | ... | 5.06 .00 | 5.07 .00 | ... |
| 1997 Aug 1.9 | +122.8 | 0.335 | 4.76 | H | ... | ... | 27.26 .00 | 27.00 .00 | 27.37 .00 | ... | 5.07 .00 | 5.12 .00 | ... |
| 1997 Aug 1.9 | +122.8 | 0.335 | 4.76 | H | ... | ... | 27.20 .00 | 26.90 .00 | 27.45 .00 | ... | 5.03 .00 | 5.03 .00 | ... |
| 1997 Sep 3.8 | +155.7 | 0.410 | 4.93 | H | ... | ... | 27.08 .00 | 26.40 .00 | 27.33 .00 | ... | 5.13 .00 | 5.09 .00 | ... |
| 1997 Sep 3.8 | +155.7 | 0.410 | 4.78 | H | ... | ... | 27.07 .00 | 26.76 .00 | 27.32 .00 | ... | 5.08 .00 | 5.09 .00 | ... |
| 1997 Oct 12.8 | +194.7 | 0.482 | 5.10 | H | ... | ... | 26.95 .00 | 26.58 .00 | 27.13 .00 | ... | 4.77 .00 | 4.80 .00 | ... |
| 1997 Oct 12.8 | +194.7 | 0.482 | 4.80 | H | ... | ... | 26.92 .00 | 26.58 .00 | 27.11 .00 | ... | 4.76 .00 | 4.79 .00 | ... |
| 1997 Oct 26.7 | +208.5 | 0.504 | 5.11 | H | ... | ... | 26.88 .00 | 26.57 .00 | 27.02 .00 | ... | 4.74 .00 | 4.76 .00 | ... |
| 1997 Oct 26.7 | +208.6 | 0.504 | 4.80 | H | ... | ... | 26.86 .00 | 26.50 .00 | 26.97 .01 | ... | 4.77 .00 | 4.79 .00 | ... |
| 1997 Dec 1.6 | +244.4 | 0.556 | 4.84 | H | ... | ... | 26.83 .00 | 26.42 .00 | 26.87 .01 | ... | 4.71 .00 | 4.78 .00 | ... |
| 1997 Dec 1.6 | +244.5 | 0.556 | 5.14 | H | ... | ... | 26.78 .00 | 26.45 .00 | 26.95 .00 | ... | 4.73 .00 | 4.76 .00 | ... |
| 1997 Dec 1.6 | +244.5 | 0.556 | 4.99 | H | ... | ... | 26.78 .00 | 26.39 .00 | 27.07 .00 | ... | 4.76 .00 | 4.78 .00 | ... |
| 1997 Dec 1.6 | +244.5 | 0.556 | 5.29 | H | ... | ... | 26.81 .00 | 26.48 .00 | 26.97 .00 | ... | 4.72 .00 | 4.75 .00 | ... |
| 1997 Dec 1.6 | +244.5 | 0.556 | 5.14 | H | ... | ... | 26.81 .00 | 26.45 .00 | 26.94 .00 | ... | 4.77 .00 | 4.78 .00 | ... |
| 1997 Dec 1.6 | +244.5 | 0.556 | 4.99 | H | ... | ... | 26.78 .00 | 26.40 .00 | 27.03 .00 | ... | 4.76 .00 | 4.77 .00 | ... |
| 1997 Dec 1.7 | +244.5 | 0.556 | 4.84 | H | ... | ... | 26.80 .00 | 26.38 .00 | 26.91 .00 | ... | 4.78 .00 | 4.80 .00 | ... |
| 1997 Dec 1.7 | +244.5 | 0.556 | 4.69 | H | ... | ... | 26.89 .00 | 26.58 .00 | 26.88 .01 | ... | 4.77 .00 | 4.79 .00 | ... |
| 1997 Dec 1.8 | +244.6 | 0.556 | 4.84 | H | ... | ... | 26.78 .00 | 26.31 .00 | 26.95 .00 | ... | 4.78 .00 | 4.78 .00 | ... |
| 1997 Dec 1.8 | +244.6 | 0.556 | 4.69 | H | ... | ... | 26.83 .00 | 26.52 .00 | 26.84 .01 | ... | 4.75 .00 | 4.77 .00 | ... |
| 1997 Dec 29.6 | +272.4 | 0.591 | 5.32 | H | ... | ... | 26.68 .00 | 26.41 .00 | 26.79 .00 | ... | 4.70 .00 | 4.73 .00 | ... |
| 1997 Dec 29.6 | +272.4 | 0.591 | 5.17 | H | ... | ... | 26.67 .00 | 26.40 .00 | 26.83 .00 | ... | 4.74 .00 | 4.76 .00 | ... |
| 1997 Dec 29.6 | +272.4 | 0.591 | 5.02 | H | ... | ... | 26.66 .00 | 26.36 .00 | 26.73 .01 | ... | 4.75 .00 | 4.78 .00 | ... |
| 1997 Dec 29.6 | +272.4 | 0.591 | 4.87 | H | ... | ... | 26.70 .00 | 26.36 .00 | 26.78 .01 | ... | 4.78 .00 | 4.81 .00 | ... |
| 1997 Dec 29.6 | +272.4 | 0.591 | 4.72 | H | ... | ... | 26.65 .00 | 26.53 .00 | 26.88 .01 | ... | 4.77 .00 | 4.80 .00 | ... |
| 1997 Dec 29.6 | +272.5 | 0.591 | 4.57 | H | ... | ... | 26.71 .00 | 26.60 .00 | 26.30 .06 | ... | 4.71 .00 | 4.82 .00 | ... |
| 1997 Dec 30.6 | +273.4 | 0.592 | 5.32 | H | ... | ... | 26.69 .00 | 26.46 .00 | 26.86 .00 | ... | 4.70 .00 | 4.73 .00 | ... |
| 1997 Dec 30.6 | +273.4 | 0.592 | 5.02 | H | ... | ... | 26.63 .00 | 26.46 .00 | 26.76 .01 | ... | 4.74 .00 | 4.77 .00 | ... |
| 1997 Dec 30.6 | +273.5 | 0.592 | 5.32 | H | ... | ... | 26.71 .00 | 26.32 .00 | 26.86 .00 | ... | 4.70 .00 | 4.71 .00 | ... |
| 1997 Dec 30.6 | +273.5 | 0.592 | 5.02 | H | ... | ... | 26.67 .00 | 26.41 .00 | 26.85 .01 | ... | 4.75 .00 | 4.77 .00 | ... |
| 1998 Jan 27.6 | +301.4 | 0.624 | 5.06 | H | ... | ... | 26.61 .00 | 26.33 .00 | 26.79 .01 | ... | 4.79 .00 | 4.81 .00 | ... |
| 1998 Jan 27.6 | +301.4 | 0.624 | 4.91 | H | ... | ... | 26.76 .00 | 26.31 .01 | 26.74 .01 | ... | 4.81 .00 | 4.83 .00 | ... |
| 1998 Jan 27.6 | +301.4 | 0.624 | 4.91 | H | ... | ... | 26.76 .00 | 26.20 .01 | 26.87 .01 | ... | 4.83 .00 | 4.84 .00 | ... |
| 1998 Jan 27.6 | +301.4 | 0.624 | 5.06 | H | ... | ... | 26.62 .00 | 26.30 .00 | 26.82 .01 | ... | 4.78 .00 | 4.80 .00 | ... |
| 1998 Mar 22.5 | +355.4 | 0.677 | 5.43 | H | ... | ... | 26.45 .00 | 26.03 .01 | 26.57 .01 | ... | 4.61 .00 | 4.65 .00 | ... |
| 1998 Mar 22.5 | +355.4 | 0.677 | 5.13 | H | ... | ... | 26.34 .00 | 26.21 .01 | 26.67 .01 | ... | 4.67 .00 | 4.69 .00 | ... |
| 1998 Mar 22.5 | +355.4 | 0.677 | 5.43 | H | ... | ... | 26.43 .00 | 26.20 .01 | 26.60 .01 | ... | 4.65 .00 | 4.68 .00 | ... |
| 1998 Mar 22.6 | +355.4 | 0.677 | 5.13 | H | ... | ... | 26.47 .00 | 26.20 .01 | 26.18 .05 | ... | 4.68 .00 | 4.73 .00 | ... |
| 1998 Apr 29.5 | +393.3 | 0.710 | 5.17 | H | ... | ... | 26.54 .01 | und | und | ... | 4.67 .00 | 4.77 .00 | ... |
| 1998 Apr 29.5 | +393.3 | 0.710 | 5.02 | H | ... | ... | 26.27 .01 | 26.41 .01 | 26.27 .07 | ... | 4.69 .00 | 4.73 .00 | ... |

Table 3—Continued

| UT Date | ΔT (day) | log $r_H$ (AU) | log $\rho$ (km) | Filt.[a] | log $Q^{b,c}$ (molecules s$^{-1}$) | | | | | log $A(\theta)f\rho^{b,c}$ (cm) | | | log $Q^b$ |
|---|---|---|---|---|---|---|---|---|---|---|---|---|---|
| | | | | | OH | NH | CN | $C_3$ | $C_2$ | UV | Blue | Green | $H_2O$ |
| 1998 Apr 29.5 | +393.4 | 0.710 | 5.17 | H | ... | ... | 26.42 .01 | 26.16 .01 | und | ... | 4.68 .00 | 4.76 .00 | ... |
| 1998 Oct 23.6 | +570.5 | 0.829 | 5.28 | H | ... | ... | 26.06 .02 | und | 26.92 .03 | ... | 4.54 .01 | 4.57 .01 | ... |
| 1998 Oct 23.6 | +570.5 | 0.829 | 5.28 | H | ... | ... | 25.97 .02 | und | 26.73 .04 | ... | 4.54 .01 | 4.59 .01 | ... |
| 1998 Oct 23.7 | +570.5 | 0.829 | 5.28 | H | ... | ... | und | und | und | ... | 4.55 .01 | 4.56 .01 | ... |
| 1998 Nov 15.7 | +593.5 | 0.841 | 5.42 | H | ... | ... | und | und | und | ... | 4.86 .00 | 4.63 .01 | ... |
| 1998 Nov 15.7 | +593.5 | 0.841 | 5.12 | H | ... | ... | 25.52 .05 | 26.15 .03 | 26.45 .09 | ... | 4.56 .01 | 4.61 .01 | ... |
| 1998 Nov 15.7 | +593.5 | 0.841 | 5.42 | H | ... | ... | 25.84 .02 | 25.80 .04 | 25.80 .18 | ... | 4.57 .01 | 4.61 .01 | ... |
| 1998 Nov 15.7 | +593.6 | 0.841 | 5.12 | H | ... | ... | und | 25.28 .16 | 25.55 .44 | ... | 4.60 .01 | 4.59 .01 | ... |
| 1998 Dec 16.6 | +624.5 | 0.857 | 5.59 | H | ... | ... | 26.04 .02 | 26.65 .02 | 25.63 .25 | ... | 4.43 .01 | 4.69 .01 | ... |
| 1998 Dec 16.6 | +624.5 | 0.857 | 5.29 | H | ... | ... | 25.81 .03 | 25.80 .06 | 26.69 .05 | ... | 4.63 .01 | 4.67 .01 | ... |
| 1998 Dec 16.7 | +624.5 | 0.857 | 5.59 | H | ... | ... | 25.92 .01 | 26.19 .02 | und | ... | 4.66 .01 | 4.81 .00 | ... |
| 1998 Dec 16.7 | +624.5 | 0.857 | 5.29 | H | ... | ... | 25.81 .02 | und | 26.93 .02 | ... | 4.67 .01 | 4.68 .00 | ... |
| 1998 Dec 20.6 | +628.4 | 0.859 | 5.14 | H | ... | ... | 25.92 .02 | 25.90 .04 | 26.74 .04 | ... | 4.53 .01 | 4.58 .01 | ... |
| 1998 Dec 20.6 | +628.4 | 0.859 | 4.99 | H | ... | ... | 24.96 .21 | 25.64 .09 | 27.00 .04 | ... | 4.48 .01 | 4.42 .01 | ... |
| 1999 Jan 18.6 | +657.4 | 0.873 | 5.45 | H | ... | ... | 25.95 .03 | 26.36 .02 | und | ... | 4.75 .01 | 4.86 .01 | ... |
| 1999 Jan 18.6 | +657.4 | 0.873 | 5.30 | H | ... | ... | und | 26.80 .02 | 26.76 .07 | ... | 4.72 .01 | 4.74 .01 | ... |
| 1999 Jan 19.6 | +658.4 | 0.874 | 5.60 | H | ... | ... | 25.95 .02 | und | und | ... | 4.79 .01 | 4.74 .01 | ... |
| 1999 Jan 19.6 | +658.4 | 0.874 | 5.30 | H | ... | ... | 26.57 .03 | und | und | ... | 4.85 .01 | 4.96 .01 | ... |
| 1999 Feb 9.6 | +679.4 | 0.884 | 5.61 | H | ... | ... | 25.91 .03 | 26.65 .02 | und | ... | 4.48 .01 | 4.80 .01 | ... |
| 1999 Feb 9.6 | +679.4 | 0.884 | 5.31 | H | ... | ... | und | 26.16 .04 | 27.15 .02 | ... | 4.71 .01 | 4.69 .01 | ... |
| 1999 Feb 10.6 | +680.4 | 0.884 | 5.46 | H | ... | ... | 25.96 .02 | 25.91 .06 | und | ... | 4.72 .01 | 4.78 .01 | ... |
| 1999 Feb 10.6 | +680.4 | 0.884 | 5.31 | H | ... | ... | 26.32 .02 | und | 27.07 .03 | ... | 4.68 .01 | 4.75 .01 | ... |
| 1999 Feb 10.6 | +680.4 | 0.884 | 5.46 | H | ... | ... | 25.89 .02 | und | und | ... | 4.74 .01 | 4.78 .01 | ... |
| 1999 Feb 10.6 | +680.5 | 0.884 | 5.31 | H | ... | ... | und | und | 27.26 .02 | ... | 4.73 .01 | 4.73 .01 | ... |
| 1999 Mar 17.5 | +715.4 | 0.900 | 5.63 | H | ... | ... | 26.85 .01 | und | 26.91 .04 | ... | 4.76 .01 | 4.62 .01 | ... |
| 1999 Mar 17.5 | +715.4 | 0.900 | 5.33 | H | ... | ... | und | 27.23 .02 | 27.14 .05 | ... | 4.70 .02 | 4.75 .01 | ... |
| 1999 Apr 17.5 | +746.3 | 0.913 | 5.64 | H | ... | ... | 26.45 .02 | und | und | ... | 4.79 .01 | 4.81 .01 | ... |
| 1999 Apr 17.5 | +746.3 | 0.913 | 5.34 | H | ... | ... | 26.07 .04 | und | 26.02 .36 | ... | 4.86 .01 | 4.81 .01 | ... |
| 1999 Apr 17.5 | +746.3 | 0.913 | 5.64 | H | ... | ... | 26.92 .01 | 24.82 .47 | und | ... | 4.71 .01 | 4.83 .01 | ... |
| 1999 Apr 17.5 | +746.4 | 0.913 | 5.34 | H | ... | ... | und | 26.08 .07 | und | ... | 4.89 .01 | 4.87 .01 | ... |
| 1999 Apr 18.5 | +747.3 | 0.914 | 5.64 | H | ... | ... | 26.28 .03 | und | und | ... | 4.72 .01 | 4.93 .01 | ... |
| 1999 Apr 18.5 | +747.4 | 0.914 | 5.34 | H | ... | ... | und | und | 27.31 .03 | ... | 4.86 .01 | 4.78 .01 | ... |
| 1999 Apr 18.6 | +747.4 | 0.914 | 5.64 | H | ... | ... | 26.73 .02 | und | 25.70 .43 | ... | 4.67 .02 | 4.84 .01 | ... |
| 1999 Apr 18.6 | +747.4 | 0.914 | 5.34 | H | ... | ... | und | 26.20 .07 | 27.19 .04 | ... | 4.85 .01 | 4.84 .01 | ... |
| 1999 Jun 9.4 | +799.3 | 0.935 | 5.66 | H | ... | ... | und | und | 27.14 .04 | ... | 5.31 .01 | 4.77 .01 | ... |
| 1999 Jun 9.4 | +799.3 | 0.935 | 5.36 | H | ... | ... | 26.15 .14 | und | und | ... | 4.57 .03 | 4.79 .01 | ... |
| 1999 Jun 13.4 | +803.3 | 0.936 | 5.66 | H | ... | ... | und | und | 26.78 .07 | ... | 5.24 .01 | 4.85 .01 | ... |
| 1999 Jun 13.5 | +803.3 | 0.936 | 5.66 | H | ... | ... | 25.07 .25 | 26.97 .03 | und | ... | 4.81 .01 | 4.96 .01 | ... |
| 1999 Jul 2.5 | +822.3 | 0.944 | 5.39 | H | ... | ... | und | 26.63 .06 | und | ... | 4.80 .02 | 4.85 .01 | ... |
| 1999 Jul 2.5 | +822.3 | 0.944 | 5.39 | H | ... | ... | und | 27.05 .04 | 25.99 .44 | ... | 4.70 .02 | 4.78 .01 | ... |
| 1999 Jul 6.5 | +826.3 | 0.945 | 5.69 | H | ... | ... | 25.04 .43 | und | 27.27 .04 | ... | 4.86 .02 | 4.42 .04 | ... |
| 1999 Jul 6.5 | +826.3 | 0.945 | 5.69 | H | ... | ... | 26.15 .06 | 27.08 .03 | 27.38 .02 | ... | 4.74 .02 | 4.58 .02 | ... |
| 1999 Sep 17.8 | +899.7 | 0.972 | 5.70 | H | ... | ... | und | und | und | ... | 4.94 .01 | 4.68 .02 | ... |
| 1999 Sep 17.8 | +899.7 | 0.972 | 5.70 | H | ... | ... | und | und | 26.93 .07 | ... | 4.89 .01 | 4.63 .02 | ... |
| 1999 Sep 17.8 | +899.7 | 0.972 | 5.70 | H | ... | ... | und | 26.96 .03 | und | ... | 4.87 .01 | 4.77 .02 | ... |

Table 3—Continued

| UT Date | ΔT (day) | log $r_H$ (AU) | log $\rho$ (km) | Filt.[a] | log $Q^{b,c}$ (molecules s$^{-1}$) OH | NH | CN | $C_3$ | $C_2$ | log $A(\theta)f\rho^{b,c}$ (cm) UV | Blue | Green | log $Q^b$ $H_2O$ |
|---|---|---|---|---|---|---|---|---|---|---|---|---|---|
| 1999 Sep 17.8 | +899.7 | 0.972 | 5.70 | H | ... | ... | und | 26.64 .04 | und | ... | 4.95 .01 | 4.80 .01 | ... |
| 1999 Dec 6.6 | +979.4 | 0.998 | 5.73 | H | ... | ... | 26.00 .07 | 26.81 .05 | 26.82 .08 | ... | 4.32 .04 | 4.63 .02 | ... |
| 1999 Dec 6.6 | +979.4 | 0.998 | 5.43 | H | ... | ... | 26.72 .03 | und | und | ... | 4.67 .02 | 4.76 .01 | ... |
| 1999 Dec 6.6 | +979.5 | 0.998 | 5.73 | H | ... | ... | 26.28 .03 | 26.49 .05 | 26.85 .04 | ... | 4.36 .03 | 4.70 .01 | ... |
| 1999 Dec 6.6 | +979.5 | 0.998 | 5.43 | H | ... | ... | 26.40 .02 | und | 26.30 .21 | ... | 4.78 .01 | 4.76 .01 | ... |
| 1999 Dec 7.5 | +980.4 | 0.998 | 5.73 | H | ... | ... | 25.48 .12 | 26.35 .04 | 27.35 .02 | ... | 4.58 .02 | 4.32 .02 | ... |
| 1999 Dec 7.6 | +980.4 | 0.998 | 5.43 | H | ... | ... | 26.52 .02 | und | 25.10 .99 | ... | 4.67 .01 | 4.77 .01 | ... |
| 1999 Dec 7.6 | +980.5 | 0.998 | 5.73 | H | ... | ... | und | und | 26.06 .29 | ... | 4.73 .02 | 4.68 .01 | ... |
| 1999 Dec 7.6 | +980.5 | 0.998 | 5.43 | H | ... | ... | und | und | 25.91 .49 | ... | 4.73 .01 | 4.77 .01 | ... |
| 2000 Feb 1.6 | +1036.4 | 1.015 | 5.47 | H | ... | ... | 26.26 .04 | und | und | ... | 4.73 .02 | 4.69 .02 | ... |
| 2000 Feb 1.6 | +1036.4 | 1.015 | 5.47 | H | ... | ... | und | 25.49 .22 | und | ... | 4.64 .02 | 4.68 .01 | ... |
| 2000 Feb 2.5 | +1037.4 | 1.016 | 5.77 | H | ... | ... | 25.39 .19 | 26.73 .04 | 26.90 .05 | ... | 4.39 .03 | 4.47 .02 | ... |
| 2000 Feb 2.5 | +1037.4 | 1.016 | 5.47 | H | ... | ... | und | und | 26.73 .11 | ... | 4.62 .02 | 4.50 .02 | ... |
| 2000 Feb 2.6 | +1037.4 | 1.016 | 5.77 | H | ... | ... | und | 26.90 .03 | 26.76 .07 | ... | 4.26 .04 | 4.55 .02 | ... |
| 2000 Feb 2.6 | +1037.4 | 1.016 | 5.47 | H | ... | ... | und | 26.18 .07 | 26.86 .08 | ... | 4.65 .01 | 4.51 .02 | ... |
| 2000 Mar 3.5 | +1067.4 | 1.024 | 5.78 | H | ... | ... | 26.63 .02 | ... | und | ... | 4.52 .01 | 4.75 .01 | ... |
| 2000 Mar 3.5 | +1067.4 | 1.024 | 5.78 | H | ... | ... | 26.63 .02 | ... | 26.63 .06 | ... | 4.52 .01 | 4.46 .01 | ... |

[a]Filter ID: I = IHW; H = HB.

[b]"und" stands for "undefined". For the gases, this means that the emission flux was measured but was less than zero following sky and continuum removal. For the continuum, this means the continuum flux was measured but following sky subtraction it was less than zero.

[c]Production rates followed by the upper, i.e. the positive, uncertainty. The "+" and "−" uncertainties are equal as percentages, but unequal in log-space; the "−" values can be computed.

Table 4. $r_H$-Dependence of the Production Rates for Comet C/Hale-Bopp

| | Slope[a] | |
|---|---|---|
| Species | pre-$q$ | post-$q$ |
| OH | −2.44 ± .06 | ... |
| NH | −1.25 ± .07 | ... |
| CN | −1.82 ± .05 | −2.16 ± .09 |
| $C_3$ | −1.79 ± .07 | −1.85 ± .23 |
| $C_2$ | −2.74 ± .07 | −2.67 ± .27 |
| Dust[b] | −2.15 ± .05 | −1.45 ± .12 |

[a]Slopes for data between 1.2 and 4.0 AU in log-log space.

[b]Phase-adjusted green continuum.